\documentclass[3p,12pt]{elsarticle}
\usepackage[fleqn]{amsmath}
\usepackage{cases}
\usepackage{booktabs}
\usepackage{multirow}
\usepackage{xcolor}
\usepackage[normalem]{ulem}
\usepackage{tabularx}
\usepackage{siunitx}
\usepackage[caption=false]{subfig}
\usepackage{comment}
\usepackage[ruled, linesnumbered]{algorithm2e}
\usepackage{algpseudocode}
\usepackage{grffile}
\usepackage{rotating}
\usepackage{lineno}
\usepackage{hyperref}
\usepackage{graphicx,enumerate}
\usepackage{amssymb}
\allowdisplaybreaks[4]
\usepackage{bm}
\usepackage[caption=false]{subfig}
\usepackage{caption,color}
\usepackage{subeqnarray}
\usepackage{multirow}
\usepackage{lscape}
\usepackage{rotating}
\usepackage{graphics}
\usepackage{epstopdf}
\usepackage{threeparttable,booktabs,tabularx}
\journal{Journal of Computational Physics}
\usepackage{floatrow}
\usepackage{floatrow}\newfloatcommand{capbtabbox}{table}[][\FBwidth]
\floatname{algorithm}{Algorithm}
\biboptions{numbers,sort&compress} % 压制引用，[1,2,3]--> [1-3]

\newcommand{\bc}{\bm c}

\newcommand{\bq}{\bm q}

\newcommand{\bu}{\bm u}

\newcommand{\bx}{\bm x}

\newcommand{\bxi}{\bm \xi}

\newcommand{\Kn}{\text{Kn}}

\newcommand{\myd}{\;\mathrm{d} }

\begin{document}
\begin{frontmatter}
\title{General synthetic iterative scheme for rarefied gas mixture flows}

\author{Jianan Zeng}
\author{Qi Li}
\author{Lei Wu\corref{mycorrespondingauthor}}
\cortext[mycorrespondingauthor]{Corresponding author}
\ead{wul@sustech.edu.cn}

\address{Department of Mechanics and Aerospace Engineering, Southern University of Science and Technology, Shenzhen 518055, China}

\begin{abstract}
The numerical simulation of rarefied gas mixtures with disparate mass and concentration is a huge research challenge. Based on our recent kinetic modelling for monatomic gas mixture flows, this problem is tackled by the general synthetic iterative scheme (GSIS), where
the mesoscopic kinetic and macroscopic synthetic equations are alternately solved by the finite-volume discrete velocity method. Three important features of GSIS are highlighted. First, 
the synthetic equations are precisely derived from the kinetic equation, naturally reducing to the Navier-Stokes equations in the continuum flow regime; in other flow regimes, the kinetic equation provides high-order closure of the constitutive relations to capture the rarefaction effects. 
Second, these synthetic equations, which can be solved quickly, help to adjust the kinetic system to relax rapidly toward the steady state. Furthermore, in such a two-way coupling, the constraint on the spatial cell size is relieved. 
Third, the linear Fourier stability analysis demonstrates that the error decay rate in GSIS is smaller than 0.5 for various combinations of mass, concentration and viscosity ratios, such that the error can be reduced by three orders of magnitude after 10 iterations. 
The efficiency and accuracy of GSIS are demonstrated through several challenging cases covering a wide range of mass ratio, species concentration, and flow speed.  % including the normal shock wave, two-dimensional supersonic flow past a cylinder, three-dimensional nozzle flow, and the two-dimensional low-speed channel flow
\end{abstract}


\end{frontmatter}
\section{Introduction}\label{sec:1}

% extreme ultraviolet

Gas mixtures are widely encountered in modern engineering problems, such as astronautics \cite{votta2013hypersonic}, vacuum technologies \cite{benschop2008extreme,fu2019euv}, and micro-electromechanical systems \cite{Beskok_book}. These flows often exhibit rarefaction effects, where the Knudsen number (i.e., the ratio of the molecular mean free path to the characteristic length) is not negligible. 
For example, in the dynamic gas lock used in EUV lithography,  clean gas is continuously injected to reduce the partial pressure of contaminants inside the vacuum optical cavity~\cite{teng2023pollutant}. The molecular masses in the mixture (e.g., $\text{H}_2$ and hydrocarbons) could differ by two orders of magnitude, while the species concentration could differ by seven orders of magnitude. Consequently, the Knudsen numbers of the flow span multiple orders of magnitude. Due to the experimental challenges in low-pressure environment, numerical simulations become crucial for understanding and optimizing rarefied gas flow problems. Under these extreme conditions,
the linear constitutive relations (i.e., Newton's law of viscosity and Fourier's law of heat conduction) in the Navier-Stokes (NS) equation are only applicable in the continuum flow regime. Instead, the Boltzmann equation from the gas kinetic theory should be used, which is applicable in the continuum, slip, transition, and free-molecular flow regimes. 

Many numerical methods have been developed to solve the Boltzmann equation, with the prevailing one being the direct simulation Monte Carlo (DSMC) method~\cite{bird1994molecular}. However, DSMC encounters several difficulties in simulating gas mixture flows. First, as in the simulation of single-species flows, the splitting of streaming and collision makes the scheme expansive or even prohibitive in the continuum flow regime, as the spatial cell size and time step must be smaller than the molecular mean free path and mean collision time, respectively.  
Second, for gas mixtures with disparate mass, the thermal speed of each species differ significantly. The simulation time step is limited by that of the lighter species, resulting in heavier species taking substantially long time to reach steady state. 
Third, for gas mixtures with disparate concentration, DSMC requires different weight factors for each species. As commented by Alves \textit{et al.}~\cite{Alves2018PSST}, this approach ``introduces problems such as enforcing conservation laws in collisions between particles of different weights, creating and destroying fractions of particles, and possibly degradation of the numerical properties''.
These limitations have prompted researchers to explore new numerical methods, such as the Fokker-Planck DSMC~\cite{Gorji2011JCP,kuchlin2018stochastic}, the exponential BGK/ESBGK~\cite{pfeiffer2022exponential}, and the time-relaxed Monte Carlo method that preserves the Navier-Stokes asymptotics in the continuum flow regime~\cite{fei2023usp}. %The Fokker-Planck method considers the influence of particle collisions on particle velocity during the particle migration process, and is more efficient than DSMC in the continuum regime. Additionally, a unified stochastic particle ESBGK (USP-ESBGK) method has been developed to address this issue. By integrating collision effects into the molecular transport step, the USP-ESBGK method also exhibits excellent multiscale characteristics.

The Boltzmann equation can also be solved by the deterministic discrete velocity method (DVM). Since the statistical noise is absent, the disparate concentration problem is naturally solved. Furthermore, when the steady-state is concerned, the time derivative can be dropped, and the problem of time step is solved. However, the Boltzmann collision integral is very hard to solve. So far, the collision integral with arbitrary inter-molecular potential has only been solved by the fast spectral method~\cite{Lei2013,lei_Jfm,wuPoF2015,SU2019573} and the conservative projection method~\cite{Tcheremissine2005,Tcheremissine2006}.  For gas mixtures, even using the fast spectral method, the numerical simulation is only carried out up to molecular mass ratio 36~\citep{Wu2015JCP}.  
Therefore, many kinetic models have been proposed to simplify the Boltzmann collision operator, such as  the McCormack model~\cite{McCormack1973}, the {Andries-Aoki-Perthame}  model~\cite{Andries2002JSP}, and other similar kinetic models \cite{Kosuge2009EJMB,Brull2015CMS,Groppi2011EPL,Todorova2019EJMB}. 
Although the Boltzmann collision operator is simplified, it is worth noting that DVM solvers are typically expensive because they need the additional discretization of velocity space. When the steady state solution is concerned, the iteration scheme is frequently used, due to the high-dimensional integro-differential nature of the of kinetic equations; the traditional DVM becomes the conventional iteration scheme (CIS). 
Similar to the DSMC method, the streaming and collision are decoupled {in CIS}, so that the restrictions on cell size and time step limit the applicability of CIS in the (near) continuum flow regime. Hence, the development of efficient numerical methods to simulate multiscale flow problems is urgently needed. 

In recent years, we have seen a huge stride in the development of multiscale numerical methods. For instances, 
the (discrete) unified gas kinetic scheme \cite{xu2010unified,guo2013discrete} handled the streaming and collision simultaneously, so that it
possesses the asymptotic preserving property and accurately recovers the NS solutions without requiring spatial grid sizes (time steps) smaller than the mean free path (mean collision time). It has been successfully applied to many single-species flow problems~\cite{yuan2020implicit,zhu2017multigrid,wang2017unified}, as well as gas mixture flows~\cite{zhang2019discrete,zhang2018discrete}.

The general synthetic iteration scheme (GSIS) is another multiscale DVM, where the kinetic equation and the corresponding synthetic equations are alternately solved~\cite{SuArXiv2019}. The constitutive relations in the synthetic equation consist of the linear ones used in the NS equation, and the higher-order terms (HoTs) that are directly calculated from the velocity distribution function. This essential idea of mesoscopic-macroscopic coupling was firstly developed in neutron transport (see the review article~\cite{DSA2002}); it was introduced to special rarefied gas flows (e.g. the planar Poiseuille flow where the flow velocity is perpendicular to the spatial coordinate~\cite{Valougeorgis:2003zr}) and finally extended to general rarefied gas flows. 
Rigorous mathematical analysis and many numerical tests~\cite{SU2020SIAM, Zhu2021JCP,su2021multiscale,zeng2023general,liu2022fast} have shown that the GSIS can significantly reduce the number of iterations. Moreover, where the Knudsen number is small, the HoTs tend to zero, ensuring the asymptotic-preserving property of the GSIS when the spatial cell size is much larger than the mean free path; when the Knudsen number is large, rarefaction effects are captured by the higher-order constitutive relations.
Although the DVM solvers are expensive due to the use of huge number of discrete velocities, in the recent work we have developed a parallel strategy and employed unstructured velocity space to further reduce the memory usage, enabling efficient simulation of rarefied gas flows~\cite{zhang2023efficient}; now this approach can be even faster than the particle method (adaptive unified gas-kinetic wave-particle method) in simulating high-speed multiscale flows, not to mention the low-speed flows~\cite{wei2023adaptive}. 

Despite of the practical engineering importance, there are not enough works on the efficient simulation of rarefied gas mixture flows. Most are limited to low-speed linearized flows \cite{siewert2001couette,naris2004discrete,Ho2016Wu} or simple two-dimensional planar and axisymmetric geometries~\cite{naris2004gaseous,  zhang2018discrete,zhang2019discrete}. The aim of this work is to extend the GSIS to gas mixture flows, across the whole range of Knudsen number, and over wide ranges of mass ratio and concentration ratio. 

The remaining paper are organized as follows. In section~\ref{sec:2} the Li kinetic model is introduced and the macroscopic synthetic equation in GSIS is constructed. In section \ref{sec:stability}, the Fourier stability analysis is performed, showing that the GSIS facilitates fast convergence. In
section \ref{sec:3}, numerical schemes for both the mesoscopic and macroscopic equations are described. The accuracy and efficiency of GSIS in validated in four challenging numerical examples in section \ref{sec:4}. Finally, conclusions are given in section \ref{sec:5}.
\section{Gas kinetic equations and macroscopic synthetic equations}\label{sec:2}

In the present work, the Li kinetic model is adopted to describe the dynamics of gas mixtures, where the intra- and inter-species collisions are modeled separately, and the model parameters are chosen to recover the transport properties of mixture gases, including the viscosity, thermal conductivity, diffusion and thermal diffusion coefficients~\cite{li2024kinetic}. This kinetic model has shown a good accuracy when the mass ratio spans three orders of magnitude, when compared to the DSMC results in several benchmark simulations.

\subsection{Kinetic model equation}
 
In order to express all the variables in the dimensionless forms, we denote $L_0, T_0, n_0, m_0$ as the reference length, temperature, number density, and molecular mass, respectively, and let $v_0 = \sqrt{k_B T_0/m_0}$ be the reference velocity with $k_B$ being the Boltzmann constant. We also define the Knudsen number $\text{Kn}_{s}$ of each species $s$ to quantify the degree of rarefaction of the gas mixture,
\begin{equation}\label{eq:Kn_define}
\text{Kn}_{s} = \frac{\mu_s(T_0)}{n_0 L_0} \sqrt{\frac{\pi}{2m_s k_B T_0} },
     % \qquad \text{Kn}_{r} =\text{Kn}_{s}\frac{\mu_{r}(T_0)}{\mu_{s}(T_0)}\sqrt{\frac{m_s}{m_r}}. 
\end{equation}
where $m_s$ is the molecular mass of species $s$, and $\mu_s$ is the species shear viscosity; when the inverse power-law potential is considered, the viscosity follows the relation $\mu_s(T_s) = \mu_s(T_0) ({T}_s/T_0)^{\omega_s}$ with $\omega_s$ being the viscosity index. For Maxwell and hard-sphere gases, we have $\omega=1$ and 0.5, respectively. 

%Note that the Knudsen number of the species $r$ is correlated as $\text{Kn}_{r}/\text{Kn}_{s}=\mu_r(T_0)/\mu_s(T_0)\sqrt{m_s/m_r}$. Thus, all the variables used in the following paper are normalized by the corresponding reference values.

Consider a mixture of monatomic gases with the velocity distribution function $f_s(\bx, \bxi, t)$ of each species describing their states, where $t$ is the time, $\bm{x}$ is the spatial coordinate, $\bm{\xi}$ is the molecular velocity. The macroscopic variables of each species, such as the number density $n_s$, mass density $\rho_s$, flow velocity $\bm{u}_s$, temperatures $T_s$, pressure tensor $\bm{P}_s$, and heat flux $\bm{q}_s$, are obtained by taking the moments of the respective distribution function $f_s$,
\begin{equation}\label{eq:getmoment}
\begin{aligned}
    &n_s = \langle \frac{1}{m_s},f_s \rangle,\quad
    \rho_s = \langle 1,f_s \rangle,\quad
    \rho_s\bm{u}_s = \langle {\bxi},f_s \rangle, \quad
    \frac{3}{2}n_s T_s=\left\langle \frac{1}{2}(\bxi-\bm{u}_s)^2, f_s \right\rangle,\\
    &\bm{P}_s =\langle (\bxi-\bm{u}_s) (\bxi-\bm{u}_s), f_s \rangle,\quad
    \bm{q}_s =\left\langle \frac{1}{2}(\bxi-\bm{u}_s)^2 (\bxi- \bm{u}_s), f_s \right\rangle,
\end{aligned}
\end{equation}
where the operator $\langle h,\psi\rangle=\int h\psi \myd \bxi$ is defined as an integral of $h\psi$ over the velocity space.

Then, the corresponding macroscopic quantities for the mixture, such as the number density $n$, mass density $\rho$ and velocity $\bu$ can be calculated as,
\begin{equation}
\begin{aligned}
n &= \sum_{s}\langle  \frac{1}{m_s} ,f_s  \rangle = \sum_{s} n_s,\quad
\rho = \sum_{s}\langle  1 ,f_s  \rangle = \sum_{s} \rho_s,\quad
\rho \bm{u}  =\sum_s\langle{\bxi} ,f_s \rangle =\sum_{s}\rho_s \bm{u}_s,
\end{aligned}
\end{equation}
and the mixture temperatures $T$, pressure tensor $\bm{P}$, and heat flux $\bq$ are given by,
\begin{equation}
\begin{aligned}
\frac{3}{2}n T &= \sum_{s} \left\langle \frac{1}{2}(\bxi-\bm{u}_s)^2 ,f_s \right\rangle = \sum_{s}\frac{3}{2}n_s T_s + \frac{1}{2}\sum_{s}\rho_s (\bm{u}_s - \bm{u})^2,\\
\bm{P} &=\sum_{s} \langle (\bxi-\bm{u}_s) (\bxi-\bm{u}_s) ,f_s  \rangle = \sum_{s} \bm{P}_s + \sum_{s} \rho_{s}(\bm{u}_s - \bm{u})(\bm{u}_s - \bm{u}),\\
\bm{q} &= \sum_{s} \left\langle  \frac{1}{2}(\bxi-\bm{u}_s)^2 (\bxi-\bm{u}_s) ,f_s \right\rangle\\
&=\sum_{s} \bm{q}_s + \sum_{s} \frac{3}{2} n_s T_s(\bm{u}_s - \bm{u}) + \frac{1}{2}\sum_{s} \rho_s (\bm{u}_s - \bm{u}) ^2(\bm{u}_s - \bm{u})  + \sum_{s} \bm{P}_s \cdot (\bm{u}_s - \bm{u}).
\end{aligned}
\end{equation}

The evolution of the velocity distribution functions in a gas mixture is modeled by the following kinetic equations~\cite{li2024kinetic}:
\begin{equation}\label{eq:generalModel}
\frac{\partial f_s}{\partial t}+\bm{\xi}\cdot \frac{\partial f_s}{\partial \bm{x}} = \sum_{r} \frac{g_{sr}-f_s}{\tau_{sr}}, 
\end{equation}
where $s=r$ indicates an intra-species collision operator, $s\ne r$ are the inter-species collisions, and $\tau_{ss}, \tau_{sr}$ are the corresponding dimensionless relaxation times, which can be written in terms of the species Knudsen number as,
\begin{equation}
     \tau_{ss}=\text{Kn}_{s}\sqrt{\frac{2 m_s}{\pi}}\frac{{T_s}^{\omega_s-1}}{{n}_s}, \qquad \tau_{sr} = \tau_{ss}\phi_{sr}^{-1}\frac{ n_s}{ n_r}.
\end{equation}
Here, $\phi_{sr}$ represents the ratio between the relaxation times of intra- and inter-species collisions for species $s$, and can be determined by matching the mixture shear viscosity, see Table~\ref{tab04:mixture_gas_perp}. The reference distribution function $g_{sr}$ is constructed as,
\begin{equation}\label{reference distribution function}
\begin{aligned}[b]
g_{sr}=&m_s\hat{n}_{sr}\left(\frac{m_s}{2\pi T_{s}}\right)^{3/2}\exp\left(-\frac{m_s\left(\bm{\xi}-\hat{\bm{u}}_{sr}\right)^2}{2T_{s}}\right) \\
&\times\left[1+\frac{\hat T_{sr} - T_s}{T_s}\left(\frac{m_s\left(\bm{\xi}-\hat{\bm{u}}_{sr}\right)^2}{2 T_{s}}-\frac{3}{2}\right)+\frac{2m_s\bm{\hat q}_{sr} \cdot \left(\bm{\xi}-\hat{\bm{u}}_{sr}\right)}{5 \hat n_{sr}T_{s}^2}\left(\frac{m_s \left(\bm{\xi}-\hat{\bm{u}}_{sr}\right)^2}{2 T_{s}}-\frac{5}{2}\right)\right],
\end{aligned}
\end{equation}
where the variables with hat accent are auxiliary macroscopic properties. %$\hat n_{sr}, \hat\rho_{sr}, \hat\bu_{sr}, \hat T_{sr}, \hat\bq_{sr}$ are defined as auxiliary number density, mass density, velocity, temperature and heat flux, respectively. 

In the intra-species collision operators ($s=r$), the auxiliary number density, flow velocity and temperature are simply the macroscopic properties of each species due to conservations of mass, momentum and energy within intra-species interactions, respectively,
\begin{equation}\label{eq:aux_para_intra}
   \hat n_{ss} = n_s,\quad
    \hat {\bm{u}}_{ss} = {\bm{u}}_s,\quad
    \hat T_{ss} = T_s.\quad
    % \bm{\hat q}_{ss} = (1-{\Pr}_{ss}) \bq_s.
\end{equation}
Thus, the reference distribution function $g_{ss}$ in the intra-species collision operator reduces to that of the Shakhov-type model of single-species gas flow.

In the inter-species collision operators ($s \ne r$), the auxiliary velocity and temperature cannot be uniquely determined by the conservation laws, but requires additional constraints. Generally speaking, the auxiliary flow velocity $\hat {\bm{u}}_{sr}$ and the auxiliary temperature $\hat T_{sr}$ are constructed to recover the correct diffusion and energy relaxation processes between species, respectively, which are calculated as,
\begin{equation}\label{eq:aux_para_inter}
\begin{aligned}
\hat n_{sr} &= n_s, \\
\hat {\bm{u}}_{sr} &= \bu_s - \frac{\rho_r \tau_{sr}}{\rho_s \tau_{rs} + \rho_r\tau_{sr} }\bm{X}_{sr},\\
\hat {T}_{sr} &= T_s - \frac{n_r \tau_{sr}}{n_s \tau_{rs} + n_r\tau_{sr} }Y_{sr} - \frac{\rho_s \rho_r \tau_{sr} \tau_{rs} \bm{X}_{sr}\cdot [\bm{X}_{sr} - 2(\bm{u_s} - \bm{u}_r)]}{3(n_s\tau_{rs} + n_r \tau_{sr})(\rho_s \tau_{rs} + \rho_r\tau_{sr})},\\
% \bm{\hat q}_{sr} &= (1-\text{Pr}_{sr} ) \bq_s + \gamma_{sr} (\bq_{sr}-\bq_s), \qquad s\neq r,
\end{aligned}
\end{equation}
with
\begin{equation}
\begin{aligned}
\bm{X}_{sr} & = a_{sr}(\bm{u}_s - \bm{u}_r) + b_{sr} (\nabla \ln T_s + \nabla \ln T_r),\\
Y_{sr} & = c_{sr}(T_s - T_r) + d_{sr} (\bm{u}_s - \bm{u}_r)^2,\\
\end{aligned}
\end{equation}
where $a_{sr} = a_{rs}, b_{sr} = -b_{rs}, c_{sr} = c_{rs}, d_{sr} = -d_{rs}$ are the adjustable parameters, describing how rapidly the equilibrium among different gas species is achieved through inter-species collisions, and can be determined by the transport properties as,
\begin{equation}
    \begin{aligned}
        a_{sr} &= \frac{T\left(\rho_s\tau_{rs}+\rho_r\tau_{sr}\right)}{m_sm_r(n_s+n_r)D_{sr}}, \\
        b_{sr} &= \frac{(n_s+n_r)T\left(\rho_s\tau_{rs} + \rho_r\tau_{sr}\right)k_{T,sr}}{2\rho_s\rho_r}, \\
        c_{sr} &= \frac{2a_{sr}\left(n_s\tau_{rs}+n_r\tau_{sr}\right)m_sm_r}{(m_s+m_r)\left(\rho_s\tau_{rs}+\rho_r\tau_{sr}\right)}, \\
        d_{sr} &= \frac{a_{sr}m_sm_r}{3\left(\rho_s\tau_{rs}+\rho_r\tau_{sr}\right)}\left[\frac{a_{sr}\left(n_r\rho_r\tau_{sr}^2-n_s\rho_s\tau_{rs}^2\right)}{\rho_s\tau_{rs}+\rho_r\tau_{sr}} - \frac{2\left(\rho_r\tau_{sr}-\rho_s\tau_{rs}\right)}{m_s+m_r}\right],
    \end{aligned}
\end{equation}
where $D_{sr}$ is the binary diffusion coefficient, and $k_{T,sr}$ is the thermal-diffusion ratio.

In addition, the auxiliary heat fluxes are constructed to recover the correct thermal conductivity,
\begin{equation}\label{eq:aux_heatflux}
    \begin{aligned}
        \hat{\bm{q}}_{ss} &= (1-\text{Pr}) \bq_s, \\
        \hat{\bm{q}}_{sr} &= (1-\varphi_{sr}\text{Pr} ) \bq_s + \gamma_{sr} (\bq_{sr}-\bq_s), 
    \end{aligned}
\end{equation}
where $\Pr$ is the Prandtl number, $\varphi_{sr}$ measures the ratio of thermal relaxation rates between the inter- and intra-species collisions, see Table~\ref{tab04:mixture_gas_perp}; $\bq_{sr}$ is defined as the heat flux of species $s$ measured relative to auxiliary velocity $\bm{\hat u}_{sr}$,
\begin{equation}
    \bm{ q}_{sr} =\left\langle \frac{1}{2}(\bxi-\bm{\hat u}_{sr})^2 (\bxi- \bm{\hat u}_{sr}), f_s \right\rangle.
\end{equation}
Finally, $\gamma_{sr} = -\gamma_{rs}$ is a dimensionless coefficient taking into account the Dufour effects, which can be regarded as an inverse process to thermal diffusion and thus yields, 
\begin{equation}
\begin{aligned}
\gamma_{sr} &=\left(\frac{1}{m_s}\frac{\tau_{ss}\tau_{sr}}{\varphi_{sr}\tau_{ss}+\tau_{sr}}+\frac{1}{m_r}\frac{\tau_{rr}\tau_{rs}}{\varphi_{rs}\tau_{rr}+\tau_{rs}}\right)^{-1}\frac{4b_{sr}}{5a_{sr}\Pr T},
\end{aligned}
\end{equation}

Given the relationship between the model parameters and transport coefficients of the mixtures in the continuum flow limit, the adjustable parameters can be uniquely determined by the properties of gas mixtures directly. In the present work, four representative binary gas mixtures will be considered, where the parameters are detailed in Table~\ref{tab04:mixture_gas_perp}.

\begin{table}[t]
    \centering
    \caption{The constituents of the four binary mixtures, and the corresponding parameters $\phi_{sr}$ and $\varphi_{sr}$ in the kinetic model fitted by matching the mixture viscosity and thermal conductivity from the DSMC with the VSS collision model, respectively \cite{li2024kinetic}. Note that the ratio of species viscosity in a mixture satisfies $\mu_2/\mu_1=\sqrt{m_2/m_1}\left(d_1/d_2\right)^2$, while the Knudsen number ratio is $\Kn_2/\Kn_1=(d_1/d_2)^2$.}
 \begin{tabular}{c c c c c c c c c c}\toprule
   Mixture&Gas type &$m_2/m_1$  & $d_2/d_1$ & $\omega_{12}$ & $\alpha_{12}$ & $\phi_{12}$& $\phi_{21}$& $\varphi_{12}$& $\varphi_{21}$\\ \hline 
   1& {Maxwell} &10& 1 &{1.0} &{2.14} & 1.214& 0.515 & 1.035 & 1.779 \\ 
   2& Maxwell &100 & 3 &{1.0} &{2.14} & 5.505& 0.078 & 0.988 & 2.093 \\ 
   3& Maxwell &1000 & 1 &{1.0} &{2.14} & 1.367& 0.057 & 0.999 & 2.259 \\ 
%   \hline
   4& {Hard-sphere} &{100} & {2} & {0.5} & {1.0} & {2.955}& {0.127} & {1.425} & {1.261}\\
   \bottomrule
   \end{tabular}
   \label{tab04:mixture_gas_perp}
\end{table}

\subsection{Macroscopic synthetic equations}

The synthetic equations for macroscopic properties, which asymptotically preserve the NS limit, are developed by combining conservation equations of species mass and relaxation equations of inter-species momentum and energy exchanges. By taking the velocity moments of the kinetic equation~\eqref{eq:generalModel}, the governing equations for macroscopic properties $\rho_s, \bm{u}_s, T_s$ are obtained,
\begin{equation}\label{eq:macro_equation}
\begin{aligned}
\frac{\partial{\rho_s}}{\partial{t}} + \nabla\cdot\left(\rho_s\bm{u}_s\right)  &= 0, \\
    \frac{\partial}{\partial{t}}\left(\rho_s\bm{u}_s\right) + \nabla\cdot\left(\rho_s\bm{u}_s\bm{u}_s\right) + \nabla\cdot (n_s T_s \bm{\mathrm{I}}+ \bm{\sigma}_s) &= \bm{Q}_s^M, \\
    \frac{\partial}{\partial{t}}\left(E_s\right) + \nabla\cdot\left(E_s\bm{u}_s\right) + \nabla\cdot\left[(n_s T_s \bm{\mathrm{I}}+ \bm{\sigma}_s)\cdot\bm{u}_s+\bm{q}_s\right] &= {Q}_s^E,
\end{aligned}
\end{equation}
where ${\bm{\sigma}_s}=\bm{P}_s-n_sT_s\bm{\mathrm{I}}$ is the species shear stress; the source terms $\bm{Q}_s^M$ and ${Q}_s^E$ represent the momentum and energy exchange between species, respectively, and satisfy $\sum_s\bm{Q}_s^M = 0$ and $\sum_s{Q}_s^E = 0$ due to the conservation of total momentum and energy. According to the kinetic equation, the source terms can be obtained as,
\begin{equation}\label{momentum_exchange}
    \begin{aligned}
        \bm{Q}_s^M = \sum_{r\ne s} \frac{\rho_s(\hat{\bm{u}}_{sr}-\bm{u}_s)}{\tau_{sr}},\quad
        \bm{Q}_s^E = \sum_{r\ne s} \frac{E_{sr}-E_s}{\tau_{sr}},
    \end{aligned}
\end{equation}
where $E_s$ and $E_{sr}$ denote the kinetic energies,
\begin{equation}
        E_s = \frac{3}{2}n_sT_s + \frac{1}{2}\rho_s u_s^2,\quad
        E_{sr} = \frac{3}{2}n_s\hat{T}_{sr} + \frac{1}{2}\rho_s \hat{u}_{sr}^2.
\end{equation}

The synthetic equations~\eqref{eq:macro_equation} are not closed since the shear stress ${\bm{\sigma}_s}$ and heat flux ${\bm{q}_s}$ are still unknown; and they in general cannot be expressed in terms of the lower-order moments such as the density, velocity, temperature, and their spatial gradients. Since in GSIS the kinetic equation is numerically solved, we can write the constitutive relations in the following form~\cite{Zhu2021JCP}:
\begin{equation}\label{eq:full_constitutive}
    \begin{aligned}
        \bm{\sigma}_s &= \underbrace{-\mu_s^{\ast} \left(\nabla\bm{u}_s+\nabla\bm{u}_s^{\mathrm{T}}-\frac{2}{3}\nabla\cdot\bm{u}_s\mathrm{I}\right)}_{ \bm{\sigma}^{\text{NS}}_s} + \text{HoT}_{\bm{\sigma}_s}, \quad
        \bm{q}_{s} =\underbrace{ -\kappa_s^{\ast} \nabla T_s}_{\bm{q}^{\text{NS}}_{s}} + \text{HoT}_{\bm{q}_{s}},
    \end{aligned}
\end{equation}
where $\bm{\sigma}^{\text{NS}}_s$ and $\bm{q}^{\text{NS}}_{s}$ are the conventional constitutive relations describing the linear dependence of shear stress and heat flux on the gradient of velocity and temperature, respectively; and the corresponding nominal shear viscosity $\mu_s^{\ast}$ and heat conductivity $\kappa_s^{\ast}$ are given by,
\begin{equation}\label{eq:nominal_NSF_constitutive}
    \begin{aligned}
        \mu_s^{\ast} & = \beta_{s,\mu} n_s T_s\tau_{ss}\frac{n_s}{\sum_r n_s\phi_{sr}}, \qquad
        \beta_{s,\mu} = \frac{1}{\text{Kn}_s}\text{min}\left(\text{Kn}_s,\frac{1}{n}\sum_r n_s\phi_{sr}\right), \\
        \kappa_s^{\ast} &= \beta_{s,\kappa} \frac{5n_s T_s \tau_{ss}}{2m_s\Pr}\frac{n_s}{\sum_r n_s\phi_{sr}\varphi_{sr}}, \qquad
        \beta_{s,\kappa} = \frac{1}{\text{Kn}_s}\text{min}\left(\text{Kn}_s,\frac{1}{n}\sum_r n_s\phi_{sr}\varphi_{sr}\right),
    \end{aligned}
\end{equation}
with $\phi_{ss}=\varphi_{ss}=1$. $\beta_{s,\mu}$ and $\beta_{s,\kappa}$ are the factors used to not only realize the fast convergence in all flow regimes, but also increase the stability and speed when solving the synthetic equations when $\text{Kn}$ is large. Note that the effect of diffusion on the shear stress and thermal conductance of a gas mixture is not included in $\mu_s^{\ast}$ and $\kappa_s^{\ast}$, but integrated into the HoTs. The HoTs can be calculated by subtracting $\bm{\sigma}^{\text{NS}}_s$ and $\bm{q}^{\text{NS}}_{s}$ from the moments of the velocity distribution functions~\cite{Zhu2021JCP},
\begin{equation}\label{eq:getHoTs}
    \begin{aligned}
        \text{HoT}_{\bm{\sigma}_s} &= \left\langle (\bxi-\bm{u}_s) (\bxi-\bm{u}_s)-\frac{1}{3}(\bxi-\bm{u}_s)^2\bm{\mathrm{I}}, f_s \right\rangle -\bm{\sigma}_s^{\text{NS}},\\
        \text{HoT}_{\bm{q}_s} &= \left\langle \frac{1}{2}(\bxi-\bm{u}_s)^2(\bxi-\bm{u}_s), f_s \right\rangle -\bm{q}_s^{\text{NS}}. 
    \end{aligned}
\end{equation}

It must be emphasised that, the NS constitutive relations in Eq.~\eqref{eq:full_constitutive} are solved at the new (current) iteration step to guide to evolution of the velocity distribution function, while the NS constitutive relations in Eq.~\eqref{eq:getHoTs} are extracted from the velocity distribution function in the previous iteration step; thus they cannot cancel out each other until the final steady state (e.g. the convergence criterion) is reached. As has been and will be demonstrated in numerical simulations, this kind of treatment facilities the fast-converging in the whole range of gas rarefaction~\cite{SU2020SIAM}. 
Moreover, it empowers the asymptotic-preserving property such that the constraint on the spatial cell size is relieved in the (near) continuum flow regime.
Finally, it should be pointed out that, although the Li model is used, the method of construction of GSIS can be applied to any kinds of kinetic models for gas mixture flows.

\section{The linear Fourier stability analysis} \label{sec:stability}

The Fourier stability analysis is performed to analyze the convergence speed of the CIS and GSIS, based on the linearized kinetic model. To make the calculation simple but keep the essential physics, the spatial coordinate is not discretized. The performance of the CIS and GSIS in discretized systems will be assessed in the numerical simulation of practical problems shown in section~\ref{sec:4}. 

\subsection{Convergence rate of CIS}

For a system of the gas mixture that slightly deviates from a global equilibrium state, the velocity distribution function $f_s$ can be written as $f_s =  f_{s}^{eq} + h_s$, where $f_{s}^{eq}$ is the equilibrium distribution function
\begin{equation}
	\begin{aligned}[b]
		f_{s}^{eq}=m_s\chi_s\left(\frac{m_s}{2\pi}\right)^{3/2}\exp{\left(-\frac{m_s\xi^2}{2}\right)},
	\end{aligned}
\end{equation}
$h_s$ is the perturbation, and $\chi_s=n_s/n$ is the mole fraction of the species $s$.
The steady-state solutions of a binary gas mixture can be obtained from CIS by iteratively solving the following linearized equations,
\begin{equation}\label{eq:steady_state_linear_kinetic_equation}
	\begin{aligned}[b]
		{\bm{\xi}} \cdot \nabla{{h}_{s}^{k+1}} &=\frac{1}{{\tau}_{ss}^{k}}\left({g'}_{ss}^{k}-{h}_s^{k+1}\right)+\frac{1}{{\tau}_{sr}^{k}}\left({g'}_{sr}^{k}-{h}_s^{k+1}\right), \quad s,r=1,2,
	\end{aligned}
\end{equation}
where the superscript $k$ represents the iteration step, and the perturbed reference distributions in the linearized collision operators are given as,
\begin{equation}\label{eq:linear_perturbed_g}
	\begin{aligned}
		g'_{ss} &= f_{s}^{eq}\left[\Delta{\rho_s} + m_s{\bm{u}}_{s}\cdot\bm{\xi} + \Delta{T_{s}}\left(\frac{1}{2}m_s\xi^2-\frac{3}{2}\right) + \frac{2m_s\hat{\bm{q}}_{ss}\cdot\bm{\xi}}{5{\chi}_{s}}\left(\frac{1}{2}m_s\xi^2-\frac{5}{2}\right)\right], \\
		g'_{sr} &= f_{s}^{eq}\left[\Delta{\rho_s} + m_s\hat{\bm{u}}_{sr}\cdot\bm{\xi} + \Delta{\hat T_{sr}}\left(\frac{1}{2}m_s\xi^2-\frac{3}{2}\right) + \frac{2m_s\hat{\bm{q}}_{sr}\cdot\bm{\xi}}{5{\chi}_{s}}\left(\frac{1}{2}m_s\xi^2-\frac{5}{2}\right)\right], 
	\end{aligned}
\end{equation}
with $\Delta{\rho_s}$, $\bm{u}_{s}$ and $\Delta{T_{s}}$ being the perturbed macroscopic variables, $\hat{\bm{u}}_{sr}$, $\Delta{\hat T_{sr}}$, $\hat{\bm{q}}_{ss}$ and $\hat{\bm{q}}_{sr}$ being the perturbed auxiliary variables. 
By taking the velocity moments of the perturbed distribution function $h_s$, the macroscopic quantities deviated from their corresponding equilibrium values are,
\begin{equation}\label{eq:linear_perturbed_macroscopic_variable}
	\begin{aligned}
		\Delta{\rho_s} &= \left\langle \frac{1}{m_s\chi_s},h_s \right\rangle, \quad
            \bm{u}_s = \left\langle \frac{1}{m_s\chi_s}\bm{\xi},h_s \right\rangle, \quad
            \Delta T_s = \left\langle \frac{1}{m_s\chi_s}\left(\frac{1}{3}m_s\xi^2-1\right),h_s \right\rangle, \\
		\bm{\sigma}_s &= \left\langle \left(\bm{\xi}\bm{\xi}-\frac{1}{3}\xi^2\mathrm{I}\right),h_s \right\rangle, \quad
            \bm{q}_s = \left\langle \frac{1}{m_s}\left(\frac{1}{2}m_s\xi^2-\frac{5}{2}\right)\bm{\xi},h_s \right\rangle.
	\end{aligned}
\end{equation}
Thus, based on Eqs. \eqref{eq:aux_para_inter} and \eqref{eq:aux_heatflux}, the perturbed auxiliary variables in \eqref{eq:linear_perturbed_g} can be determined as, 
\begin{equation}\label{eq:linear_perturbed_auxiliary_variable}
	\begin{aligned}
		\hat{\bm{u}}_{sr} &= \left(1-A_{sr}\right){\bm{u}}_{s} + A_{sr}{\bm{u}}_{r}, \quad
		\Delta{\hat T_{sr}} = \left(1-C_{sr}\right)\Delta{T}_{s} + C_{sr}\Delta{T}_{r}, \\ 
		\hat{\bm{q}}_{ss} &= \left(1-\text{Pr}\right)\bm{q}_s, \quad
		\hat{\bm{q}}_{sr} = \left(1-\varphi_{sr}\text{Pr}\right)\bm{q}_s,
	\end{aligned}
\end{equation}
where
\begin{equation}\label{eq:A_C}
 	\begin{aligned}
		A_{sr} = \left({\frac{\phi_{sr}\text{Kn}_r}{\phi_{rs}\text{Kn}_s}\sqrt{\frac{m_s}{m_r}}+1}\right)^{-1}a_{sr}, \\
		C_{sr} = \left({\frac{\phi_{sr}\text{Kn}_r}{\phi_{rs}\text{Kn}_s}\sqrt{\frac{m_r}{m_s}}+1}\right)^{-1}c_{sr}.
 	\end{aligned}
\end{equation}
Note that all the parameters, including $a_{sr}$, $c_{sr}$, $\phi_{sr}$ and $\varphi_{sr}$, are evaluated at the equilibrium state due to linearization.

To calculate the convergence rate, we define the error functions $Y_s\left(\bm{x},\bm{\xi}\right)$ between distribution functions at two consecutive iteration steps as~\cite{SU2020SIAM},
\begin{equation}\label{eq:error_function_h}
	\begin{aligned}[b]
		Y_s^{k+1}\left(\bm{x},\bm{\xi}\right) \equiv h_s^{k+1}\left(\bm{x},\bm{\xi}\right)-h_s^{k}\left(\bm{x},\bm{\xi}\right).
	\end{aligned}
\end{equation}
Similarly, the error functions $\Phi_s\left(\bm{x}\right)$ for macroscopic quantities $M_s=\left[\Delta{\rho_s},\bm{u}_s,\Delta T_s,\bm{q}_s\right]$ between two consecutive iteration steps are,
\begin{equation}\label{eq:error_function_M}
	\begin{aligned}[b]
		\Phi_s^{k+1}\left(\bm{x}\right) &= \left[\Phi_{s,\Delta{\rho}}^{k+1},\Phi_{s,\bm{u}}^{k+1},\Phi_{s,\Delta{T}}^{k+1},\Phi_{s,\bm{q}}^{k+1}\right] \\
		&\equiv M_s^{k+1}\left(\bm{x}\right)-M_s^{k}\left(\bm{x}\right) = 
  \int{\varPhi_s\left(\bm{\xi}\right) Y_s^{k+1}\left(\bm{x},\bm{\xi}\right)}\mathrm{d}\bm{\xi},
	\end{aligned}
\end{equation}
where $\varPhi_s\left(\bm{\xi}\right)$ is a $1\times8$ vector,
\begin{equation}\label{eq:error_function_M_varphi}
	\begin{aligned}[b]
		\varPhi_s\left(\bm{\xi}\right)=\frac{1}{m_s\chi_s}\left[1,\bm{\xi},\left(\frac{1}{3}m_s\xi^2-1\right),\chi_s\left(\frac{1}{2}m_s\xi^2-\frac{5}{2}\right)\bm{\xi}\right].
	\end{aligned}
\end{equation}

On substituting Eqs. \eqref{eq:error_function_h} and \eqref{eq:error_function_M} into Eq.~\eqref{eq:steady_state_linear_kinetic_equation}, the error functions $Y_s\left(\bm{x},\bm{\xi}\right)$ are found to satisfy,
\begin{equation}\label{eq:error_function_h_equation}
	\begin{aligned}[b]
		&\left[1+\frac{\chi_r}{\chi_s}\phi_{sr}+\frac{\text{Kn}_s}{\chi_s}\left(\frac{2m_s}{\pi}\right)^{1/2}\bm{\xi}\cdot\nabla\right] Y_s^{k+1} = \\
		&f_{s}^{eq} \left\{ \Phi_{s,\Delta{\rho}}^{k}\left(1+\frac{\chi_r}{\chi_s}\phi_{sr}\right) \right. \\ 
		&\left. + \left[\Phi_{s,\bm{u}}^{k}\left(1+\frac{\chi_r}{\chi_s}\phi_{sr}\left(1-A_{sr}\right)\right) +\Phi_{r,\bm{u}}^{k}\frac{\chi_r}{\chi_s}\phi_{sr}A_{sr} \right] \cdot m_s\bm{\xi} \right. \\ 
		&\left. + \left[\Phi_{s,\Delta{T}}^{k}\left(1+\frac{\chi_r}{\chi_s}\phi_{sr}\left(1-C_{sr}\right)\right) +\Phi_{r,\Delta{T}}^{k}\frac{\chi_r}{\chi_s}\phi_{sr}C_{sr} \right] \left(\frac{1}{2}m_s\xi^2-\frac{3}{2}\right) \right. \\ 
		&\left. + \left[\Phi_{s,\bm{q}}^{k}\left(\left(1-\text{Pr}\right) + \frac{\chi_r}{\chi_s}\phi_{sr}\left(1-\varphi_{sr}\text{Pr}\right)\right) \right] \cdot \frac{2m_s}{5\chi_s}\left(\frac{1}{2}m_s\xi^2-\frac{5}{2}\right)\bm{\xi} \right\}.
	\end{aligned}
\end{equation}

Next, we perform the Fourier stability analysis to determine the convergence rate, by seeking the eigenvalue $e$ and eigenfunctions $y_s\left(\bm{\xi}\right)$ and $\alpha_{s,M}=\left[\alpha_{s,\Delta{\rho}},\alpha_{s,\bm{u}},\alpha_{s,\Delta{T}},\alpha_{s,\bm{q}}\right]$ of the forms
\begin{equation}\label{eq:error_function_fourier}
	\begin{aligned}
		Y_s^{k+1}\left(\bm{x},\bm{\xi}\right) &= e^k y_s\left(\bm{\xi}\right)\exp{\left(i\bm{\theta}\cdot\bm{x}\right)}, \\
		\Phi_s^{k+1}\left(\bm{x}\right) &= e^{k+1}\alpha_{s,M}\exp{\left(i\bm{\theta}\cdot\bm{x}\right)},
	\end{aligned}
\end{equation}
where $i$ is the imaginary unit, and $\bm{\theta}$ is the wave vector of perturbation. Note that the factor $e^k$, rather than $e^{k+1}$, emerges in the expression of error function $Y_s^{k+1}$, which is determined by macroscopic quantities in the $k$-th iteration step.

Therefore, by substituting Eq.~\eqref{eq:error_function_fourier} into Eq.~\eqref{eq:error_function_h_equation}, we have,
\begin{equation}\label{eq:eigenfunction_y}
	\begin{aligned}[b]
		y_s\left(\bm{\xi}\right) &= \left[1+\frac{\chi_r}{\chi_s}\phi_{sr}+\frac{\text{Kn}_s}{\chi_s}\left(\frac{2m_s}{\pi}\right)^{1/2}i\bm{\theta}\cdot\bm{\xi}\right]^{-1}f_{s}^{eq} \\
		& \times \left\{ \alpha_{s,\Delta{\rho}}\left(1+\frac{\chi_r}{\chi_s}\phi_{sr}\right) \right. \\ 
		&\left. + \left[\alpha_{s,\bm{u}}\left(1+\frac{\chi_r}{\chi_s}\phi_{sr}\left(1-A_{sr}\right)\right) +\alpha_{r,\bm{u}}\frac{\chi_r}{\chi_s}\phi_{sr}A_{sr} \right] \cdot m_s\bm{\xi} \right. \\ 
		&\left. + \left[\alpha_{s,\Delta{T}}\left(1+\frac{\chi_r}{\chi_s}\phi_{sr}\left(1-C_{sr}\right)\right) +\alpha_{r,\Delta{T}}\frac{\chi_r}{\chi_s}\phi_{sr}C_{sr} \right] \left(\frac{1}{2}m_s\xi^2-\frac{3}{2}\right) \right. \\ 
		&\left. + \left[\alpha_{s,\bm{q}}\left(\left(1-\text{Pr}\right) + \frac{\chi_r}{\chi_s}\phi_{sr}\left(1-\varphi_{sr}\text{Pr}\right)\right) \right] \cdot \frac{2m_s}{5\chi_s}\left(\frac{1}{2}m_s\xi^2-\frac{5}{2}\right)\bm{\xi} \right\}.
	\end{aligned}
\end{equation}
Multiplying Eq.~\eqref{eq:eigenfunction_y} with $\varPhi_s\left(\bm{\xi}\right)$ and integrating the resultant equations over velocity space, we obtain the following 16 linear algebraic equations for a binary gas mixture using Eqs.~\eqref{eq:error_function_M} and~\eqref{eq:error_function_fourier},
\begin{equation}\label{eq:eigenfunction_cis}
	\begin{aligned}[b]
	e\left[\alpha_{1,M},\alpha_{2,M}\right]^{\mathsf{T}} &= C_{16\times16}\left[\alpha_{1,M},\alpha_{2,M}\right]^{\mathsf{T}} \\
 &= \int{\left[\varPhi_1\left(\bm{\xi}\right)y_1\left(\bm{\xi}\right),\varPhi_2\left(\bm{\xi}\right)y_2\left(\bm{\xi}\right)\right]^{\mathsf{T}}}\mathrm{d}\bm{\xi},
	\end{aligned}
\end{equation}
Then, the convergence rate can be obtained by numerically computing the spectral radius $e$ of the matrix $C_{16\times16}$ and taking the maximum absolute value of $e$ (i.e., the spectral radius). Clearly, the numerical scheme is stable when $e\leq 1$, and the iteration converges fast when $e$ is close to zero. However, it should be noted that, when $|e|$ is close to one, the error between two consecutive steps rarely decreases and hence the iteration converges extremely slowly. Worse still, it may suffers the false convergence problem~\cite{SU2020SIAM}.

\begin{figure}[!t]\centering
    {\includegraphics[width=0.45\textwidth,clip = true]{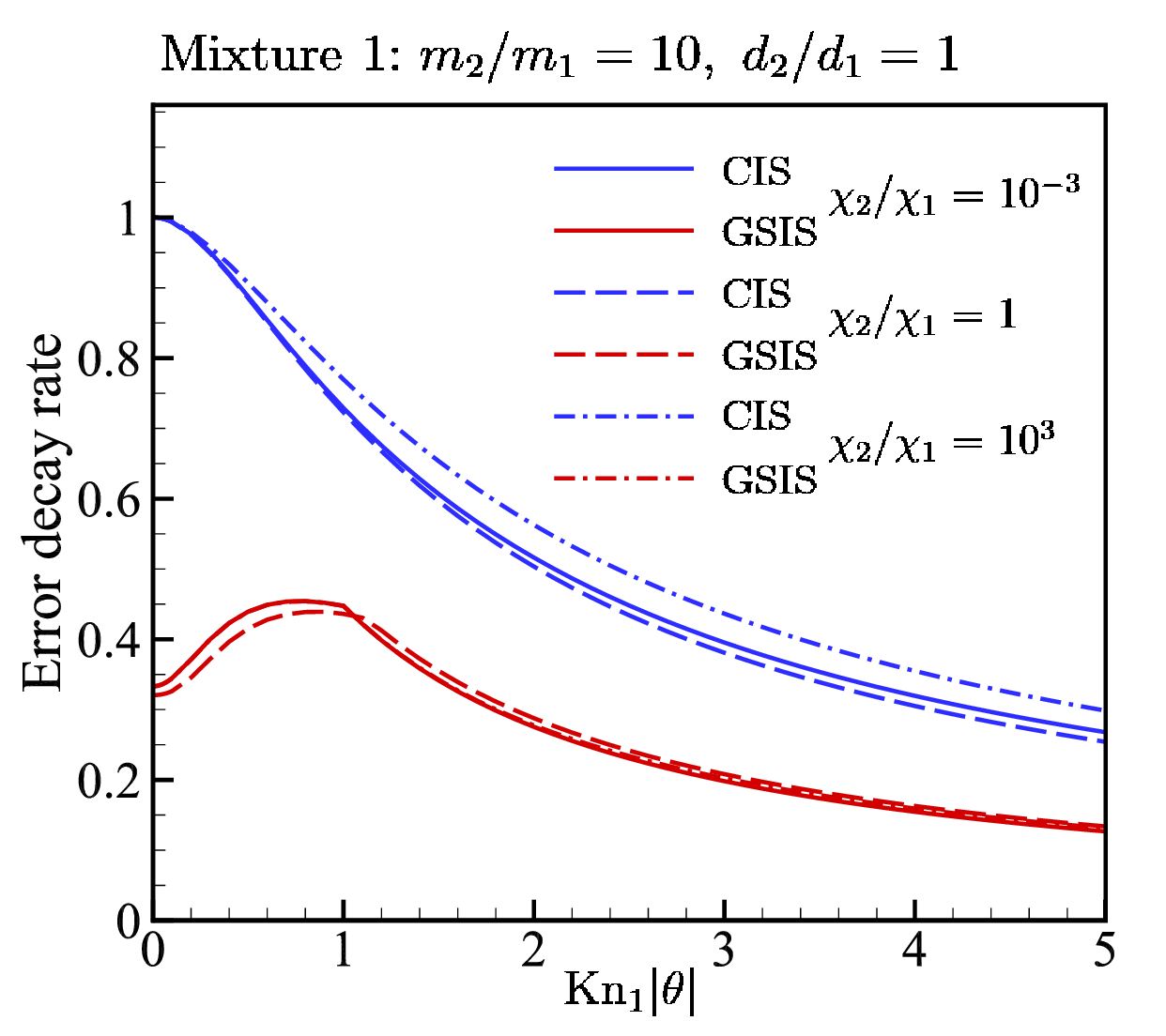}}
    {\includegraphics[width=0.45\textwidth,clip = true]{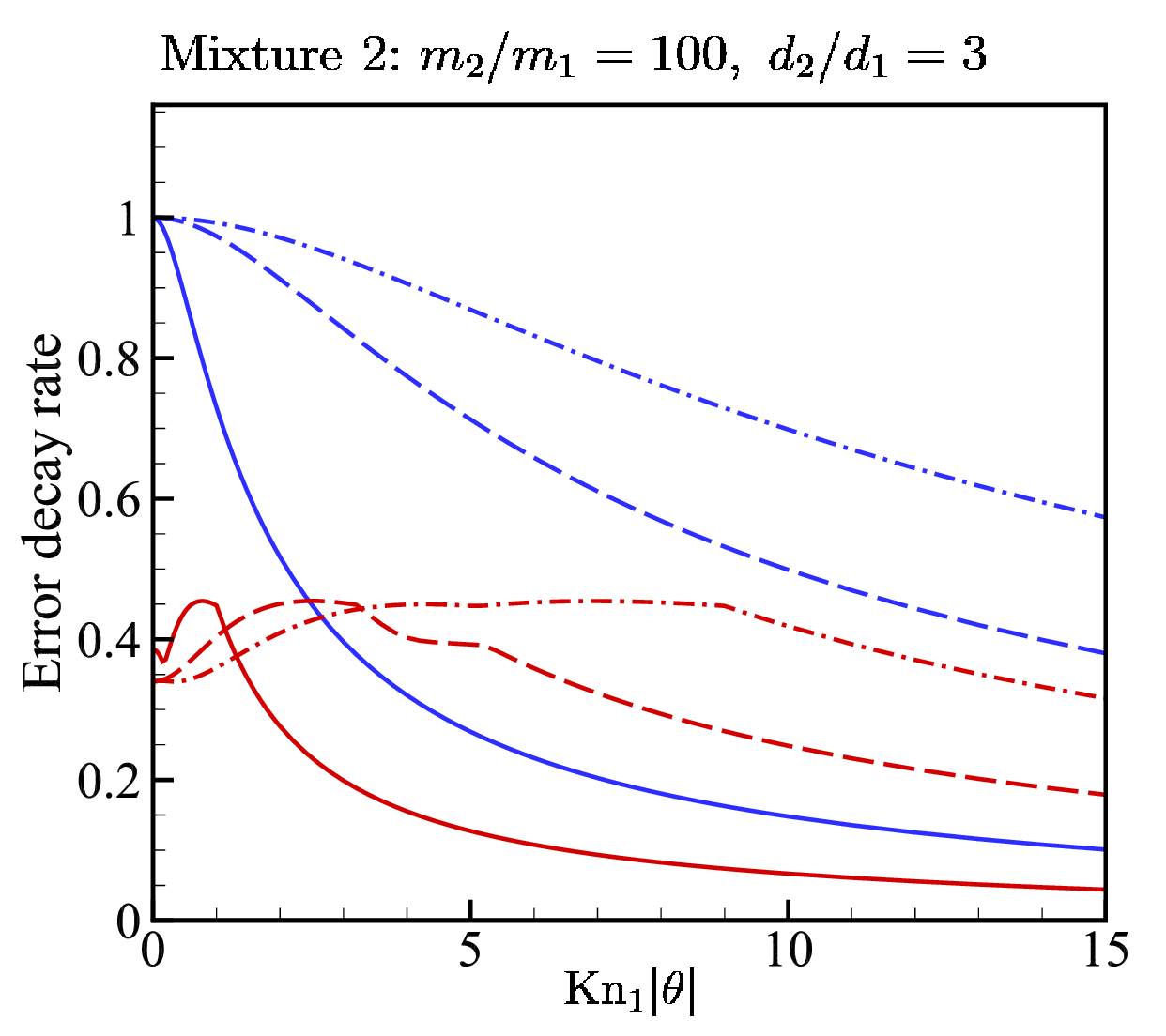}}
    {\includegraphics[width=0.45\textwidth,clip = true]{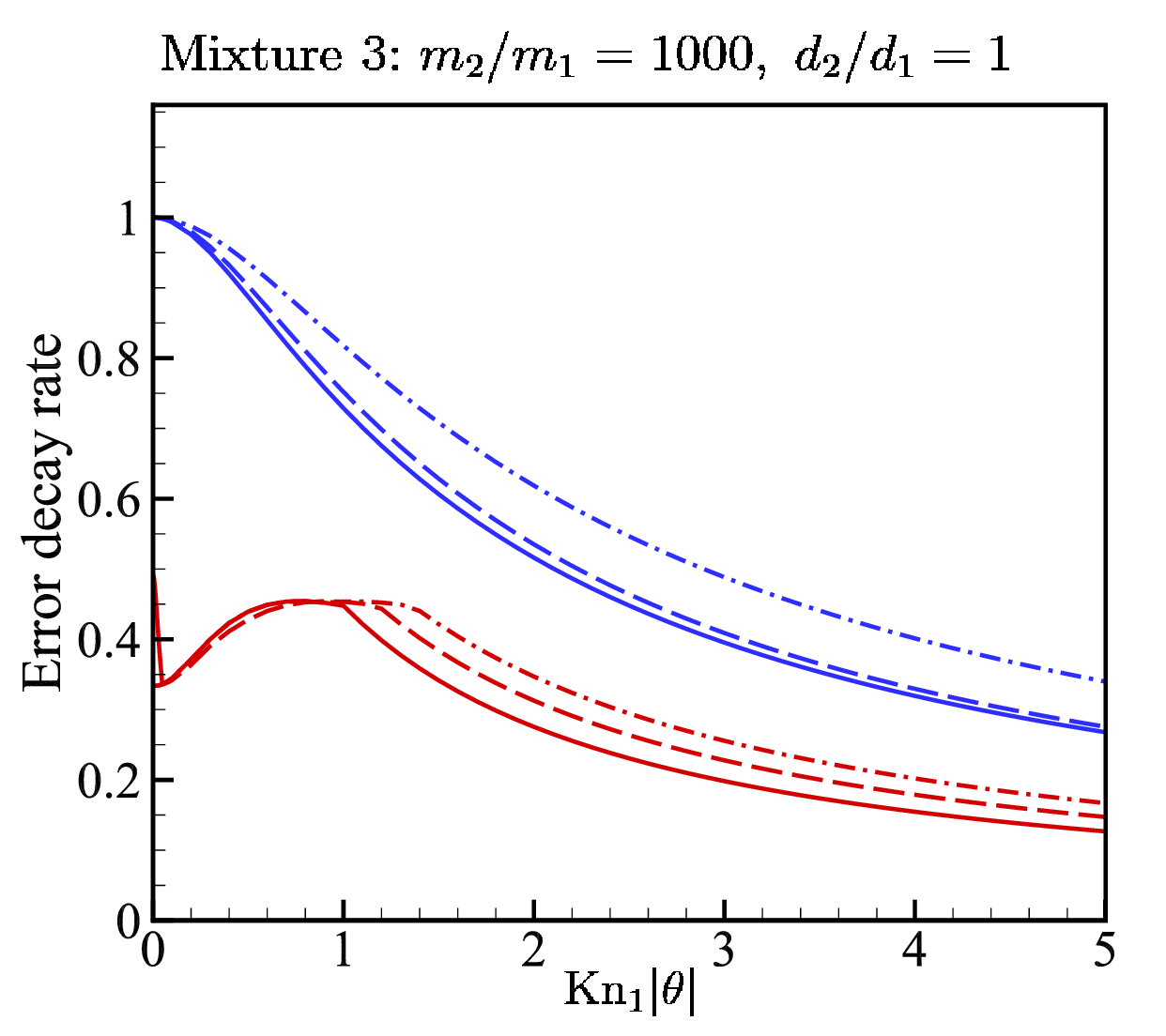}}\\
\caption{The error decay rate as a function of the $\text{Kn}_1|\bm{\theta|}$ in CIS and GSIS, where the model parameters are given in Table~\ref{tab04:mixture_gas_perp}.  The concentration ratio is $\chi_2/\chi_1=n_2/n_1=0.001,1$, and 1000. } 
	\label{fig:error_decay_compare}
\end{figure}

Figure~\ref{fig:error_decay_compare} shows the convergence rate of CIS as a function of $\text{Kn}_1|\bm{\theta}|$ for the binary mixtures of Maxwell molecules (see Table~\ref{tab04:mixture_gas_perp}), over a wide range of concentration ratio ($\chi_2/\chi_1=10^{-3},1,10^3$). For all the mixtures, the CIS spectral radius approaches one when $\text{Kn}_1|\bm{\theta}|$ goes to zero, indicating the convergence of CIS is extremely slow in the continuum flow limit. Physically, this is due to the massively inefficient information exchange across the length scale that is much larger than the mean free path. Fortunately, the spectral radius decreases as the $\text{Kn}_1|\bm{\theta}|$ increases and hence  the CIS can be efficient for highly rarefied gas flows. Meanwhile, we notice that the convergence rate of CIS does not monotonically change with the concentration ratio in Mixture 1 and 3 ($d_2=d_1$, $\text{Kn}_1=\text{Kn}_2$), but generally, the convergence is slower for a mixture with a significantly larger proportion of heavier molecules, and becomes faster when the concentration are comparable. On the other hand, for Mixture 2 with different molecular diameters ($d_2=3d_1$, $\text{Kn}_1=9\text{Kn}_2$), the convergence rate becomes significantly slower as the concentration of the heavier species increases, due to its much shorter species mean free path.

\subsection{Convergence rate of GSIS}

In GSIS, the solutions of macroscopic synthetic equations are used to accelerate the evolution of the gas kinetic equation. To be specific, when the velocity distribution function $h^k$ has been obtained from a previous iteration step $k$, $h_s^{k+1/2}$ can be solved from the kinetic equation \eqref{eq:steady_state_linear_kinetic_equation} with $k+1$ replaced by $k+1/2$. Then the macroscopic quantities $M_s^{k+1}$ at the $(k+1)$-th iteration step are obtained by solving the following linearized synthetic equations,
\begin{equation}\label{eq:steady_state_linear_macroscopic_equation}
	\begin{aligned}
		\nabla \cdot \bm{u}_s^{k+1} &= 0, \\
		\chi_s\nabla\left(\Delta{\rho}_s^{k+1}\right) +\chi_s\nabla\left(\Delta{T}_s^{k+1}\right) +\nabla\cdot\bm{\sigma}_s^{k+1} &= \frac{1}{\tau_{sr}}A_{sr}m_s\chi_s\left(\bm{u}_{r}^{k+1}-\bm{u}_s^{k+1}\right), \\
		\nabla\cdot\bm{q}_s^{k+1} &= \frac{3}{2\tau_{sr}}C_{sr}\chi_s\left(\Delta{T}_r^{k+1}-\Delta{T}_s^{k+1}\right),
	\end{aligned}
\end{equation} 
with
\begin{equation}\label{eq:linear_constitutive}
	\begin{aligned}
		\bm{\sigma}_s^{k+1} &= \int{\left(\bm{\xi}\bm{\xi}-\frac{1}{3}\xi^2\mathrm{I}\right)h_s^{k+1/2}}\mathrm{d}\bm{\xi} \\
		&+ \beta_{s,\mu} \left(\frac{2m_s}{\pi}\right)^{1/2}\text{Kn}_s\left(1+\frac{\chi_r}{\chi_s}\phi_{sr}\right)^{-1}\left(\nabla\bm{u}_s^{k+1/2}+\left(\nabla\bm{u}_s^{k+1/2}\right)^{\mathrm{T}}-\frac{2}{3}\nabla\cdot\bm{u}_s^{k+1/2}\bm{\mathrm{I}}\right) \\
		&- \beta_{s,\mu} \left(\frac{2m_s}{\pi}\right)^{1/2}\text{Kn}_s\left(1+\frac{\chi_r}{\chi_s}\phi_{sr}\right)^{-1}\left(\nabla\bm{u}_s^{k+1}+\left(\nabla\bm{u}_s^{k+1}\right)^{\mathrm{T}}-\frac{2}{3}\nabla\cdot\bm{u}_s^{k+1}\bm{\mathrm{I}}\right), \\
        \bm{q}_s^{k+1} &= \frac{1}{m_s}\int{\left(\frac{1}{2}m_s\xi^2-\frac{5}{2}\right)\bm{\xi}h_s^{k+1/2}}\mathrm{d}\bm{\xi} \\
		&+ \beta_{s,\kappa} \frac{5}{\text{Pr}}\left(\frac{1}{2\pi m_s}\right)^{1/2}\text{Kn}_s\left(1+\frac{\chi_r}{\chi_s}\phi_{sr}\varphi_{sr}\right)^{-1}\nabla\left(\Delta{T}_s^{k+1/2}\right) \\
		&- \beta_{s,\kappa} \frac{5}{\text{Pr}}\left(\frac{1}{2\pi m_s}\right)^{1/2}\text{Kn}_s\left(1+\frac{\chi_r}{\chi_s}\phi_{sr}\varphi_{sr}\right)^{-1}\nabla\left(\Delta{T}_s^{k+1}\right).
	\end{aligned}
\end{equation}
To calculate the convergence rate of GSIS, the error function for the velocity distribution function is redefined as~\cite{SU2020SIAM},
\begin{equation}\label{eq:error_function_gsis}
	\begin{aligned}[b]
		Y_s^{k+1/2}\left(\bm{x},\bm{\xi}\right) &\equiv h_s^{k+1/2}\left(\bm{x},\bm{\xi}\right)-h_s^{k-1/2}\left(\bm{x},\bm{\xi}\right) = e^k y_s\left(\bm{\xi}\right)\exp{\left(i\bm{\theta}\cdot\bm{x}\right)}, 
		% \Phi_s^{k+1}\left(\bm{x}\right) &= \left[\Phi_{s,\Delta{\rho}}^{k+1},\Phi_{s,\bm{u}}^{k+1},\Phi_{s,\Delta{T}}^{k+1},\Phi_{s,\bm{q}}^{k+1}\right] \\
		% &\equiv M_s^{k+1}\left(\bm{x}\right)-M_s^{k}\left(\bm{x}\right) = \int{\varPhi_s\left(\bm{\xi}\right) Y_s^{k+1}\left(\bm{x},\bm{\xi}\right)}\mathrm{d}\bm{\xi},
	\end{aligned}
\end{equation}
where $y_s\left(\bm{\xi}\right)$ is given by Eq.~\eqref{eq:eigenfunction_y}. The definition of error function $\Phi_s\left(\bm{x}\right)$ remains unchanged, while in GSIS they are calculated from the macroscopic synthetic equations \eqref{eq:steady_state_linear_macroscopic_equation} and \eqref{eq:linear_constitutive} with $\Delta{\rho}_s$, $\bm{u}_s$, $\Delta{T}_s$ and $\bm{q}_s$ replaced by $\Phi_{s,\Delta{\rho}}$, $\Phi_{s,\bm{u}}$, $\Phi_{s,\Delta{T}}$ and $\Phi_{s,\bm{q}}$, respectively, and $h_s\left(\bm{x},\bm{\xi}\right)$ replaced by $Y_s\left(\bm{x},\bm{\xi}\right)$. Then, we obtain the following linear algebraic equations,
\begin{equation}\label{eq:eigenfunction_gsis}
	\begin{aligned}
		e&\left[i\bm{\theta}\cdot{\alpha_{s,\bm{u}}}\right] = S_{s,1}, \\		e&\left[i\bm{\theta}\chi_s\left(\alpha_{s,\Delta{\rho}}+\alpha_{s,\Delta{T}}\right)+\beta_{s,\mu} \left(\frac{2m_s}{\pi}\right)^{1/2}\text{Kn}_s\left(1+\frac{\chi_r}{\chi_s}\phi_{sr}\right)^{-1}\theta^2{\alpha_{s,\bm{u}}} \right. \\ 
		&\left.+\left(\frac{\pi m_s}{2}\right)^{1/2}\frac{\chi_s\chi_r}{\text{Kn}_s}A_{sr}\phi_{sr}\left({\alpha_{s,\bm{u}}}-{\alpha_{r,\bm{u}}}\right)\right] = S_{s,2-4}, \\
		e&\left[i\bm{\theta}\cdot{\alpha_{s,\bm{q}}} +\frac{3}{2}\left(\frac{\pi}{2m_s}\right)^{1/2}\frac{\chi_s\chi_r}{\text{Kn}_s}C_{sr}\phi_{sr}\left({\alpha_{s,\Delta{T}}}-{\alpha_{r,\Delta{T}}}\right)\right] = S_{s,5}, \\
		e&\left[i\bm{\theta}\beta_{s,\kappa} \frac{5}{\text{Pr}}\left(\frac{1}{2\pi m_s}\right)^{1/2}\text{Kn}_s\left(1+\frac{\chi_r}{\chi_s}\phi_{sr}\varphi_{sr}\right)^{-1}\alpha_{s,\Delta{T}}+\alpha_{s,\bm{q}}\right] = S_{s,6-8},
	\end{aligned}
\end{equation}
where the source terms $S_{s,1-8}$ are,
\begin{equation}\label{eq:eigenfunction_gsis_S}
	\begin{aligned}
		S_{s,1} &= 0, \\
		S_{s,2-4} &= \int {\left[-i\bm{\theta}\cdot\left(\bm{\xi}\bm{\xi}-\frac{1}{3}\xi^2\mathrm{I}\right)+ \right.}\\
		&\qquad \left.\beta_{s,\mu} \left(\frac{2}{\pi m_s}\right)^{1/2}\frac{\text{Kn}_s}{\chi_s}\left(1+\frac{\chi_r}{\chi_s}\phi_{sr}\right)^{-1}\left(\theta^2\bm{\xi}+\frac{1}{3}\bm{\theta}\left(\bm{\theta}\cdot\bm{\xi}\right)\right)\right]y_s \mathrm{d}\bm{\xi}, \\
		S_{s,5} &= 0, \\
		S_{s,6-8} &= \int \left[\frac{1}{m_s}\left(\frac{1}{2}m_s\xi^2-\frac{5}{2}\right)\bm{\xi}+ \right.\\
		&\qquad \left.  i\bm{\theta}\beta_{s,\kappa} \frac{5}{\text{Pr}}\left(\frac{1}{2\pi m_s}\right)^{1/2}\frac{\text{Kn}_s}{m_s\chi_s}\left(1+\frac{\chi_r}{\chi_s}\phi_{sr}\varphi_{sr}\right)^{-1}\left(\frac{1}{3}m_s\xi^2-1\right)\right]y_s \mathrm{d}\bm{\xi}.
	\end{aligned}
\end{equation}

Equation \eqref{eq:eigenfunction_gsis} with the source terms \eqref{eq:eigenfunction_gsis_S} can be rewritten in the matrix form as,
\begin{equation}\label{eq:eigenfunction_gsis_matrix}
	\begin{aligned}[b]
		eL_{16\times16}\left[\alpha_{1,M},\alpha_{2,M}\right]^{\mathsf{T}} = R_{16\times16}\left[\alpha_{1,M},\alpha_{2,M}\right]^{\mathsf{T}}.
	\end{aligned}
\end{equation}
Therefore, the convergence rate can be obtained by solving the eigenvalues $e$ of matrix $(L^{-1}_{16\times16})R_{16\times16}$ and finding its spectral radius.

The convergence rate of GSIS is calculated for Mixture 1,2 and 3 with a wide range of concentration ratio ($\chi_2/\chi_1=10^{-3},1,10^3$), as the red lines shown in Fig.~\ref{fig:error_decay_compare}. It is observed that the spectral radius of GSIS can be kept lower than 0.5 in the entire range of Knudsen numbers, and for wide ranges of the concentration ratio and Knudsen number ratio. This means that, although adopting the simple GSIS constitutive relation in Eq.~\eqref{eq:full_constitutive} and \eqref{eq:nominal_NSF_constitutive}, the error can be reduced by three orders of magnitude after 10 iterations in all the flow regimes.

\section{Numerical schemes of the GSIS} \label{sec:3}

The GSIS solver involves the discrete solutions of two sets of governing equations, namely, the mesoscopic kinetic equation and the macroscopic synthetic equations; the latter can be considered as the general NS equations with source terms from the HoTs. Both sets of governing equations can be solved by traditional computational fluid dynamics methods. Here, the unstructured finite volume method is used.

\subsection{Finite volume method}

Since the two kinetic equations~\eqref{eq:generalModel} differ only in the form of collision terms, in order to keep the presentation simple, $f$ and $g$ will be used to represent the velocity distribution functions and the reference velocity distribution functions, respectively, without distinguishing each species. The same is applied to the two macroscopic equations~\eqref{eq:macro_equation}.  

The unstructured cell-centered finite volume method is adopted for spatial discretization. 
Using the Gauss theorem to transfer the volume integration to sum of fluxes through surfaces, the semi-discrete forms of the mesoscopic and macroscopic governing equations are as
\begin{subequations}
\begin{align}
    \frac{\partial f_i}{\partial t} + \frac{1}{V_i}\sum_{j\in N(i)} \xi_n f_{ij} S_{ij} &= \frac{g_i-f_i}{\tau_i},\\
    \frac{\partial \bm{W}_i}{\partial t} + \frac{1}{V_i}\sum_{j\in N(i)} \bm{F}_{ij} S_{ij} &= \bm{Q},
\end{align}
\end{subequations}
where $f_i$ and  $\bm{W}_i$ are the averaged distribution function and conservative variables for the discrete cell $i$, respectively. $N(i)$ denotes the set of neighboring cells of $i$, and cell $j$ is one of the neighbors, the interface connecting the cell $i$ and cell $j$ is denoted as subscript $ij$. $V_i$ is the volume of the cell $i$, $S_{ij}$ represents the area of the cell interface $ij$, and $\xi_n = \bm{\xi}\cdot \bm{n}_{ij}$ is the vector of the molecular velocity $\bm{\xi}$ along normal direction $\bm{n} = \bm{S}/|\bm{S}|$. $\xi_n f_{ij}$ and $\bm{F}_{ij}$ are the interface flux of the velocity distribution functions and macroscopic governing equation, respectively. $g$ and $\bm{Q}$ correspond to the equilibrium and source terms, respectively, which are determined by the macroscopic variables $\bm{W}$, see the detailed expressions in \ref{apdex:source_term}. 

For a given numerical time step $\Delta t = t^{k+1} - t^{k}$ in an implicit method process, the governing equations are discretized by the backward Euler formula:
\begin{subequations}
\begin{align}
    \label{eq:dis_micro}
  \frac{f_i^{k+1}-f_i^k}{\Delta t_i} + \frac{1}{V_i}\sum_{j\in N(i)} \xi_n f_{ij}^{k+1}A_{ij}&=\frac{g^{k}_{i}-f^{k+1}_{i}}{\tau^{k}_i},\\ 
  \label{eq:dis_macro}
  \frac{\bm{W}_i^{k+1}-\bm{W}_i^k}{\Delta t_i} + \frac{1}{V_i}\sum_{j\in N(i)}\bm{F}_{ij}^{k+1}A_{ij}&=\bm{Q}^{k+1}.
\end{align}
\end{subequations}

To solve the above two equations, it is common to construct an incremental form in the implicit process. First, when $f^{k}$ and $\bm{W}^{k}$ are known, the incremental variable $\Delta f_i^k=f_i^{k+1} - f_i^{k}$ and $\Delta\bm{W}_i^k=\bm{W}_i^{k+1} - \bm{W}_i^{k}$ in time step $\Delta t_i^k$ is introduced. Second, the delta-form governing equations of Eq.~\eqref{eq:dis_micro} and Eq.~\eqref{eq:dis_macro} for implicit iterative algorithm can be written as:
\begin{subequations}\label{eq:delta_form}
\begin{align}
\label{eq:delta_form_micro}
\left( \frac{1}{\Delta t_i} + \frac{1}{\tau_i^k}\right)\Delta f_i^{k} + \frac{1}{V_i}\sum_{j\in N(i)} \xi_n\Delta f_{ij}^{k}A_{ij}=\underbrace{\frac{g^{k}_{i}-f^{k}_{i}}{\tau^{k}_i}-\frac{1}{V_i}\sum_{j\in N(i)} \xi_nf_{ij}^{k}A_{ij}}_{r_i^k},\\
\label{eq:delta_form_macro}
\left(\frac{1}{\Delta t_i}- \frac{\partial \bm{Q}_i}{\partial \bm{W}_i}\right)\Delta \bm{W}_i^{k} + \frac{1}{V_i} \sum_{j\in N(i)} \Delta\bm{F}_{ij}^{k}A_{ij}=\underbrace{-\frac{1}{V_i}\sum_{j\in N(i)} \bm{F}_{ij}^{k}S_{ij}+\bm{Q}_i^{k}}_{\bm{R}_i^n}.
\end{align}
\end{subequations}
Here, $ r_i^k$ and $ \bm{R}_i^k $ represent the mesoscopic and macroscopic residuals in the $ k $-th step, respectively. Since the delta-forms will not affect the accuracy of steady solution, the mesoscopic and macroscopic fluxes on the cell interface $ij$ on the left-hand side of the above two equations can be constructed using the first-order upwind scheme. When converged, the accuracy is completely determined by the calculation of the residual terms. That is, the fluxes are fully evaluated by the following numerical scheme: 
\begin{equation}\label{eq:interfaceflux}
\begin{aligned}
    \xi_n \Delta f_{ij}^k&= \frac{1}{2}\xi_n^+\Delta f_i^k + \frac{1}{2}\xi_n^-\Delta f_j^k, \quad
    \xi_n f_{ij}=\frac{1}{2}\xi_n^+f_L + \frac{1}{2}\xi_n^-f_R,\\
    \Delta \bm{F}_{ij}^k &= \frac{1}{2}[\Delta \bm{F}_i^k + \Delta \bm{F}_j^k+\Gamma_{ij}(\Delta\bm{W}_i^k - \Delta\bm{W}_j^k)], \quad 
    F_{ij} = F(W_L, W_R, S_{ij}).
\end{aligned}
\end{equation}
The term $\xi_n^{\pm}=[1\pm \text{sign}(\xi_n)]$ represents the interface sign directions relative to the cell center value. To achieve second-order spatial accuracy in the mesoscopic equation, the interface flux values on the right side of Eq.~\eqref{eq:delta_form_micro} are fully computed using gradient calculations and interpolation. Specifically, we define $f_{L/R} = f_{i/j} + \phi\nabla f_{i/j} \cdot \bm{x}$, where $\phi$ is calculated using the Venkatakrishnan limiter.
The macroscopic flux can be decomposed into convective and viscous fluxes as $\bm{F}_{ij} = \bm{F}_c + \bm{F}_v$, see Eq.~\eqref{macro_var_define}. The viscous flux $\bm{F}_v$ is computed using central difference. For the reconstruction of the convective flux $\bm{F}_c$, the Rusanov scheme is employed for the high-speed outflow case, while the AUSM+UP scheme is used for the low-speed pressure-driven case.
The reconstructed macroscopic variables of the left and right sides of the interface can be obtained as $\bm{W}_{L/R} = \bm{W}_{i/j} + \phi\nabla \bm{W}_{i/j} \cdot \bm{x}$, while the limiter $\phi$ is consistent with that in the mesoscopic equations. To simplify the numerical fluxes on the left-hand side of Eq.~\eqref{eq:delta_form_macro}, the first-order upwind scheme is employed. Additionally, for the macroscopic terms on the left-hand side of delta-form flux $\Delta \bm{F}_{ij}$, Euler equations-based fluxes are adopted to simplify the implicit increment of macroscopic fluxes. Here, $\Gamma_{ij} = |u_{n}| + c_s + {2\mu}/{\rho|\bm{n}_{ij}\cdot(\bm{x}_j)-\bm{x}_i|}$ represents the approximate spectral radius for each species, with $u_n$ being the scalar product of the macro velocity vector and the unit normal vector of the face, and $c_s$ the sound of speed.
Given that the control volume satisfies $\sum_{j\in N(i)}\bm{n}_{ij} A_{ij}=0$, $\sum_{j\in N(i)}\bm{F}_{i} A_{ij}=0$, and the flux can be directly represented by the convective flux, the subscript $j$ flux is expressed in a matrix-free form as $\Delta \bm{F}_j^k=\bm{F}_c(\bm{W}_j^k + \Delta \bm{W}_j^k) - \bm{F}_c(\bm{W}_j^k)$.

Substituting Eq.~\eqref{eq:interfaceflux} into Eq.~\eqref{eq:delta_form}, the implicit governing equations for mesoscopic and macroscopic equation are
\begin{subequations}
\begin{align}
    d_i\Delta f_i^{k} + \frac{1}{2V_i}\sum_{j\in N(i)}\bm{\xi}_n^-\Delta f_j^{k}S_{ij}&=r_i^k,\\
    D_i\Delta \bm{W}_i^{k} + \frac{1}{2V_i}\sum_{j\in N(i)} \left(\Delta\bm{F}_{j}^{k} - \Gamma_{ij}\Delta \bm{W}_j^k\right)S_{ij}&=\bm{R}_i^k,
\end{align}
\end{subequations}
where the matrix elements are
\begin{subequations}
\begin{align}
d_i&=\frac{1}{\Delta t_i} + \frac{1}{\tau_i^k} + \frac{1}{2V_i}\sum_{j\in N(i)}\bm{\xi}_n^+A_{ij},\\
D_i &= \frac{1}{\Delta t_i}+\frac{1}{2V_i}\sum_{j\in N(i)}\Gamma_{ij}S_{ij}- \left(\frac{\partial \bm{Q}}{\partial \bm{W}}\right)^k_i.
\end{align}
\end{subequations}
The above equations can be solved by the standard LU-SGS method, which can be described by two steps:
\begin{equation}
    \begin{aligned}
        \Delta \bm{W}^* &= D_i^{-1}\left(R_i^k - \frac{1}{2V_i}\sum_{j\in L(i)} \left(\Delta\bm{F}_{j}^{*} - \Gamma_{ij}\Delta \bm{W}_j^*\right)S_{ij}\right),\\
        \Delta \bm{W}^{k} &= \Delta W_i^* -  D_i^{-1}\left(\frac{1}{2V_i}\sum_{j\in U(i)} \left(\Delta\bm{F}_{j}^{k} - \Gamma_{ij}\Delta \bm{W}_j^k\right)S_{ij}\right).
    \end{aligned}
\end{equation}
where $L(i)$ and $U(i)$ are subsets of $N(i)$. $L(i)$ is the set of cells around cell $i$ with index number less than $i$, while $U(i)$ is the set of cells around cell $i$ with index number greater than $i$. In contrast to the synthetic equations for polyatomic gases, the implicit solutions of two species macroscopic equations requires an additional derivative of the source term with respect to the momentum equation. The derivation is detailed in \ref{apdex:source_term}.

\subsection{Wall boundary conditions}

% \leir{
% describe the boundary conditions for VDF first and then for macroscopic equations\\
% }

In this work, a fully diffusive boundary condition is applied to the wall, where gas molecules reflect diffusely from the wall in thermodynamic equilibrium. The velocity distribution of gas molecules at the wall is given by:
\begin{equation}
    f_{\text{wall}} = \left\{
    \begin{array}{cc}
        f_{\text{in}}, &  \xi_n \geq 0 \\
        g_{\text{wall}}^{\text{eq}}, & \xi_n < 0 
    \end{array}
    \right.
\end{equation}
Here, $ f_{\text{in}} $ is the velocity distribution function of gas molecules incident on the wall, computed using a second-order interpolation method. $g_{\text{wall}}^{\text{eq}}$ is the fully thermally equilibrium of gas molecules reflected from the wall under specified wall temperature and velocity, with the density determined by the non-penetration condition:
\begin{equation}
    \int_{\xi_n\geq 0} \xi_n f_{\text{in}} \mathrm{d}\bm{\xi} = -\int_{\xi_n < 0} \xi_n g_{\text{wall}}^{\text{eq}} \mathrm{d}\bm{\xi}.
\end{equation}
This boundary condition can be easily implemented in the computer code. 

However, the implementation of the boundary condition in the macroscopic solvers is not easy, as in rarefied gas flows the conventional non-velocity-slip and non-temperature jump conditions are not valid any more. After a few attempts~\cite{Zhu2021JCP,zeng2023general}, we finally developed a consistent boundary treatment between the mesoscopic and macroscopic solvers~\cite{liu2024further}. In gas mixture flow, the boundary condition for individual species is handled in a similar way described in Ref.~\cite{liu2024further}.

\subsection{The flowchart of GSIS}

To achieve fast convergence and asymptotic-preserving, GSIS alternately solves the mesoscopic kinetic equations and macroscopic synthetic equations, where the solutions of the former provide constitutive relations and the solutions of the latter guide the evolution of distribution functions towards the steady-state solution. The algorithm of GSIS is elaborated below:
% \leir{replace n by k}
\begin{enumerate}[Step 1.]
    \item When both the macroscopic properties $\bm{W}_s^{k}=\left[\rho_s^k,(\rho_s\bm{u}_s)^k,E_s^k\right]^\mathbf{T}$ and the distribution functions $f_s^{k}~(s=1,2)$ are obtained in a previous $k$-th step, $f_s^{k+1/2}$ can be solved according to the kinetic equations~\eqref{eq:dis_micro}, by replacing $k+1$ with $k+1/2$. Note that the superscript $k+1/2$ means that the distribution functions solved at this stage have not been corrected by the solutions of synthetic equations.
    \item For a given velocity distribution functions $f_s^{k+1/2}$, the corresponding macroscopic variables $\bm{W}_s^{k+1/2}$ and high order term $\text{HoTs}^{k+1/2}$ are calculated based on Eq.~\eqref{eq:getmoment} and~\eqref{eq:getHoTs}, respectively. Then, the synthetic equations~\eqref{eq:dis_macro} with the constitutive relations~\eqref{eq:full_constitutive} can be solved to obtain $\bm{W}_s^{k+1}$.
    % where the associated macroscopic boundary conditions are evaluated from $f_s^{n+1/2}$ as well.
    \item The converged solution of macroscopic variables of the synthetic equations $\bm{W}^{k+1}$ is used in the next DVM step to calculate the equilibrium statement $g_s^{\text{eq}}$. The distribution functions are modified to incorporate the change of macroscopic properties by changing the equilibrium part from $g_s^{\text{eq}}(\bm{W}_s^{k+1/2})$ to $g_s^{\text{eq}}(\bm{W}_s^{k+1})$:
    \begin{equation}
        \begin{aligned}[b]\label{eq:updatef}
            &f_s^{k+1} = f_s^{k+1/2} + \left[g_s^{\text{eq}}\left(\bm{W}_s^{k+1}\right) - g_s^{\text{eq}}\left(\bm{W}_s^{k+1/2}\right)\right], \quad s=1,2,\\
            &\text{with:} \quad
            g_s^{\text{eq}} = {\rho_s}\left(\frac{m_s}{2\pi  T_s}\right)^{3/2}\exp\left(-\frac{m_s(\xi - u_s)^2}{2 T_s}\right).
        \end{aligned}
    \end{equation}
\end{enumerate}
% \leir{note that $ f_s^{\text{eq}}$ is defined as the linear equilibrium vdf before! you should introduce the nonlinear equilibrium vdf in section 2.1  }

\begin{figure}[!t]
    \centering
    \subfloat[]{\includegraphics[width=0.55\textwidth,clip = true]{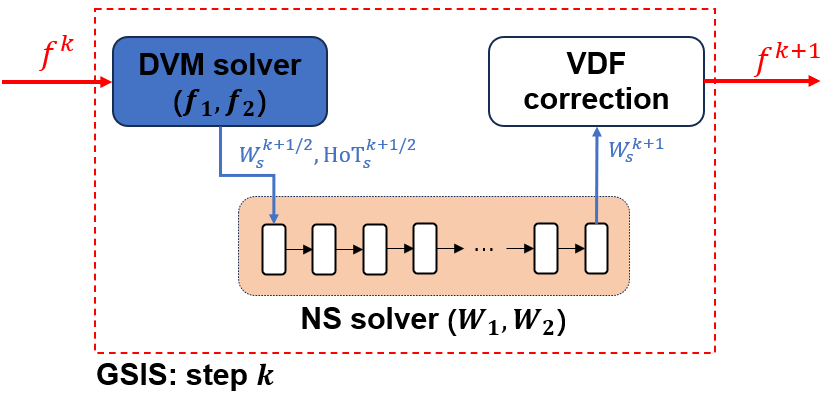}\label{fig:flowchart1}}
    \quad
    \subfloat[]{\includegraphics[width=0.35\textwidth,clip = true]{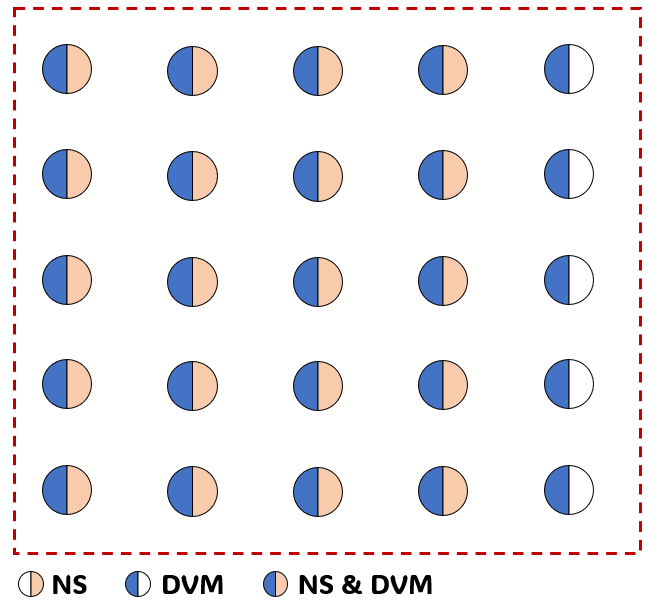}\label{fig:flowchart2}}
    \caption{ (a) GSIS flowchart. The blue blocks and orange blocks represent the mesoscopic and macroscopic solvers, respectively.  (b) Parallel computing strategy in GSIS. By utilizing MPI communication across nodes, the macroscopic solver can be parallelized across a subset or all the CPU cores.}
    % "HoT" denotes the high-order terms obtained after completing each mesoscopic solver iteration.
    \label{fig:flowchart}
\end{figure}

The above steps are repeated until convergence is achieved.
The overall computing process is illustrated in Fig.~\ref{fig:flowchart1}.  In a single GSIS step, this involves solving the DVM once and iterating the NS solver a specified number of times (e.g. 1000 times). The outer loop solves the kinetic equation using the DVM, while the inner loop solves the macroscopic equations using the LU-SGS technique. Each inner step starts from the current time step with the latest macroscopic state, HoTs, and boundary conditions from the outer loop. In order to increase the stability of the algorithm, pre-conditionings of the macroscopic and kinetic equations are adopted. Namely, before calling the GSIS, the Euler equation is simulated for 5000 steps, then the kinetic equation is solved for 10 steps.

For large scale problems, partitions in the physical domain velocity space are utilized. The macroscopic solver uses physical domain partitioning for parallel acceleration, while the mesoscopic solver utilizes both physical domain and velocity space partitioning for parallel acceleration. As illustrated in Fig.~\ref{fig:flowchart2}, the physical grid and velocity grid can be divided into $N_{\text{phy}}$ and $N_{\text{vel}}$ segments, respectively. The total number of computational cores is $N_{\text{total}} = N_{\text{phy}} \times N_{\text{vel}}$. For different computational scales, appropriate partitioning criteria can be selected, see our recent work~\cite{zhang2023efficient}. In this study, we employed a criterion of assigning approximately 1000 physical grid cells per core for physical domain partitioning.

\section{Numerical tests}\label{sec:4}

In this section, four examples are simulated to assess the GSIS for gas mixtures, especially when the mass and concentration ratios are large; they are the one-dimensional normal shock wave, two-dimensional supersonic flow past a cylinder, the three-dimensional nozzle flow, and the two-dimensional pressure driven channel flow. 

%Four types of binary gas mixtures are considered: the Mixture 1, 2, and 3 consist of Maxwell gas molecules with a mass ratio of 10, 100, 1000, respectively, while the Mixture 4 consists of hard-sphere molecules with a mass ratio of 100. All the parameters associated with relaxation ratios are given in Table~\ref{tab04:mixture_gas_perp}. 

The convergence criterion for the DVM is that the volume-weighted relative change of the moments (density, velocity, and total temperature) between two consecutive iterations
\begin{equation}
    E^k = \frac{\sqrt{\sum_i(\phi_i^k - \phi_i^{k-1})^2\myd \Omega}}{\sqrt{\sum_i(\phi_i^{k-1})^2\myd \Omega}}\Big|_{max}, \quad \phi\in(\rho, \bm{u}, T)
\end{equation}
is smaller than $\epsilon$. The error for the one-dimensional normal shock wave is set to $\epsilon=5\times 10^{-6}$, and the error for all other test cases is $\epsilon=10^{-6}$. For the macroscopic solver, the iteration stops when $\epsilon\le10^{-8}$, or when a maximum iteration step of 1000 is reached. The Courant number in the micro and macro solver are $10^8$ and $10^4$, respectively. All those tests are performed in double precision on a workstation with Intel (R) Core(TM) i7-9700K CPU@3.60GHz processors. 

For one- and two-dimensional simulation, the three-dimensional velocity space is reducted to two-dimensional by introducing the reduced velocity distribution functions, see \ref{sec:reduction}.

\subsection{Normal shock wave}

The normal shock wave of binary gas mixture is a good test case to verify the accuracy and efficiency of GSIS, due to the absence of wall boundary conditions. 
The mass of the lighter species (denoted as species 1), the mixture number density $n_u$ and the temperature $T_u$ of the upstream flow are taken as reference values, i.e., $m_0 = m_1, n_0 = n_u, T_0 = T_u $. The characteristic length $L_0$ is set to be the mean free path of the lighter species $\ell_0$ in the upstream, so that the spatial Knudsen number is $\Kn_{1} = 1$. The computational domain ranges from $100\ell_0\sim 1000\ell_0$, depending on the species mass ratio. The Mach number (Ma) is calculated based on the speed of sound $ v_{s} = \sqrt{{5 k_B T_u}/{3 m_{\text{mix}}}} $ in the upstream mixture, where $ m_{\text{mix}} = \sum m_s \chi_s $ is the average mass. Dirichlet boundary conditions are used in the upstream and downstream, where the macroscopic quantities are determined by the Rankine-Hugoniot relations, when the Mach number is $\text{Ma} = 3$.
We consider the equimolar mixture with the mass ratio $\beta_m = {m_2}/{m_1}=10,100,1000$.

\begin{figure}[p]
\centering
\begin{minipage}{0.3\linewidth}
\vspace{3pt}
\centerline{\includegraphics[scale=0.22,clip=true]{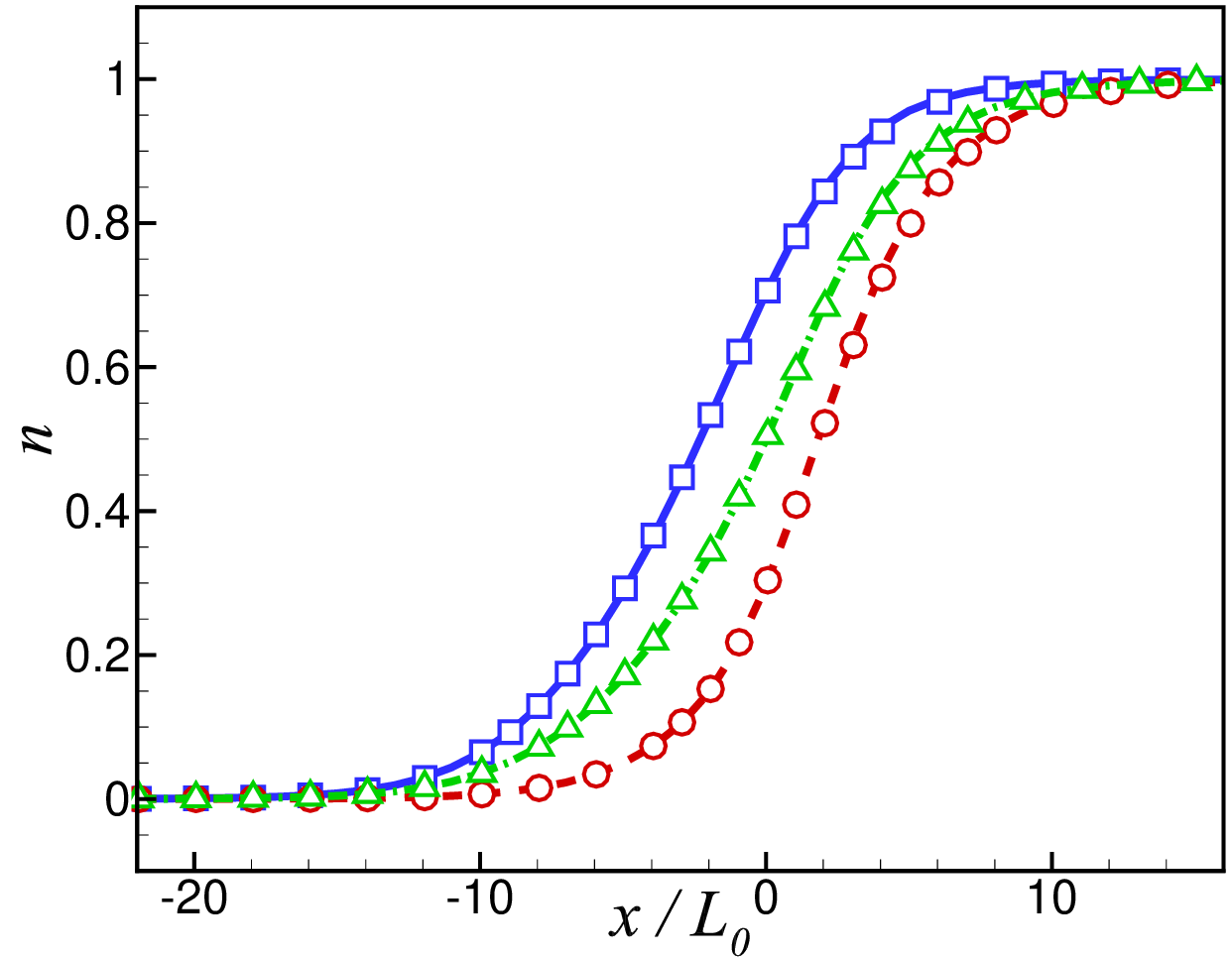}}
\vspace{3pt}
\centerline{\includegraphics[scale=0.22,clip=true]{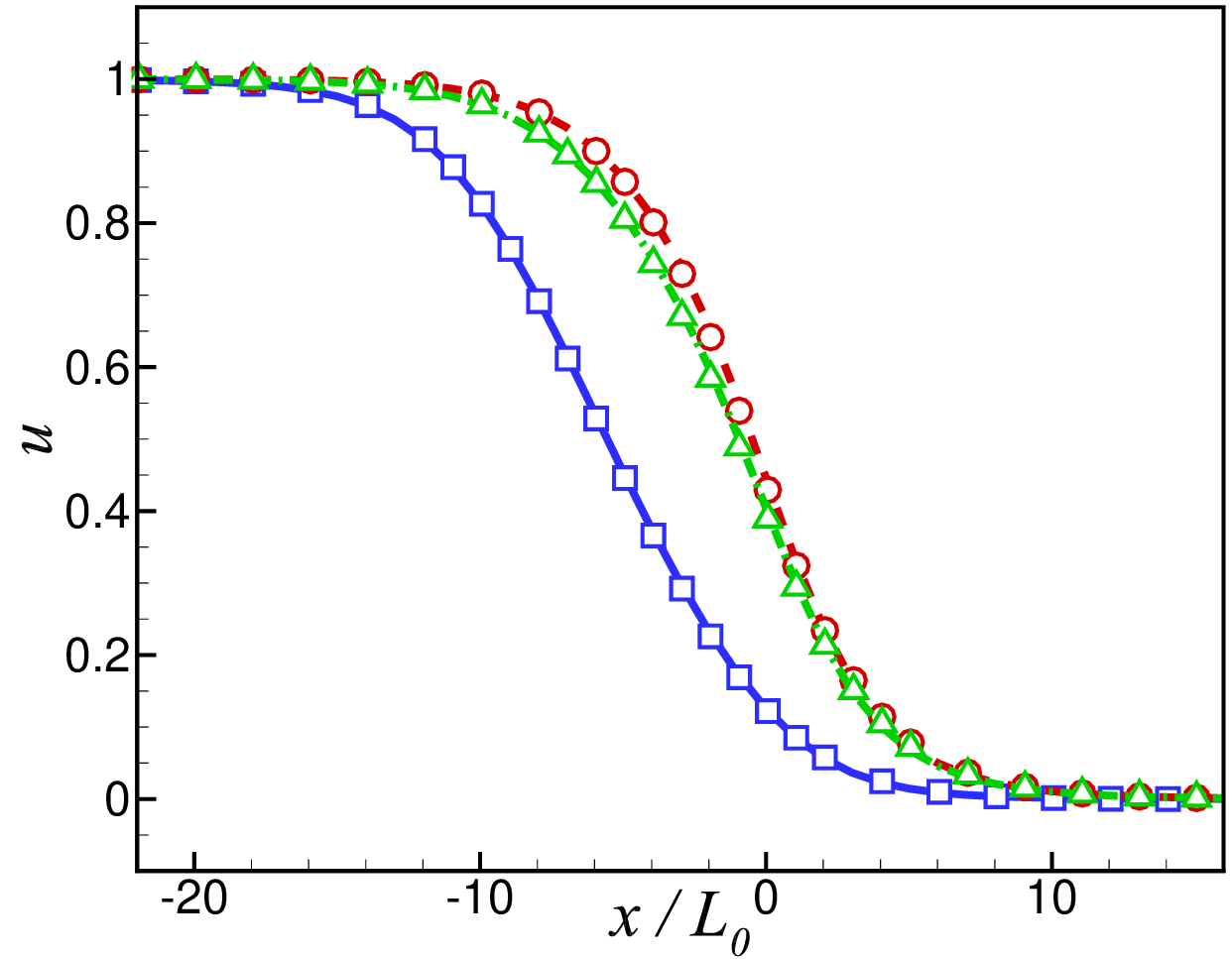}}
\vspace{3pt}
\centerline{\includegraphics[scale=0.22,clip=true]{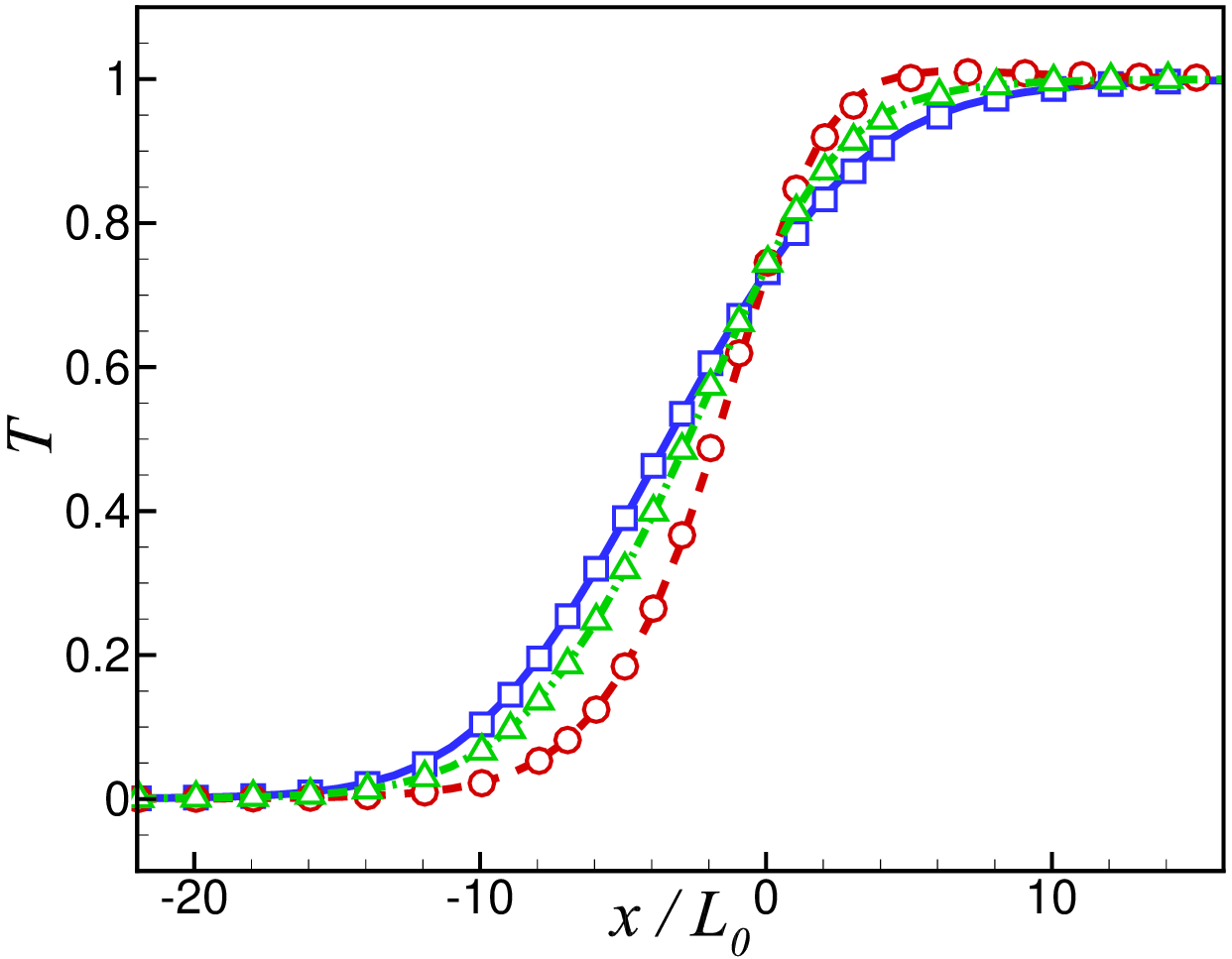}}
\vspace{3pt}
\centerline{\includegraphics[scale=0.22,clip=true]{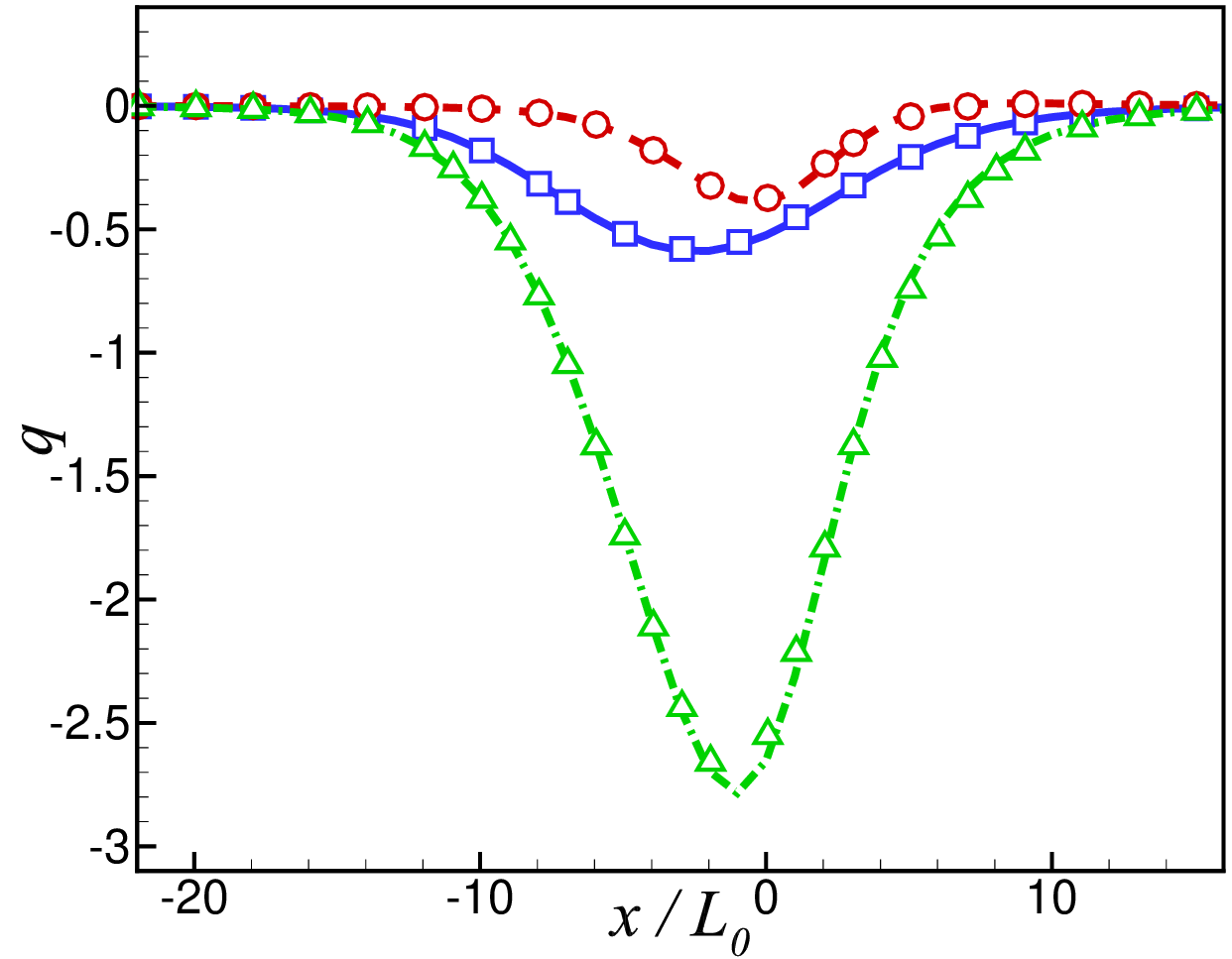}}
\vspace{3pt}
\centerline{\includegraphics[scale=0.22,clip=true]{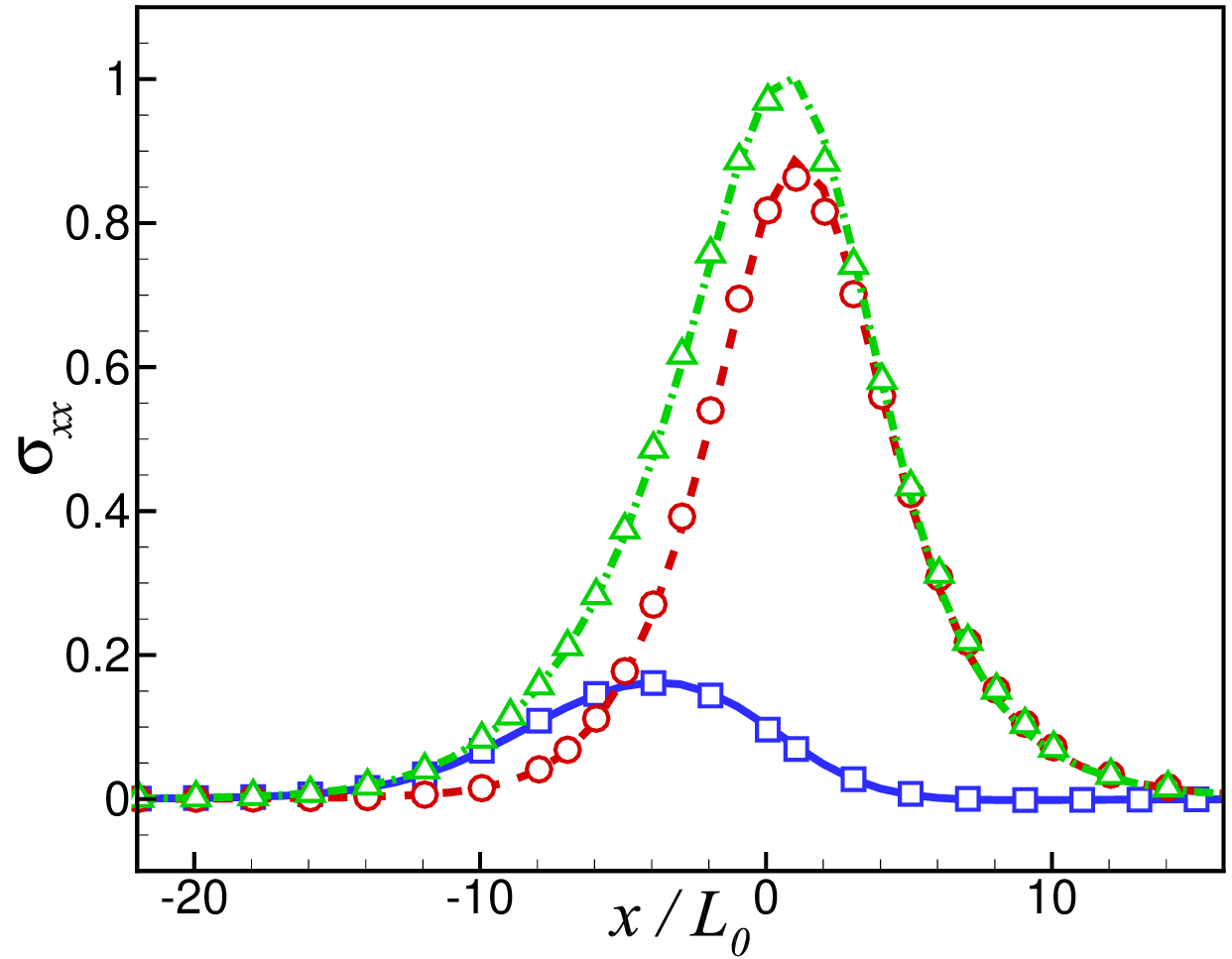}}
\end{minipage}
\begin{minipage}{0.3\linewidth}
\vspace{3pt}
\centerline{\includegraphics[scale=0.22,clip=true]{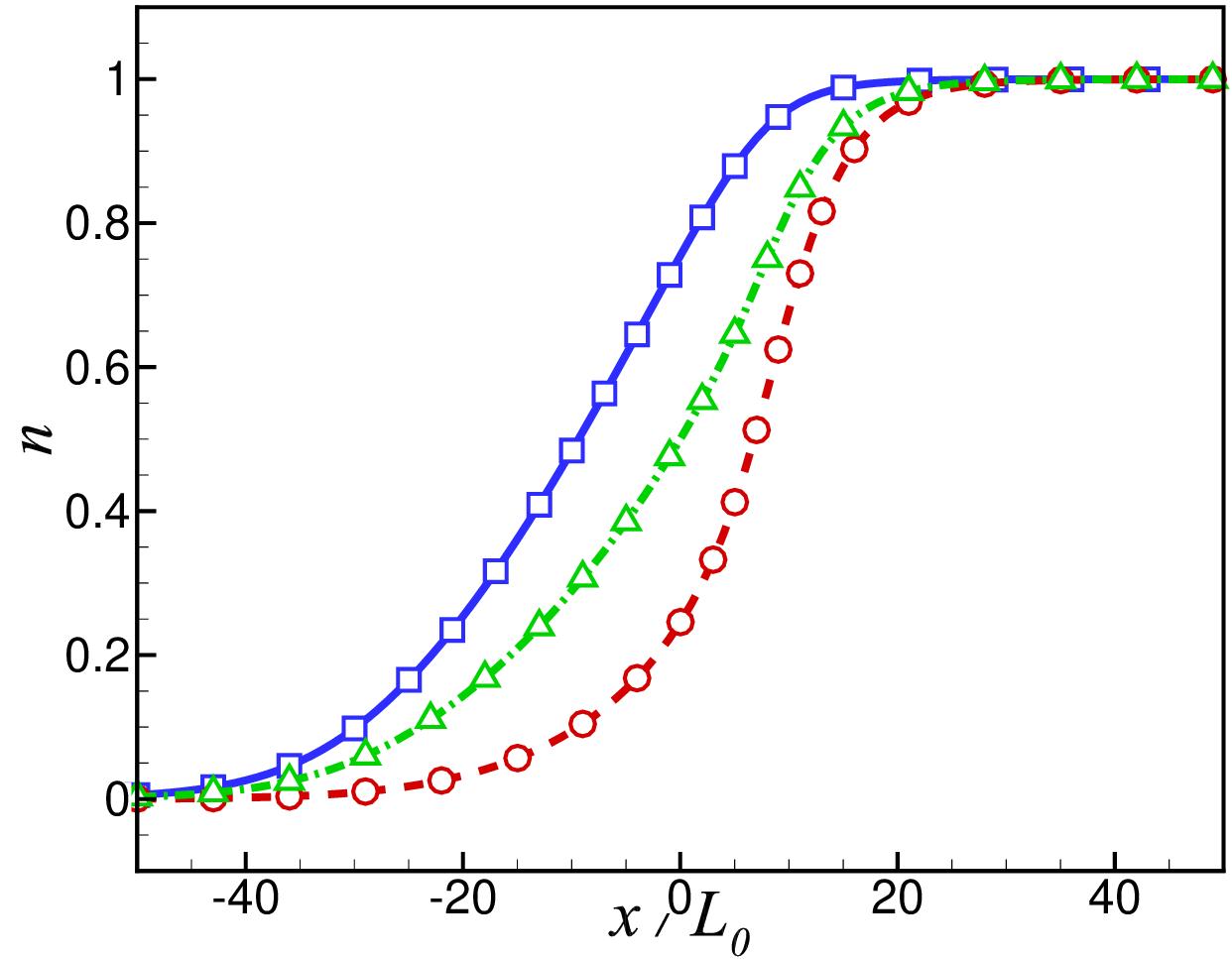}}
\vspace{3pt}
\centerline{\includegraphics[scale=0.22,clip=true]{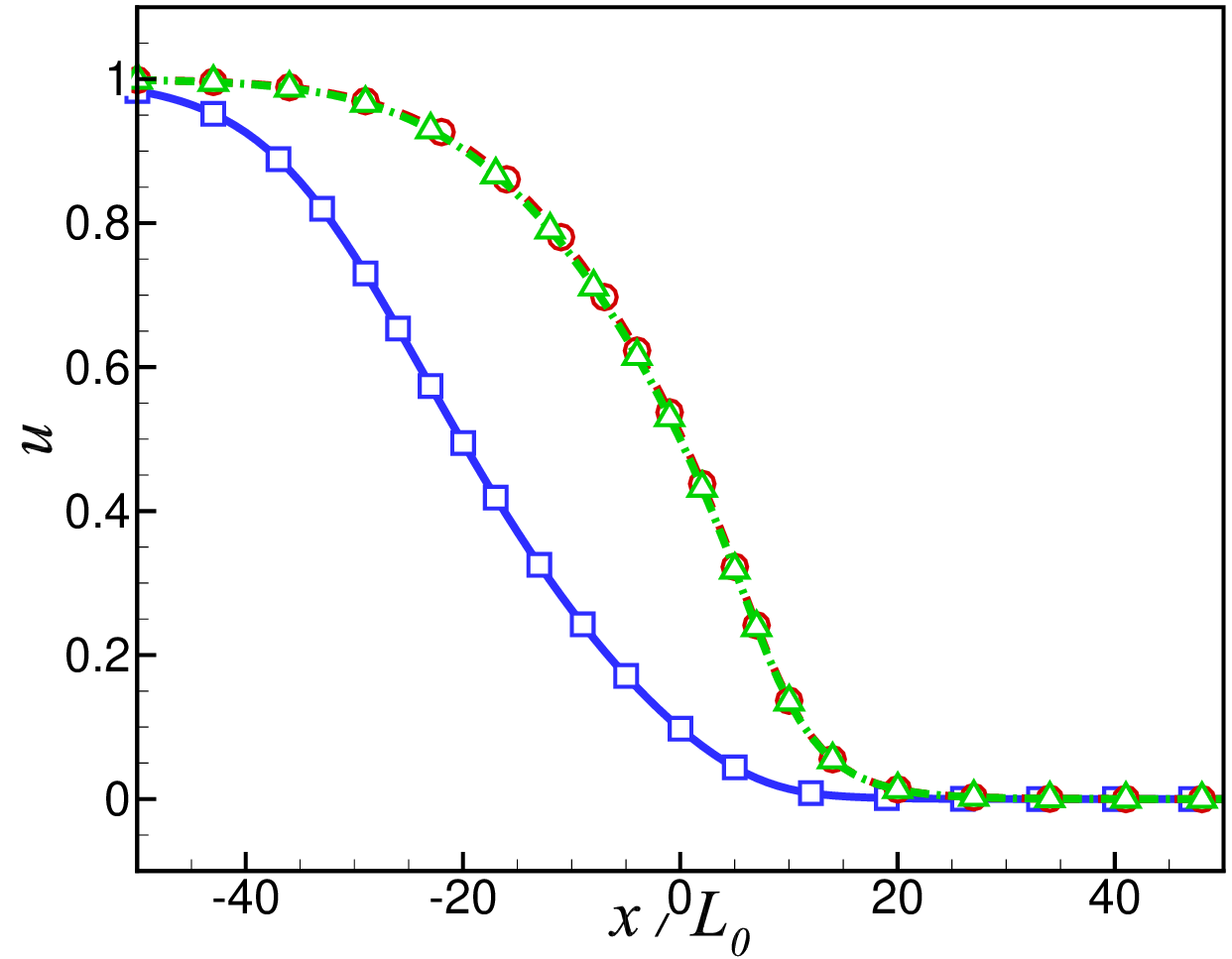}}
\vspace{3pt}
\centerline{\includegraphics[scale=0.22,clip=true]{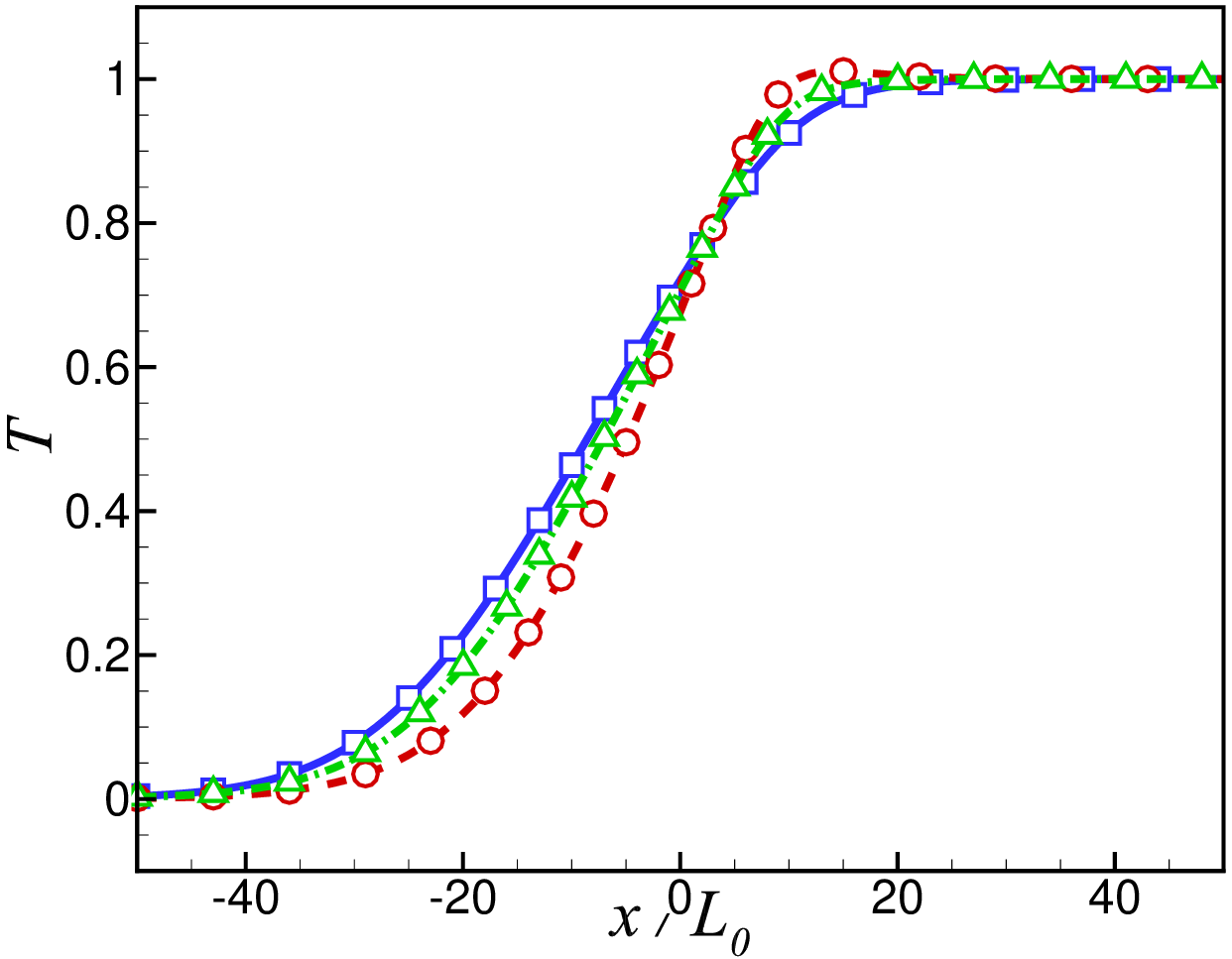}}
\vspace{3pt}
\centerline{\includegraphics[scale=0.22,clip=true]{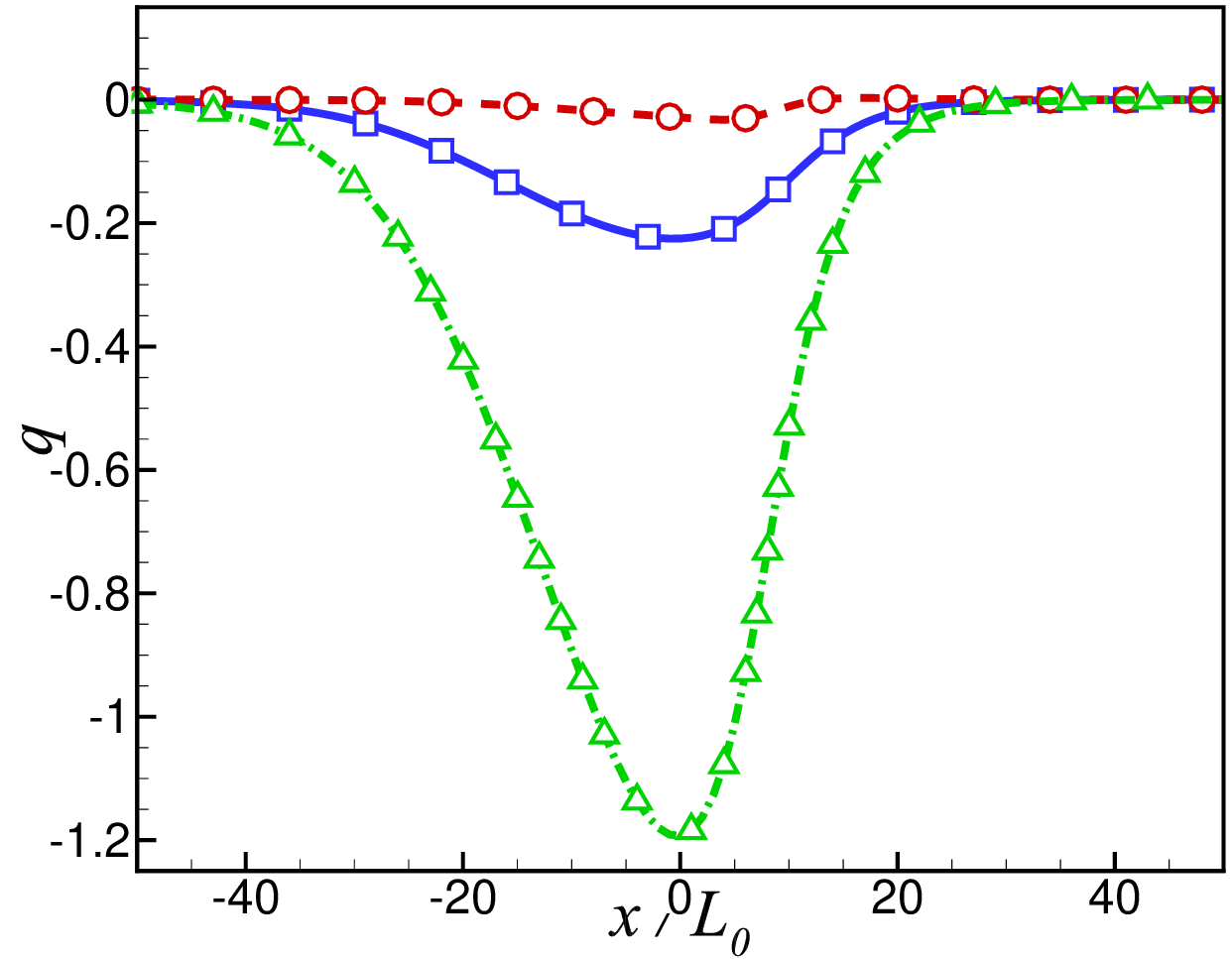}}
\vspace{3pt}
\centerline{\includegraphics[scale=0.22,clip=true]{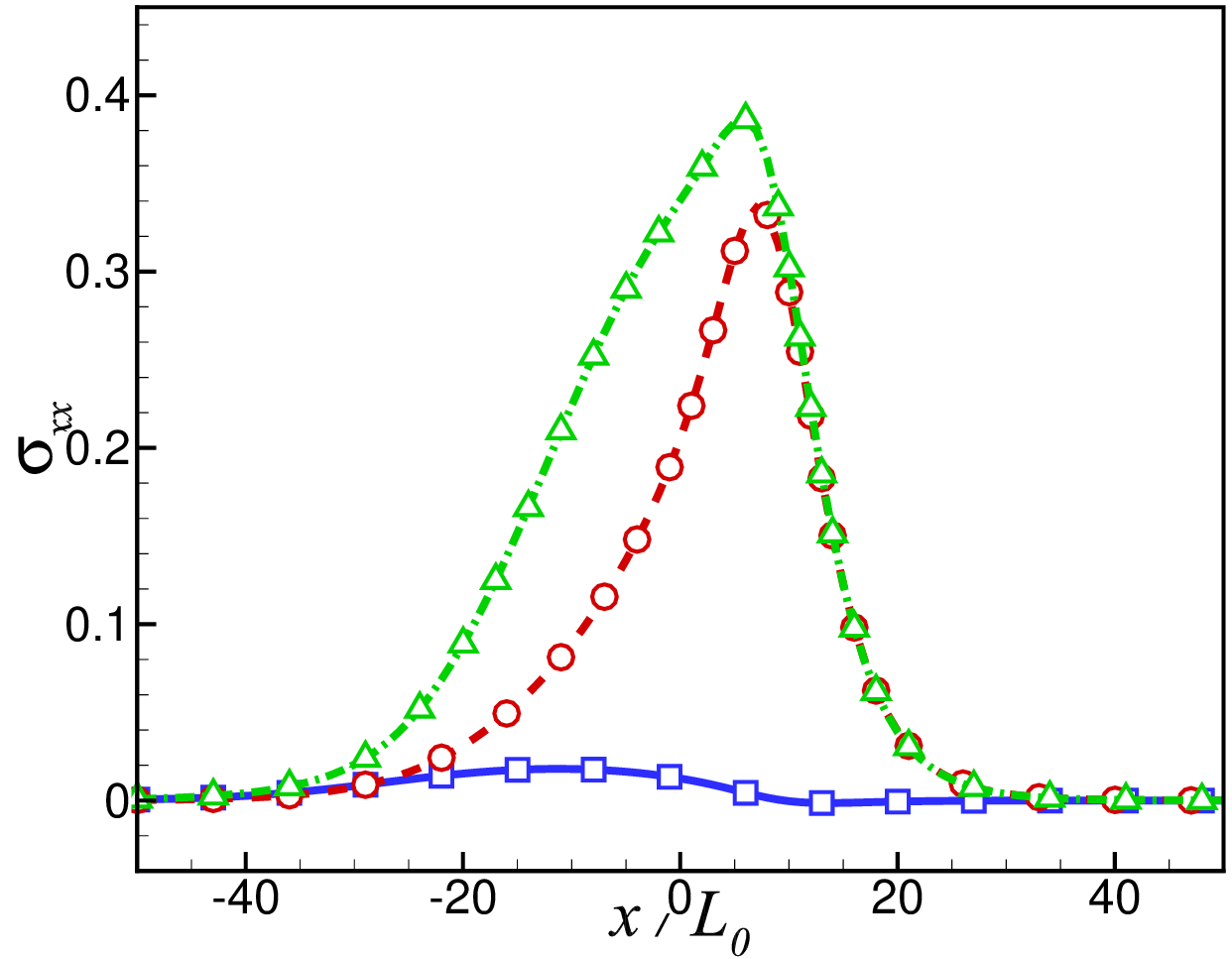}}
\end{minipage}
\begin{minipage}{0.3\linewidth}
\vspace{3pt}
\centerline{\includegraphics[scale=0.22,clip=true]{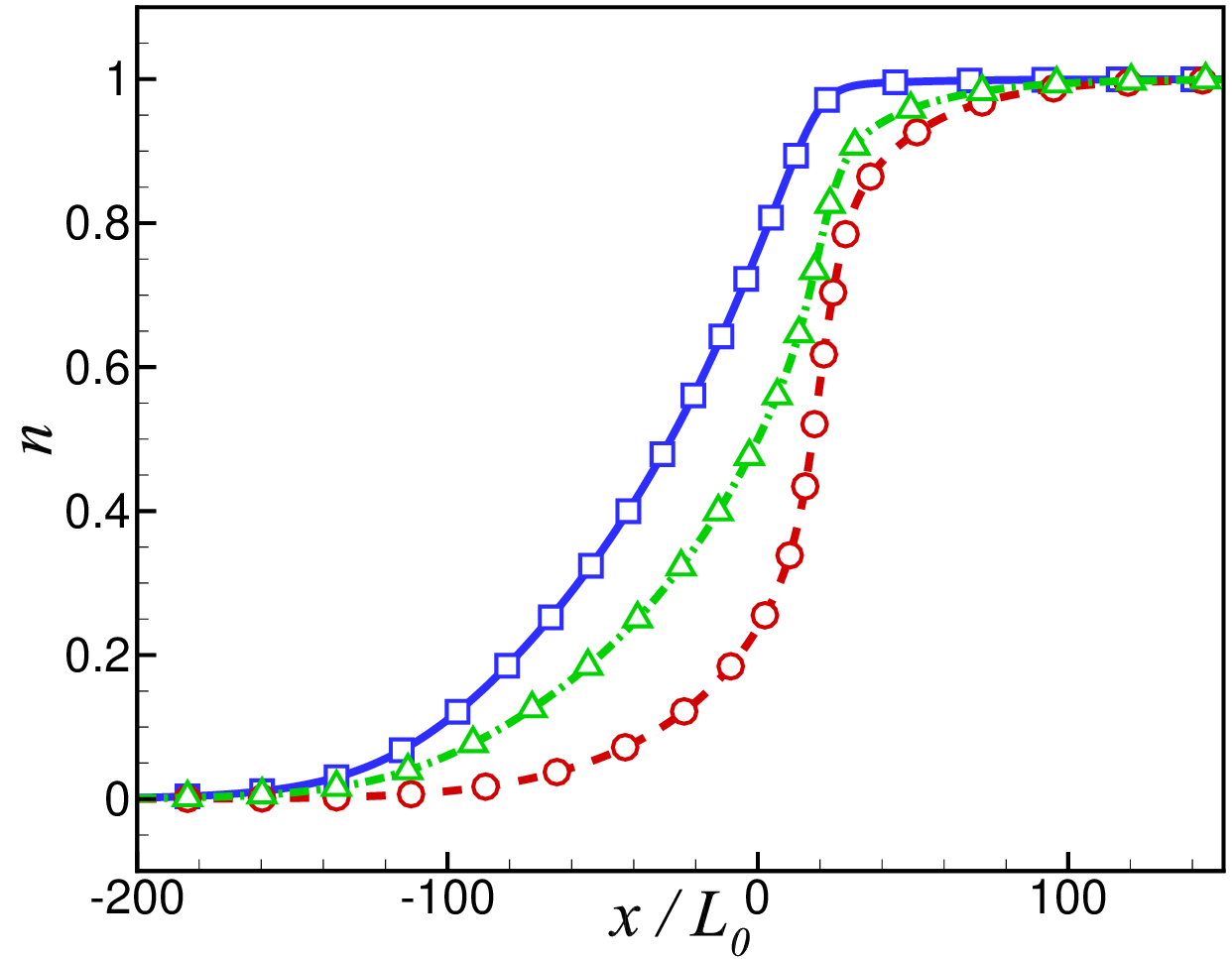}}
\vspace{3pt}
\centerline{\includegraphics[scale=0.22,clip=true]{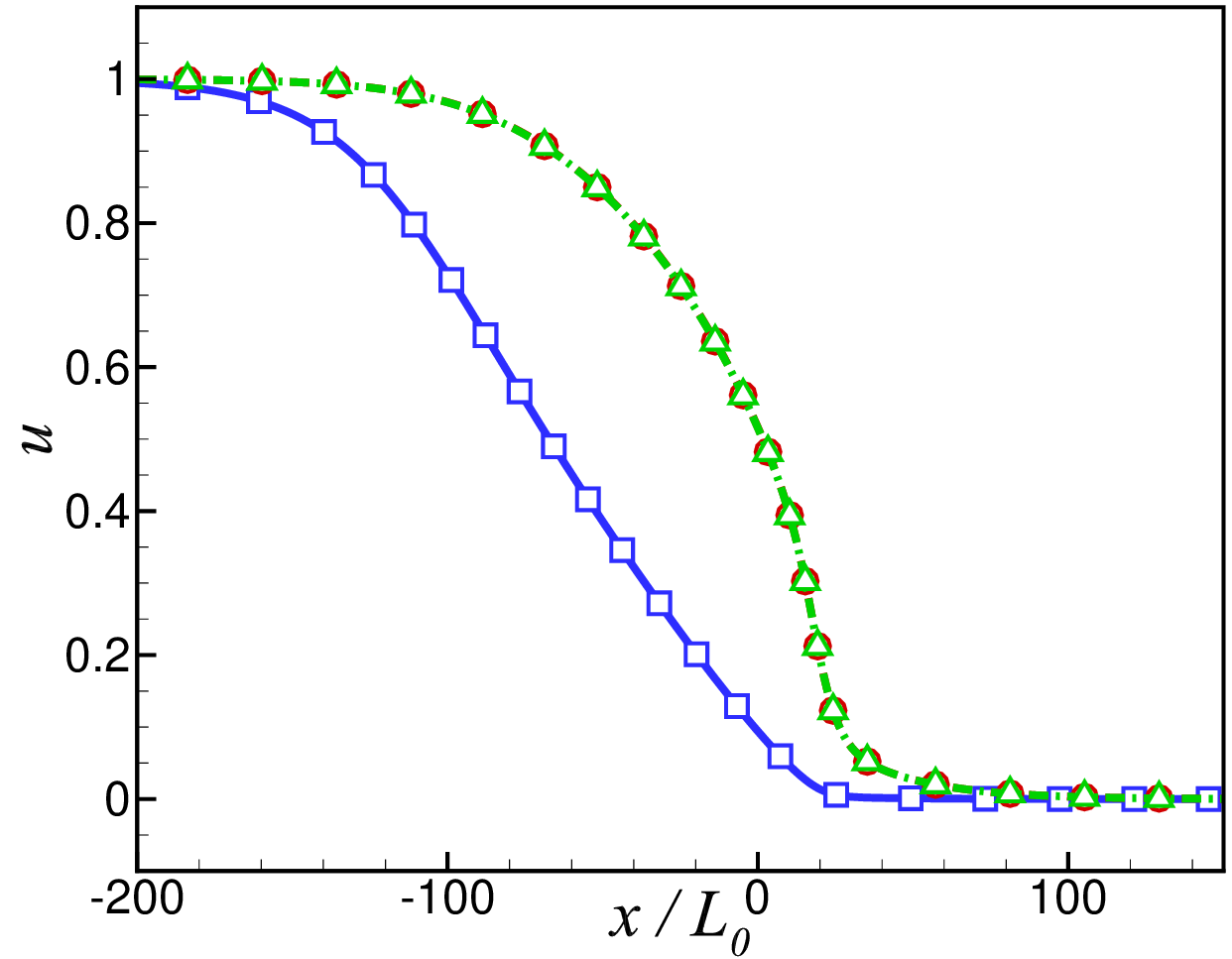}}
\vspace{3pt}
\centerline{\includegraphics[scale=0.22,clip=true]{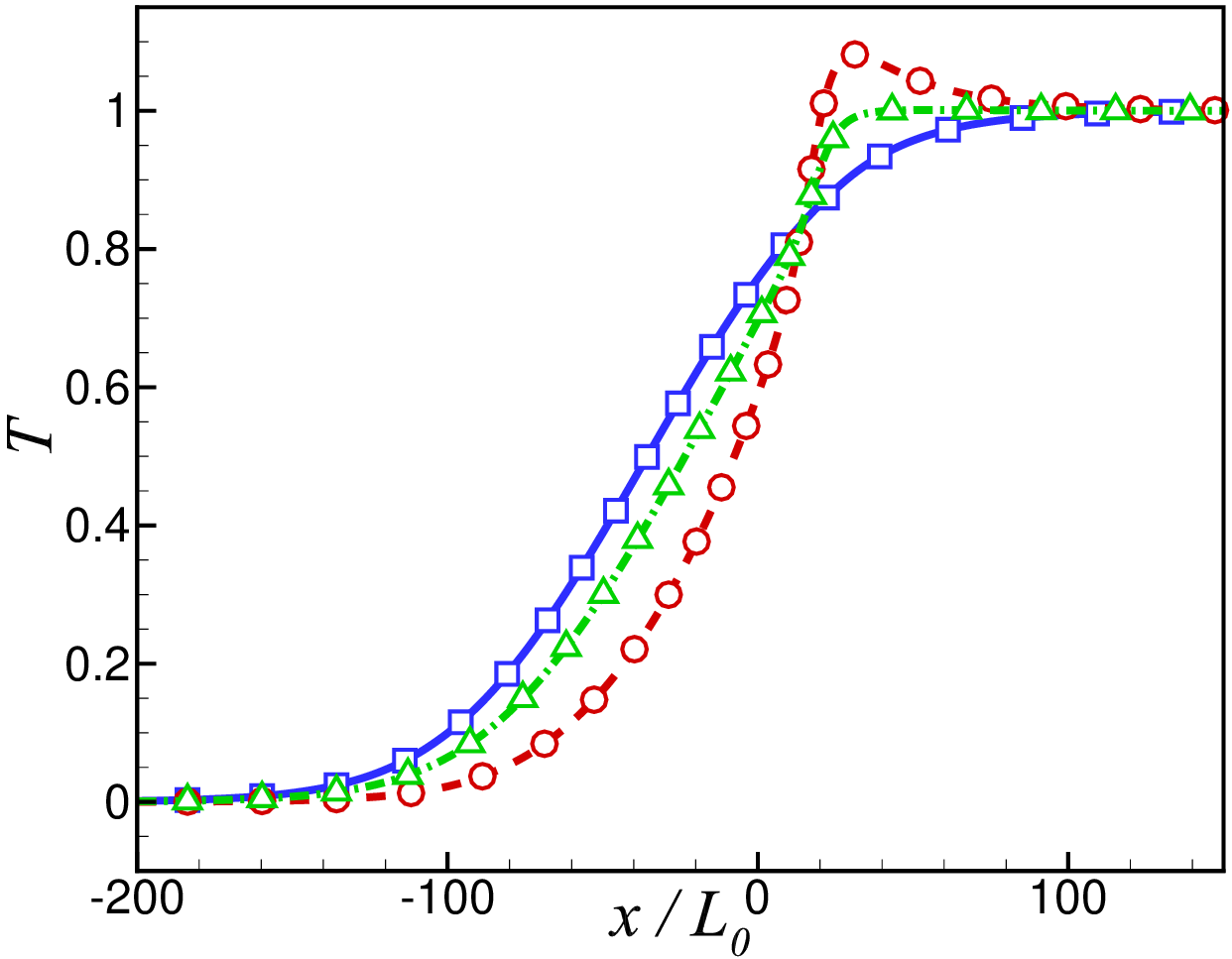}}
\vspace{3pt}
\centerline{\includegraphics[scale=0.22,clip=true]{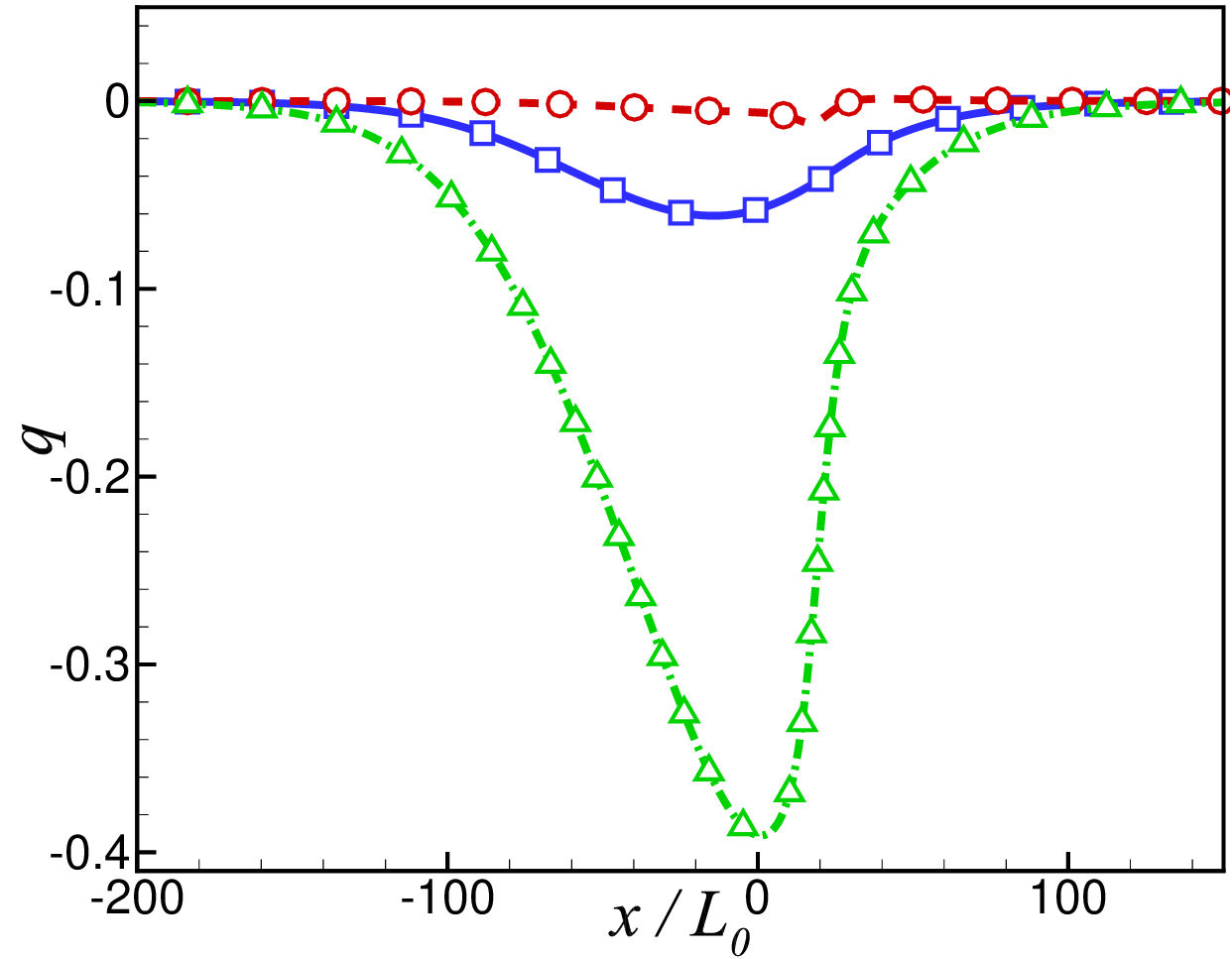}}
\vspace{3pt}
\centerline{\includegraphics[scale=0.22,clip=true]{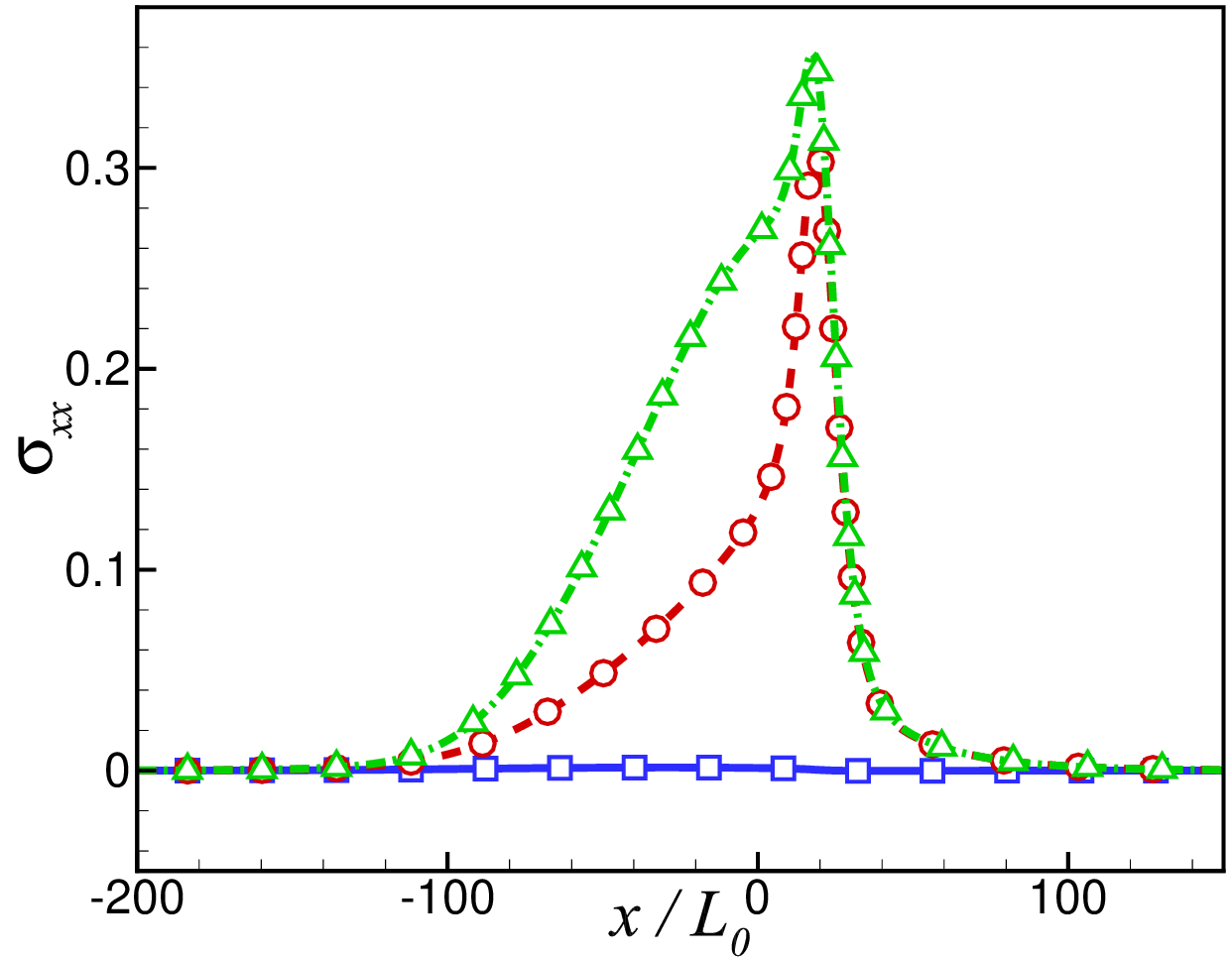}}
\end{minipage}\\
\centerline{\includegraphics[scale=0.35,clip=true]{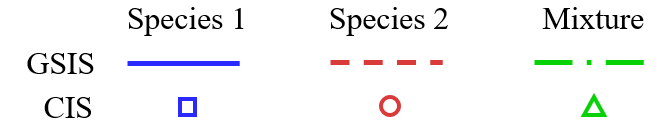}}
\caption{Comparison of the normalized number density, velocity, temperature, dimensionless heat flux and shear stress of the gas mixture between the CIS (symbols) and GSIS (lines) for the normal shock wave at $\text{Ma}=3$. From the left to right column, the  mass ratio is $\beta_m =10$, 100, and 1000, respectively. The reference pressure is defined as $ p_0 = n_0 T_0 $, and the reference heat flux is given by $ q_0 = p_0 v_{\text{mix}} $.
}
\label{fig:nswMa5}
\end{figure}

The physical space is discretized by a uniform mesh with a grid size of $\Delta x =\ell_0$. Due to the mass imparity, the velocity truncation for the two sets of distribution functions are not the same. For the light gas the velocity space in each direction is truncated to $[-12, 12]$. For the heavy species, the velocity regions in each direction are truncated to $[-5,5]$, $[-1.8,1.8]$, $[-0.7,0.7]$ when the mass ratios are $10, 100, 1000$, respectively.
The truncated velocity space is discretized uniformly, where the number of discrete velocity points for the light and heavy species are $48\times 32$ and $80 \times 64$, respectively.

The accuracy of the Li model has been extensively verified by the DSMC~\cite{li2024kinetic}, therefore here we only validate the accuracy and efficiency of the GSIS against the CIS.
Fig.~\ref{fig:nswMa5} shows the profile of number density, velocity, temperature, shear stress, and heat flux, where the GSIS results are consistent with CIS. When the mass ratio is 10, the upstream and downstream structures of the shock wave are nearly symmetric. With the increase of mass ratio, the overall thickness of the shock wave increases, mainly due to the stretched thickness in the upstream, and the asymmetric shock wave structure appears.

\begin{figure}[t]
     \centering
     \includegraphics[scale=0.4,clip = true]{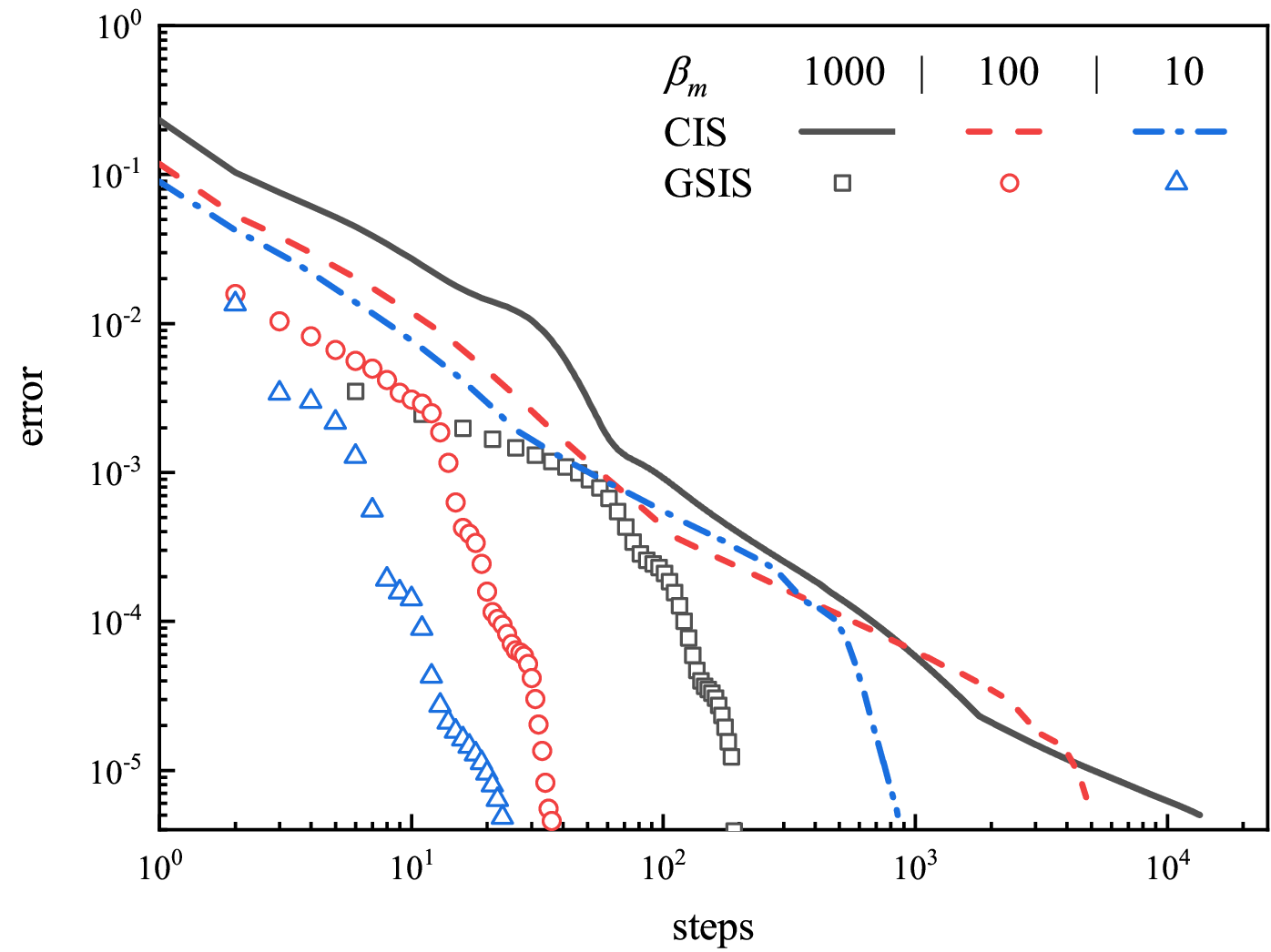}
     \caption{Error decay history of the normal shock wave at three different mass ratios $\beta_m = 10, 100, 1000$, when $\text{Ma}=3$.  
     }
     \label{fig:nsw_step_cmp}
\end{figure}

\begin{figure}[!t]
\centering
{\includegraphics[scale=0.34,clip=true]{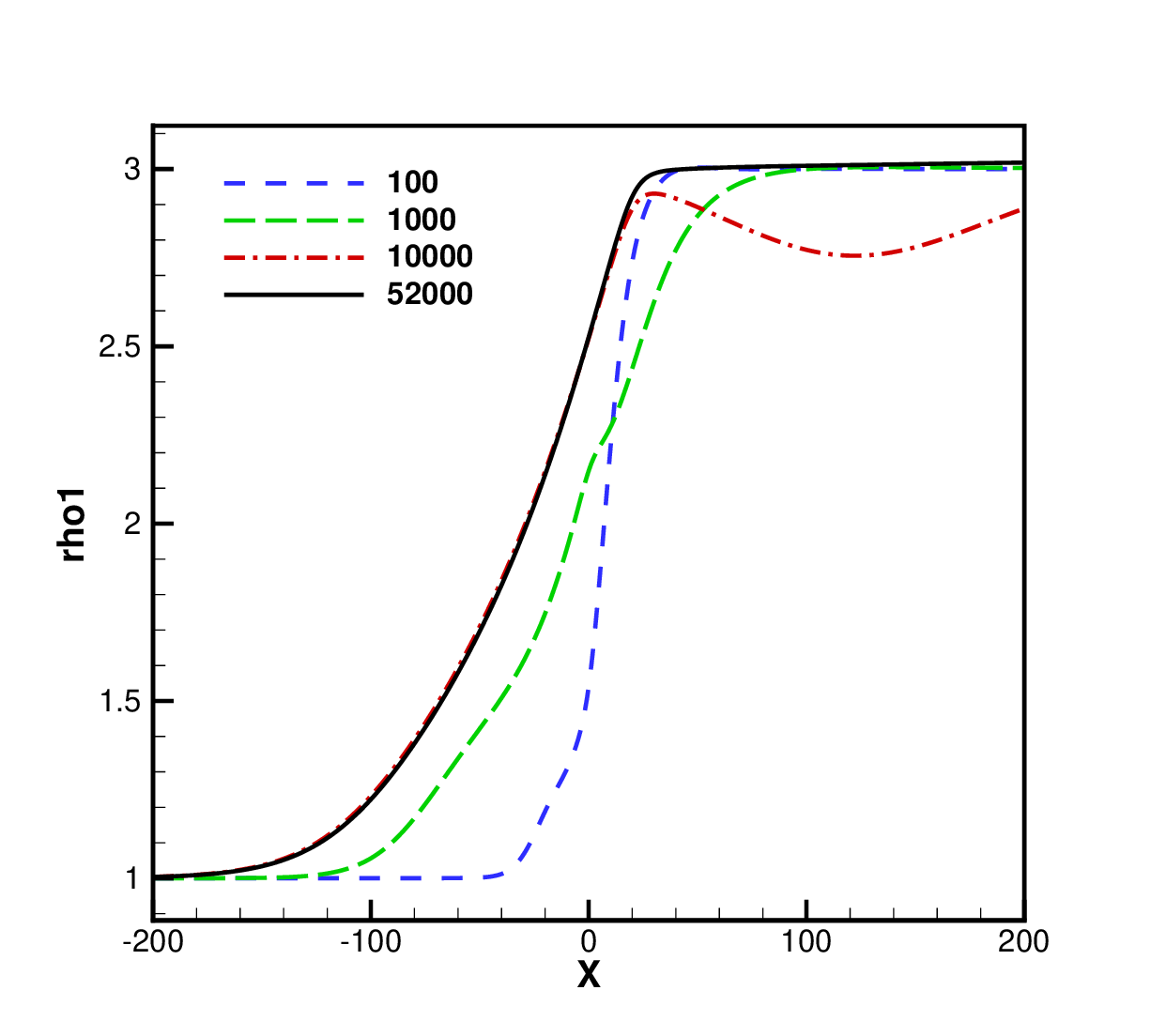}}
{\includegraphics[scale=0.34,clip=true]{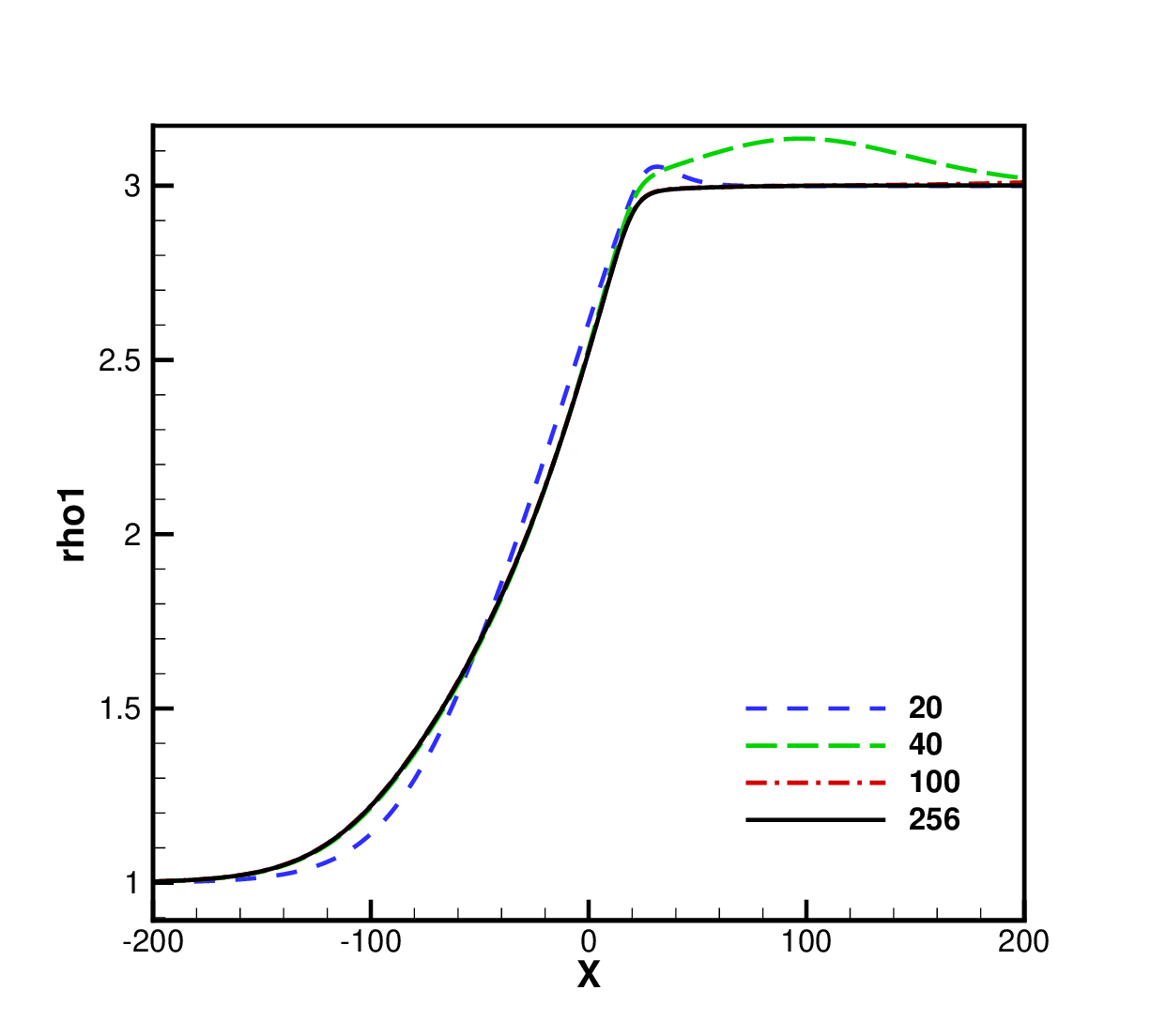}}\\
{\includegraphics[scale=0.34,clip=true]{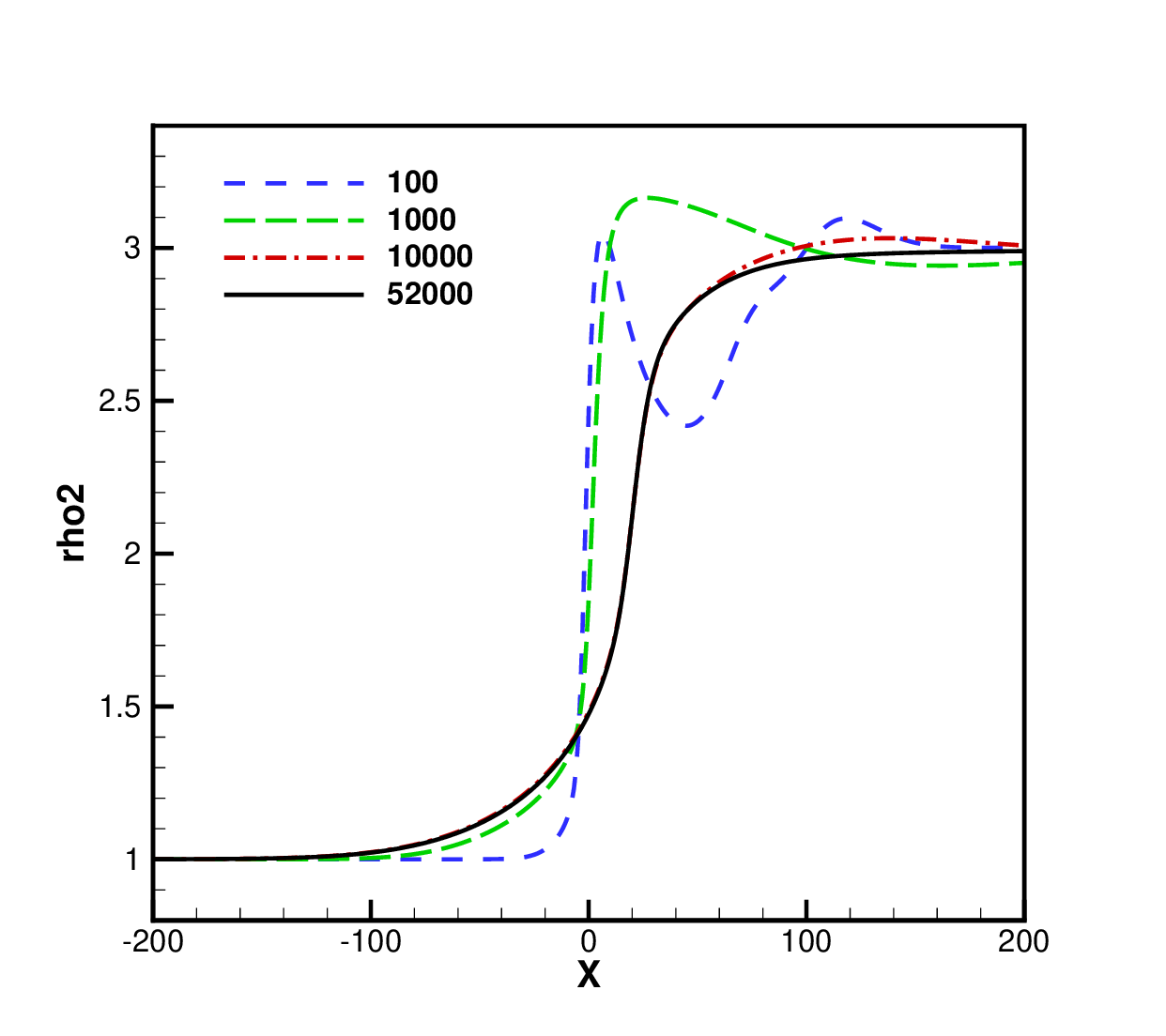}}
{\includegraphics[scale=0.34,clip=true]{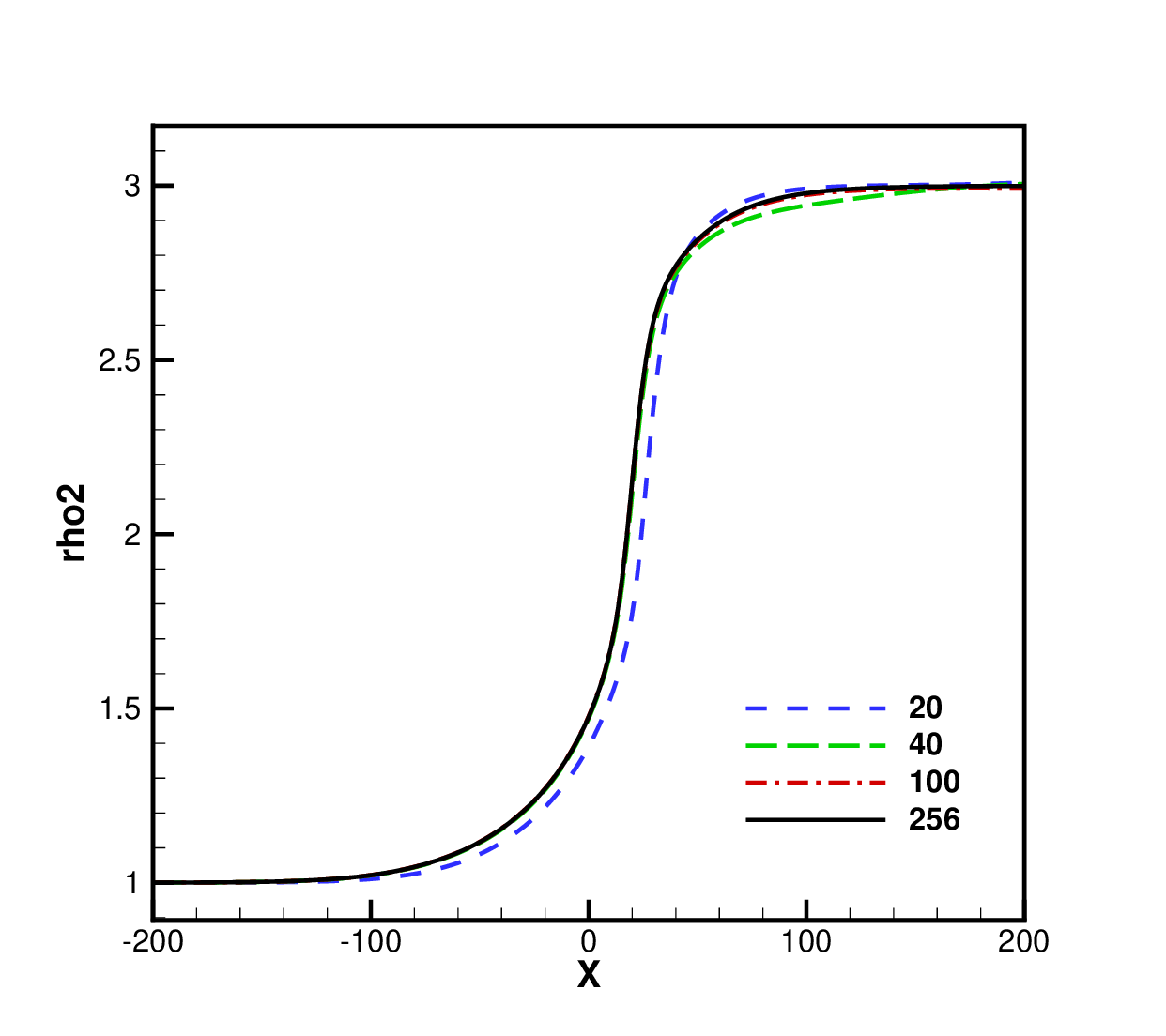}}\\
{\includegraphics[scale=0.34,clip=true]{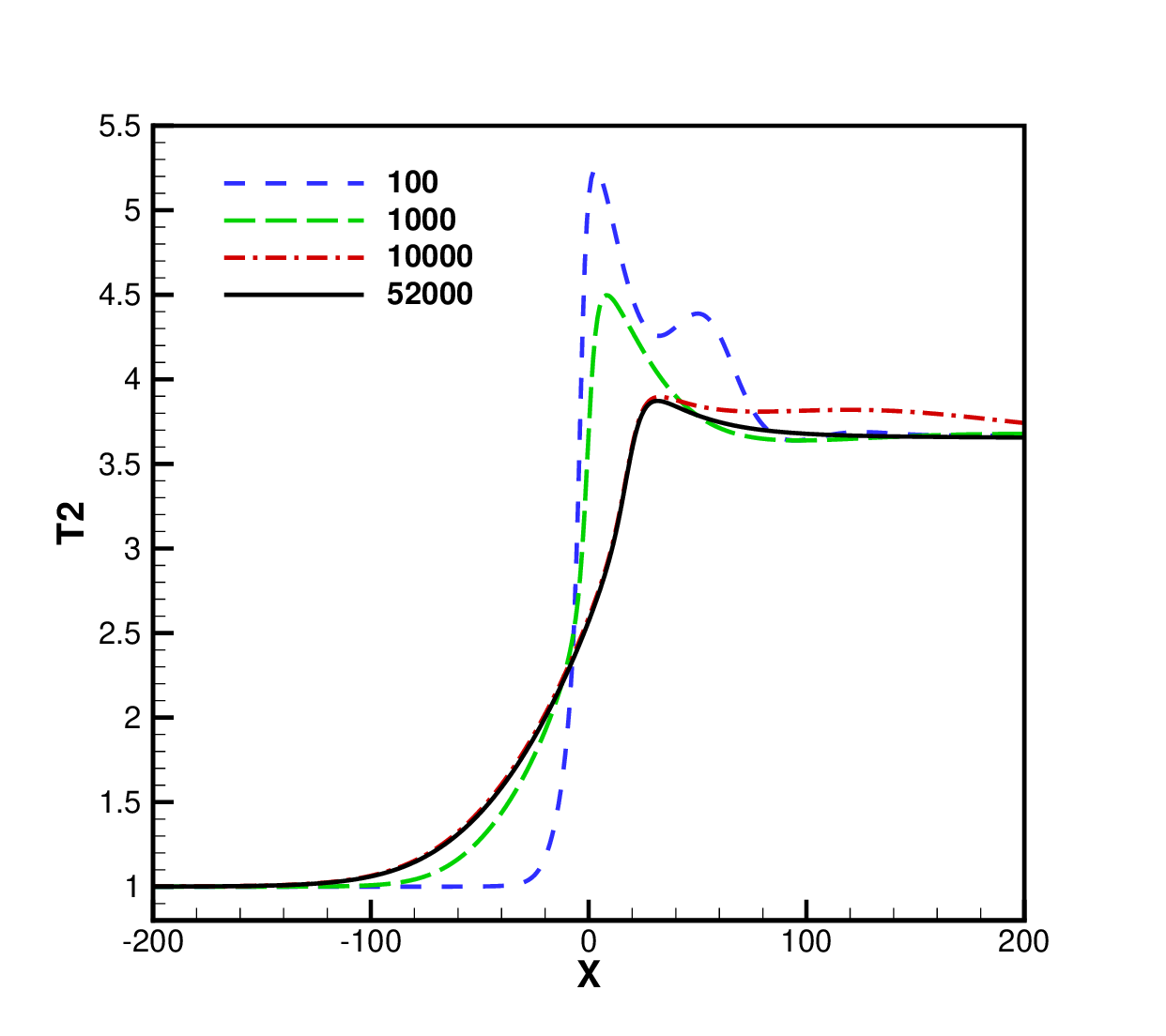}}
{\includegraphics[scale=0.34,clip=true]{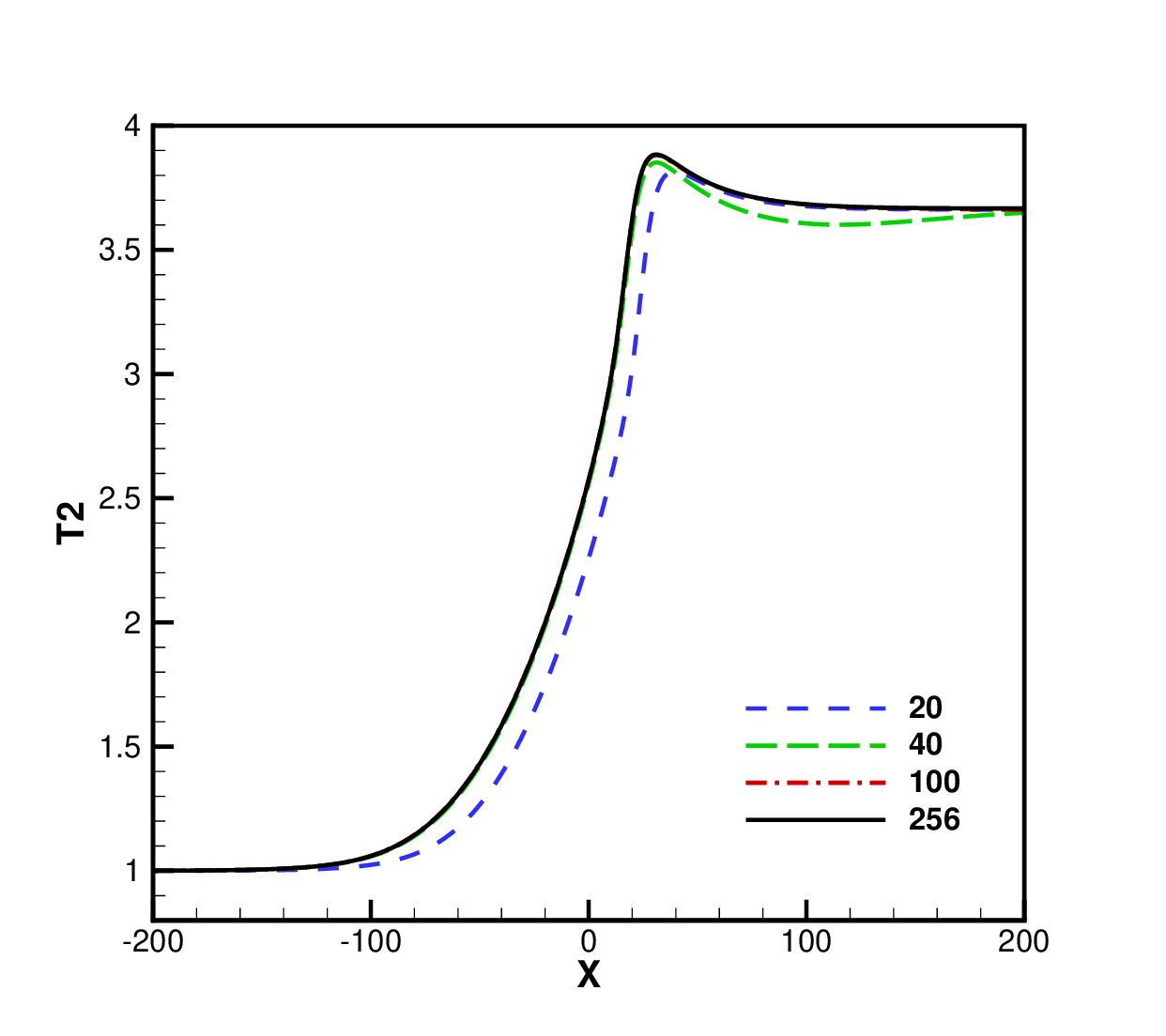}}
\caption{Histories of convergence of the mass density of light gas (first row), mass density of heavy gas, and temperature of heavy gas at different iteration steps (see the numbers in legends) obtained from the CIS (left column) and GSIS (right column), when $\beta_m = 1000$ and $Ma = 3$.}
    \label{fig:nswMa3_evo}
\end{figure}

% \begin{figure}[]
% \centering
% \begin{minipage}{0.3\linewidth}
% \vspace{3pt}
% \centerline{\includegraphics[scale=0.25,clip=true]{fig/evo_rho1_dvm.eps}}
% \vspace{3pt}
% \centerline{\includegraphics[scale=0.25,clip=true]{fig/evo_rho2_dvm.eps}}
% \vspace{3pt}
% \centerline{\includegraphics[scale=0.25,clip=true]{fig/evo_t1_dvm.eps}}
% \vspace{3pt}
% \centerline{\includegraphics[scale=0.25,clip=true]{fig/evo_t2_dvm.eps}}
% \end{minipage}
% \begin{minipage}{0.3\linewidth}
% \vspace{3pt}
% \centerline{\includegraphics[scale=0.25,clip=true]{fig/evo_rho1_gsis.eps}}
% \vspace{3pt}
% \centerline{\includegraphics[scale=0.25,clip=true]{fig/evo_rho2_gsis.eps}}
% \vspace{3pt}
% \centerline{\includegraphics[scale=0.25,clip=true]{fig/evo_t1_gsis.eps}}
% \vspace{3pt}
% \centerline{\includegraphics[scale=0.25,clip=true]{fig/evo_t2_gsis.eps}}
% \end{minipage}
% \caption{Histories of convergence of the density and temperature profiles at different iteration steps (see the numbers in legends) obtained from the CIS (left column) and GSIS (right column), when $\beta_m = 1000$ and $Ma = 3$.}
%     \label{fig:nswMa3_evo}
% \end{figure}

Figure \ref{fig:nsw_step_cmp} plots the error decay history of both CIS and GSIS. To achieve a convergence criterion of $\epsilon=5\times 10^{-6}$,
GSIS requires 21, 36, and 191 steps when $\beta_m=10,100$, and 1000, respectively, while CIS requires 833, 4991, and 13465 steps. Note that the iteration steps for GSIS exclude the initial 10 pre-conditioning steps of CIS.
The iteration steps for CIS gradually increase with the mass ratio, primarily due to the larger computational domain that makes the effective Knudsen number much smaller. Notably, for mass ratios exceeding 100, although the residuals for CIS meet the convergence criterion, downstream profiles (not depicted on the right side of Fig. \ref{fig:nsw_step_cmp}) do not fully converge. In our experience, an additional $50\%$ more iteration steps are necessary for complete convergence.

% Fig.~\ref{fig:nsw_step_cmp} shows GSIS (markers) requires 21, 36, and 191 steps, respectively, to achieve convergence criteria $5\times 10^{-6}$. In contrast, CIS (lines) requires 833, 4991, and 51980 steps under the same conditions. Since the total iteration steps of GSIS include 20 steps of DVM preprocessing, all the iteration steps given in the table are those with the 20 pre-steps removed. It can be seen that the iteration steps of CIS gradually increase with the increase of the mass ratio, mainly due to the slower progress downstream of the excitation wave, and the evolution process gradually increases with the increase of the computational domain, and it is worth mentioning that in the case of the mass ratio greater than 100, although the residuals of CIS have already reached the convergence criterion, the profiles in the downstream (not shown in the right side of Fig.2 ) do not fully converge, and in our experience, $50\%$ more iteration steps are needed to fully converge. In addition, the acceleration ratio of the computational overhead is not the same as the acceleration ratio of the iteration step, because the DVM uses ten cores for velocity space parallelism, while the macroscopic solver uses only a single core.

The fast convergence of GSIS is further illustrated when $\text{Ma} = 3, \beta_m = 1000$. Due to the larger mass ratio, the computational domain is extended to $[-500, 500]$ and 1000 uniformly spaced cells are used. Fig.~\ref{fig:nswMa3_evo} shows the  evolution history of the density and temperature in both CIS and GSIS simulations. A significant number of iterative steps is required in CIS for the perturbation to propagate to both ends and be reflected. 
To be specific, after 100 iterations, the density field of the heavy species remains largely unchanged in the upstream region, and the temperature evolution in the downstream region has only propagated approximately 100 mean free paths. After 10,000 iterations, the macroscopic quantities in the upstream region have converged; however, in the downstream region, approximately 40,000 additional steps are still required to achieve convergence.
In contrast, the macroscopic synthetic equations used in GSIS rapidly propagate and exchange information throughout the computational domain, enabling the macroscopic quantities to converge quickly. For instance, the density and temperature after 40 steps closely approximate those achieved by CIS after 10,000 steps. Full convergence is reached after 236 iterations, which is 220 times fewer than that of the CIS (52,000 steps).
Note that we focus on steady-state problems, where both CIS and GSIS employ an implicit scheme with a pseudo-time step. It's important to clarify that the evolution in the pseudo-time step does not represent the actual physical evolution.

% In contrast, the macroscopic synthetic equations used in GSIS can rapidly propagate and exchange information throughout the computational domain, allowing the physical quantities to converge rapidly after 40 steps; for example, the density and temperature after 40 steps are close to those in CIS after $10,000$ steps. The final convergence is reached after $236$ iterations, which is 220 times less than that of CIS (52000 steps).

%Furthermore, 

% \begin{figure}[h]
%     \centering
%    {\includegraphics[width=0.4\textwidth,clip = true]{fig/parallel_set1.png}}
%     \caption{Parallel computing strategy. \leir{More descriptions} }
%     \label{fig:parallel}
% \end{figure}

\subsection{Supersonic flow over cylinder}

A two-dimensional hypersonic flow with number density $n_0$ past a cylinder at $\text{Ma}_{\infty}=3$ is simulated for $\text{Kn}_1=0.005, 0.05, 0.5$, where the reference length is $L_0 = r$. The temperature of both the freestream $T_{\infty}$ and isothermal surfaces $T_w$ are maintained at $T_0$. The reference pressure and heat flux are $p_0 = n_0T_0$ and $q_0 = p_0 v_0$, respectively. The simulation is conducted in the upper half-domain $[-L, L] \times [0, L]$ due to symmetry, with $L/L_0 = 6, 10$ for Mixture 1 and 4, respectively. The spatial domain is discretized by $240\times 300$ cells, and the height of the first mesh layer on the surface is $0.02r$.  
The truncation range and number of discretization points in velocity space are set the same as those used in the normal shock wave problem.
The isothermal surface with fully diffuse gas-wall interaction is adopted. Again, equimolar mixtures are considered.

% as shown in Fig.~\ref{fig:cylindermesh}

% Consider a two-dimensional high-speed rarefied flow past a cylinder with the Mach number $Ma = 5$. The cylinder, which is placed at the origin with radius $r_{in} = 0.5$, has diffuse reflection boundary condition and temperature $T_w = 1$, while the flow domain has a radius of $r_{out} = 10r_{in}$ and temperature of $T_{\infty} = 1$. The physical domain is discretized by a structured mesh with $64\times 100$ cells. The mesh size in the normal direction is refined near the cylinder, and the height of the first mesh layer on the surface is $0.004r$. The velocity space $[-12, 12]$ is discretized as $48\times 48$ in $\xi_x, \xi_y$.

% \begin{figure}[h]
%     \centering
%     \subfloat[Whole mesh]{\includegraphics[scale=0.2,clip = true]{fig/cyd_mesh1.png}}
%     \subfloat[local enlargement]{\includegraphics[scale=0.2,clip = true]{fig/cyd_mesh2.png}}
%     \caption{The computational mesh for the cylinder case.}
%     \label{fig:cylindermesh}
% \end{figure}

\begin{figure}[!t]
    \centering
    \subfloat[number density]{\includegraphics[scale=0.2,clip = true]{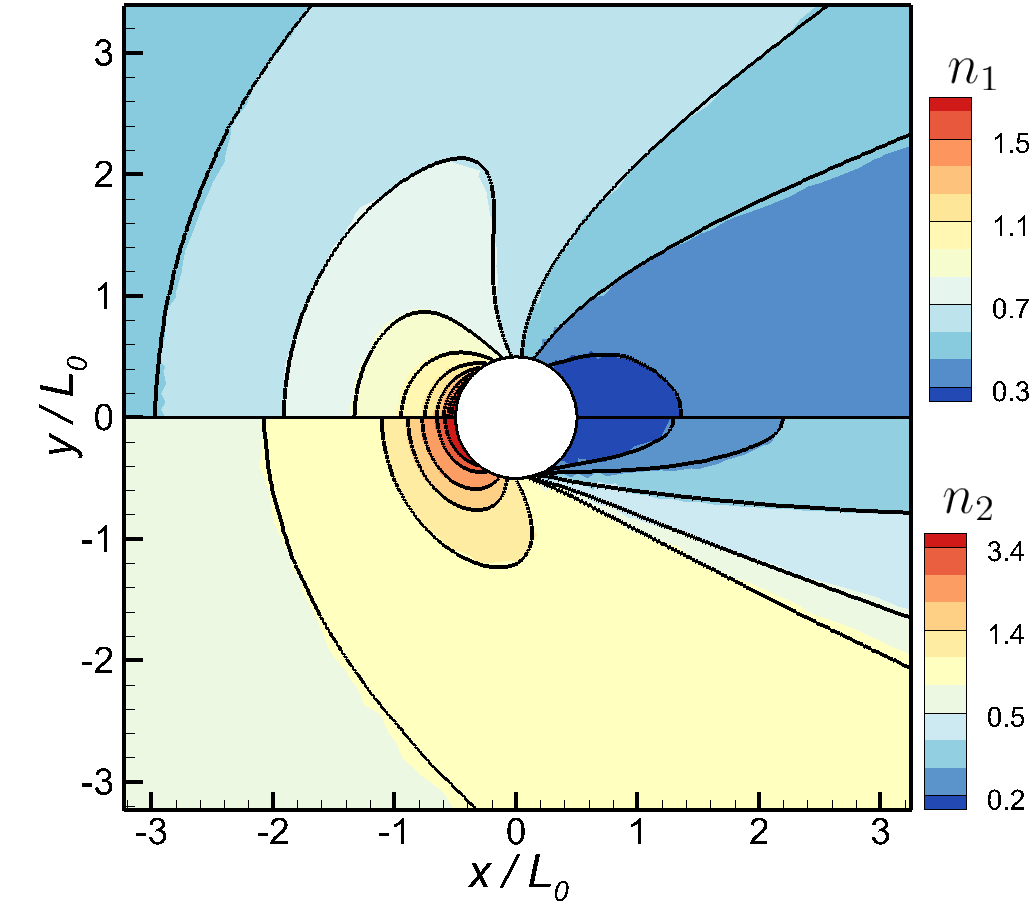}}\;
    \subfloat[velocity]{\includegraphics[scale=0.2,clip = true]{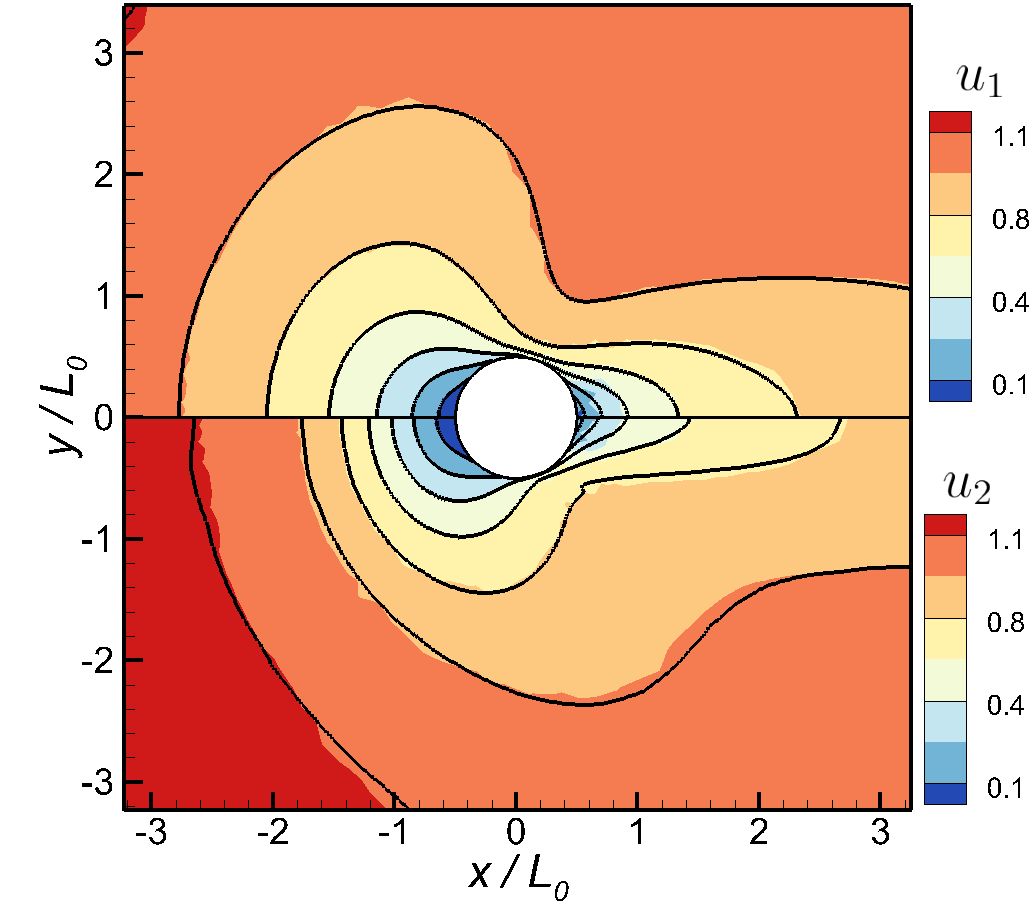}} \\
    \subfloat[temperature]{\includegraphics[scale=0.2,clip = true]{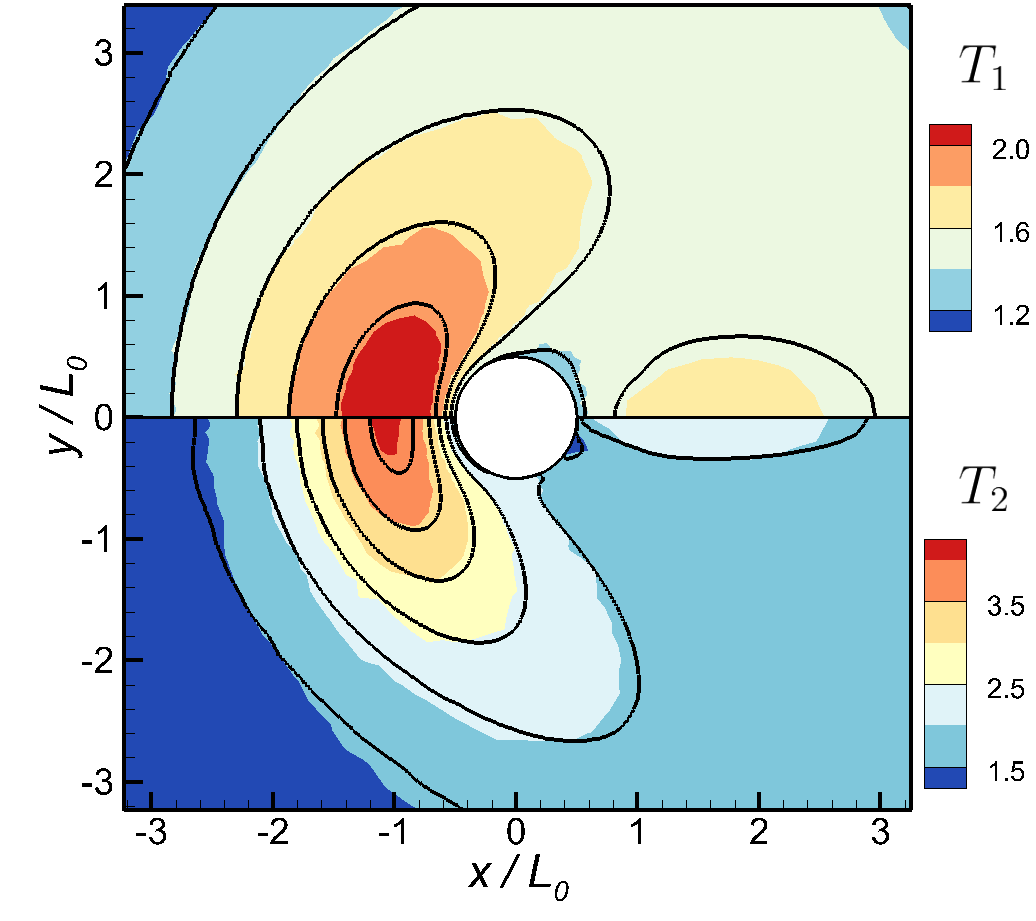}}\;
    \subfloat[heat flux in x-direction]{\includegraphics[scale=0.2,clip = true]{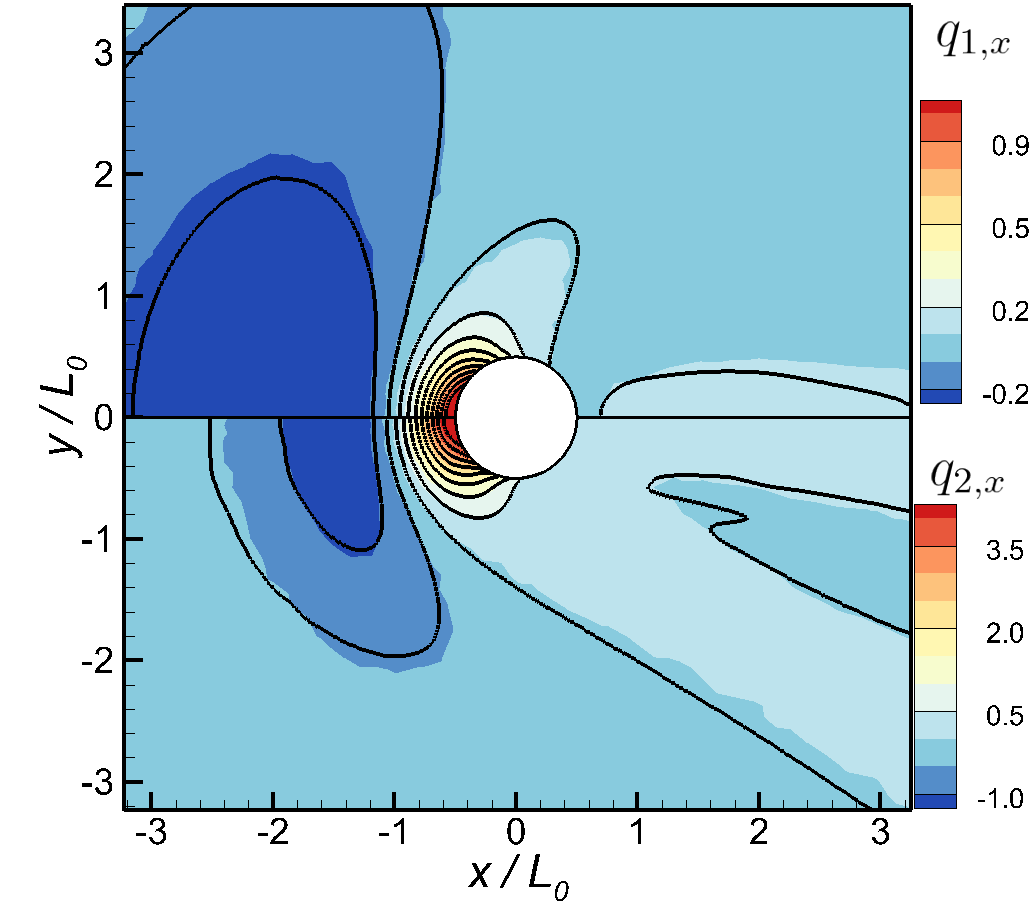}}
    \caption{Comparisons of the dimensionless (a) number density, (b) horizontal velocity, (c) temperatures and (d) heat flux in the cylindrical flow at $\text{Ma}_{\infty}=3, \text{Kn}_1=0.5$. The top and bottom half domains show the results for the light and heavy species, respectively. The background contours show the DSMC results and the black solid lines denote the GSIS results. The binary mixture consists of Mixture 1, consisting of Maxwell gases.
    }
    \label{fig:cyd_den}
\end{figure}

The GSIS is compared to DSMC in Fig.~\ref{fig:cyd_den}. The normalized density, velocity, and temperature of each species are presented under the Mixture 1 flow conditions when $\text{Kn} = 0.5$. Both species exhibit supersonic flow, as depicted by the velocity contour. Near the leading edge of the wall, the maximum density for the lighter species is approximately $1.8n_{\infty}$, whereas for the heavier species, it is around $5.8n_{\infty}$, due to its larger effective Mach number. Overall, excellent agreement between the GSIS and CIS is observed.

% For the mesoscopic equation, parallelization with 4 cores in velocity space was employed, while for the synthetic equation in GSIS, parallelization with 4 cores in physical space partitioning was utilized. 

\begin{table}[t]
 \centering
 \caption{Computational overhead of CIS and GSIS for supersonic flow past a cylinder at different Knudsen numbers for Mixture 1, using 4 cores.}
  \begin{tabular}{c c c|  c c| c c}\toprule
  & \multicolumn{2}{c}{CIS} & \multicolumn{2}{c}{GSIS} & \multicolumn{2}{c}{Speedup ratio}\\ %\cline{2-5}
{Kn} & steps & Wall time (min) & steps & Wall time (min) & in steps & in time \\ \hline
0.5 & 218 & 50.5 & 90 & 36.3 &2.4 & 1.40\\ 
0.05 & 1395 & 323.6 & 83 & 26.9 & 16 & 12.1\\ 
0.005 & 11254 & 2610.9 & 62 & 17.9 & 181 & 145.8\\ 
\bottomrule
\end{tabular}
 \label{tab04:Ma3_compare_time}
\end{table}

Table~\ref{tab04:Ma3_compare_time} shows the computational time for simulating the two-species flow over a cylinder. The GSIS achieves convergence within 100 steps across the entire flow regimes. Specifically, when $\text{Kn}=0.005$, GSIS requires only 62 steps to find the steady state. In contrast, even if CIS employs an implicit algorithm with a CFL of $10^8$ to solve the pure mesoscopic equations, it still requires 11,254 iteration steps. Therefore, the GSIS reduces the iteration steps by approximately 181 times.
It is worth noting that, compared to our previous work in polyatomic gas flow~\cite{zeng2023general}, the computational iteration steps of GSIS-mixture have slightly increased. This is mainly attributed to the enlargement of the spectral radius in the implicit iteration of macroscopic synthetic equations for the mixture gas, to handle the additional momentum source terms in Eq.~\eqref{momentum_exchange}. This adjustment reduces the equivalent time step in the near-continuum flow, slowing down the convergence speed of the internal loop of the macroscopic synthetic equations, thereby diminishing the convergence speed of GSIS.

% \begin{figure}[!t]
%     \centering
%     {\includegraphics[scale=0.35,clip = true]{fig/2D_cyd_ms10_line_n.eps}}
%     {\includegraphics[scale=0.35,clip = true]{fig/2D_cyd_ms10_line_u.eps}}\\
%    {\includegraphics[scale=0.35,clip = true]{fig/2D_cyd_ms10_line_t.eps}}
%    {\includegraphics[scale=0.35,clip = true]{fig/2D_cyd_ms10_line_qx.eps}}
%     \caption{Comparison of the number density, horizontal velocity, temperature, and heat flux along the stagnation line for Mixture 4 cylinder flow, consisting of hard sphere gas. Lines and symbols  represent the GSIS and DSMC results, respectively. Squares and triangles denotes the lighter and heavier species, respectively. Flow conditions are $\text{Kn}_1=0.5, \text{Ma}_{\infty}=3$ and $ \beta_n = 1$.}
%     \label{fig04:cyd_ms100}
% \end{figure}

\begin{figure}[]
\centering
\subfloat[$\text{Kn}=0.5$]{
% \centering
\includegraphics[scale=0.14,clip=true]{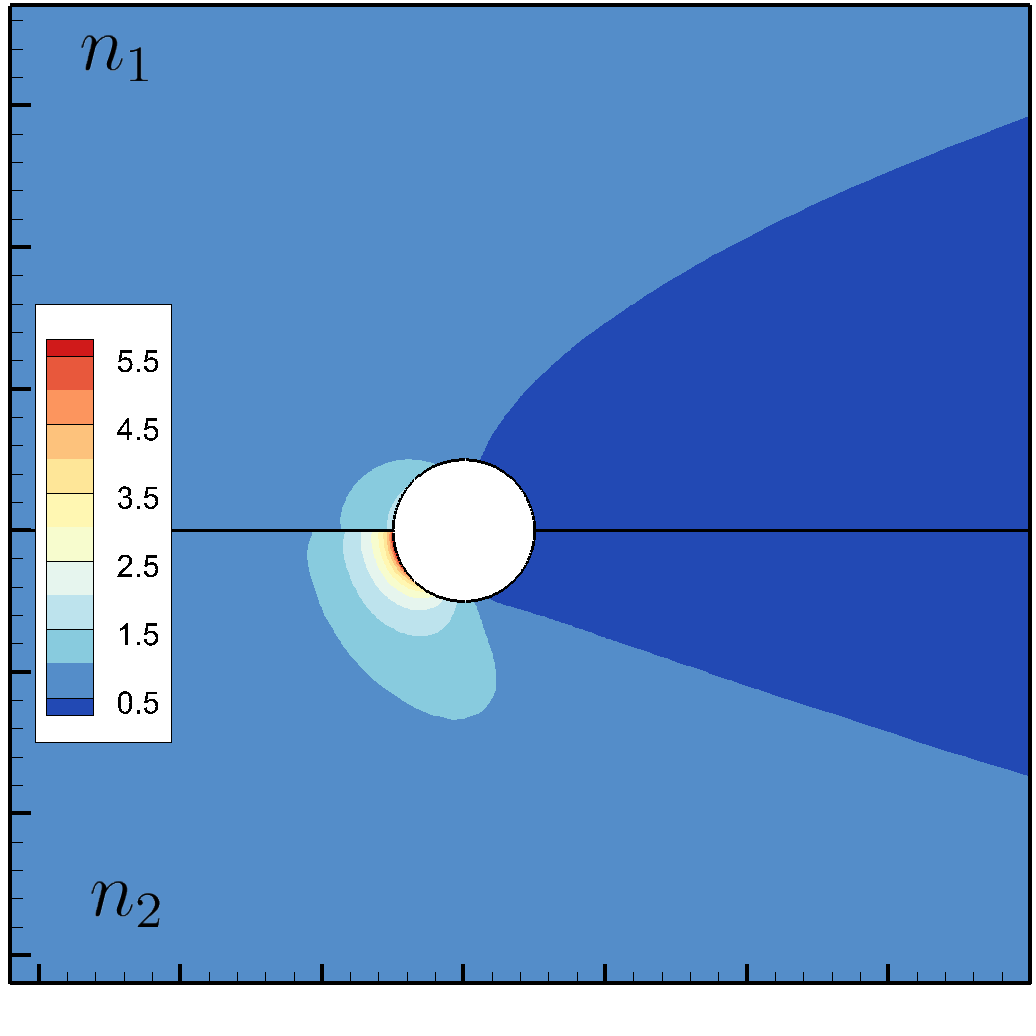}\;
\includegraphics[scale=0.14,clip=true]{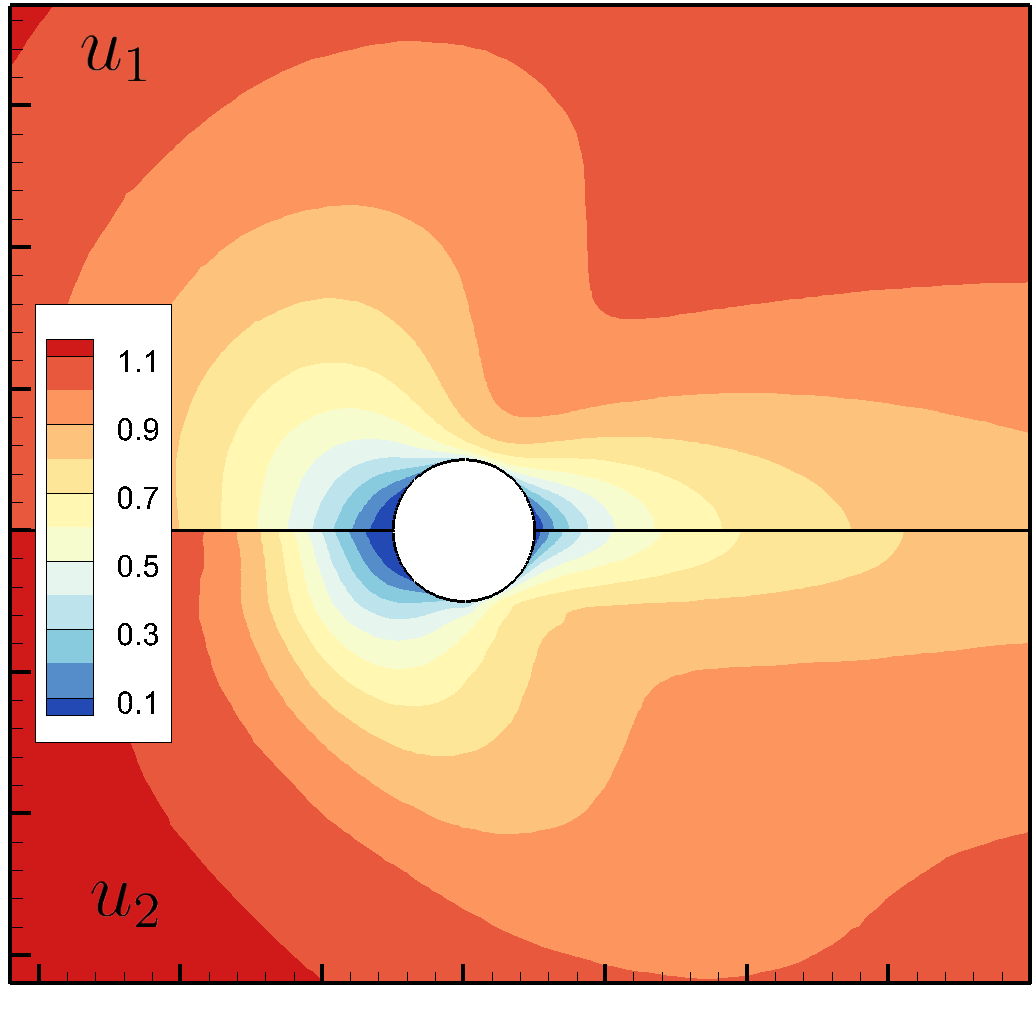}\;
\includegraphics[scale=0.14,clip=true]{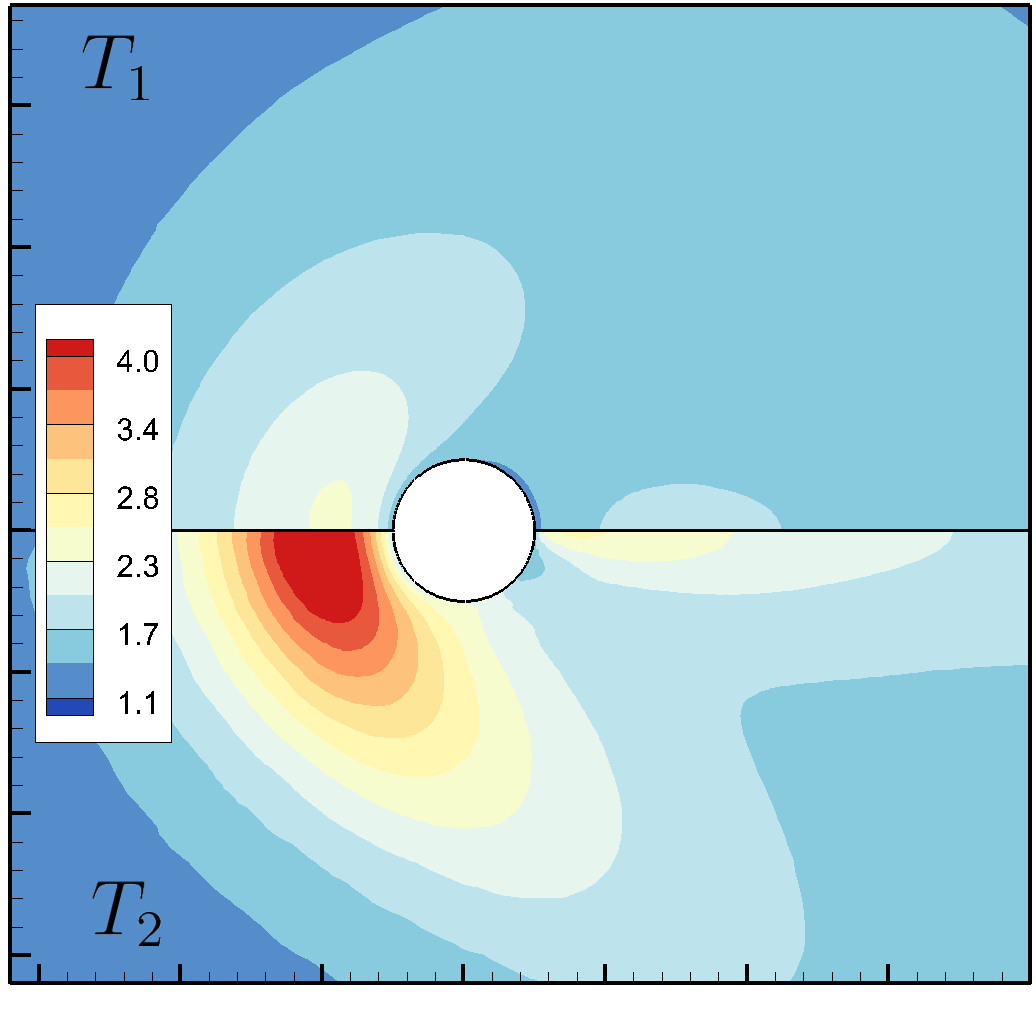}
}\\
\subfloat[$\text{Kn}=0.05$]{
% \centering
\includegraphics[scale=0.14,clip=true]{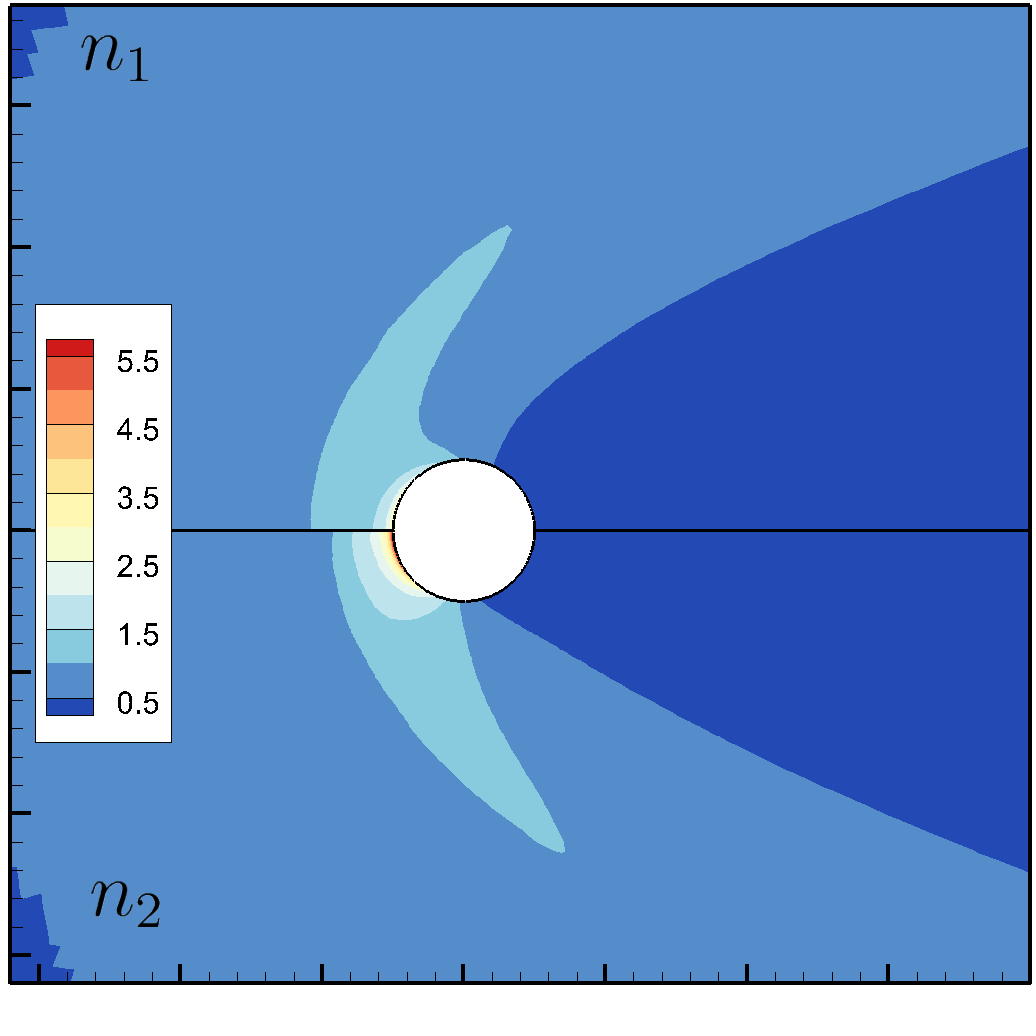}\;
\includegraphics[scale=0.14,clip=true]{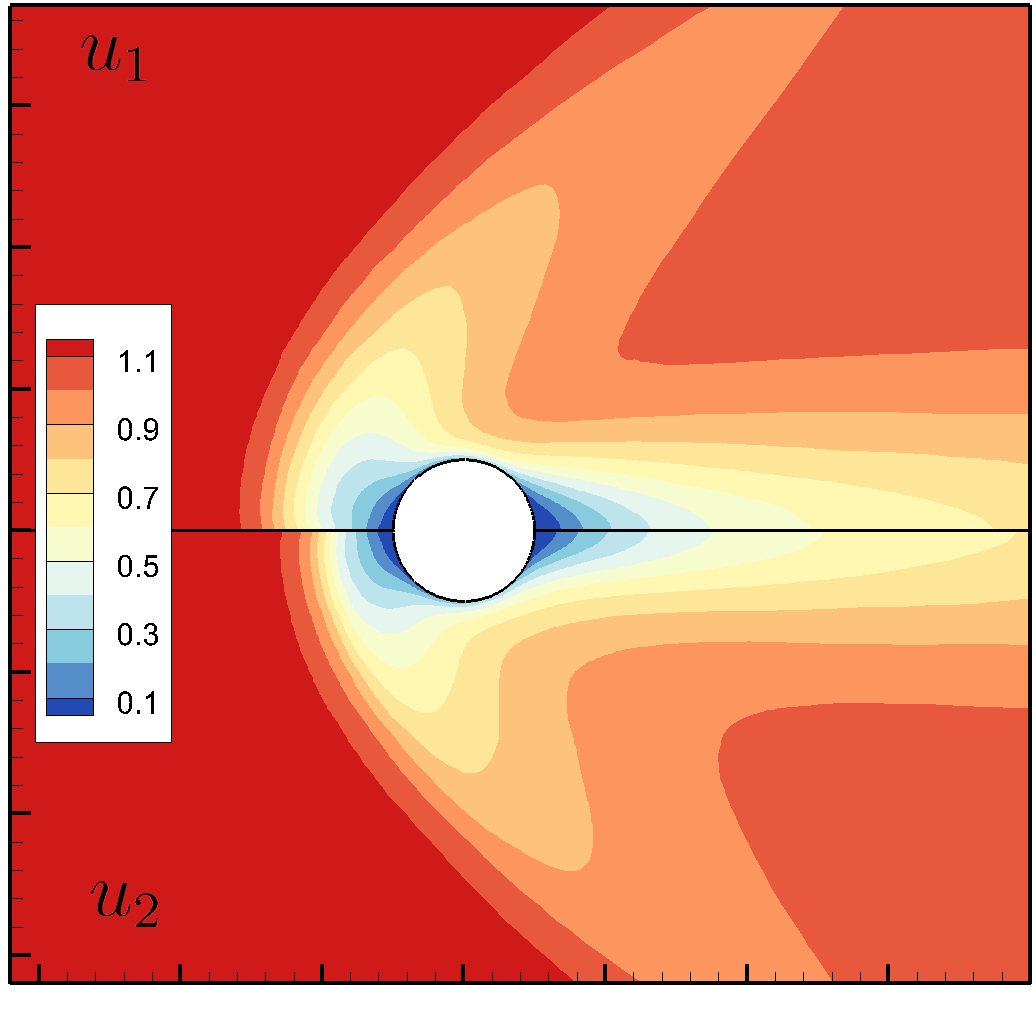}\;
\includegraphics[scale=0.14,clip=true]{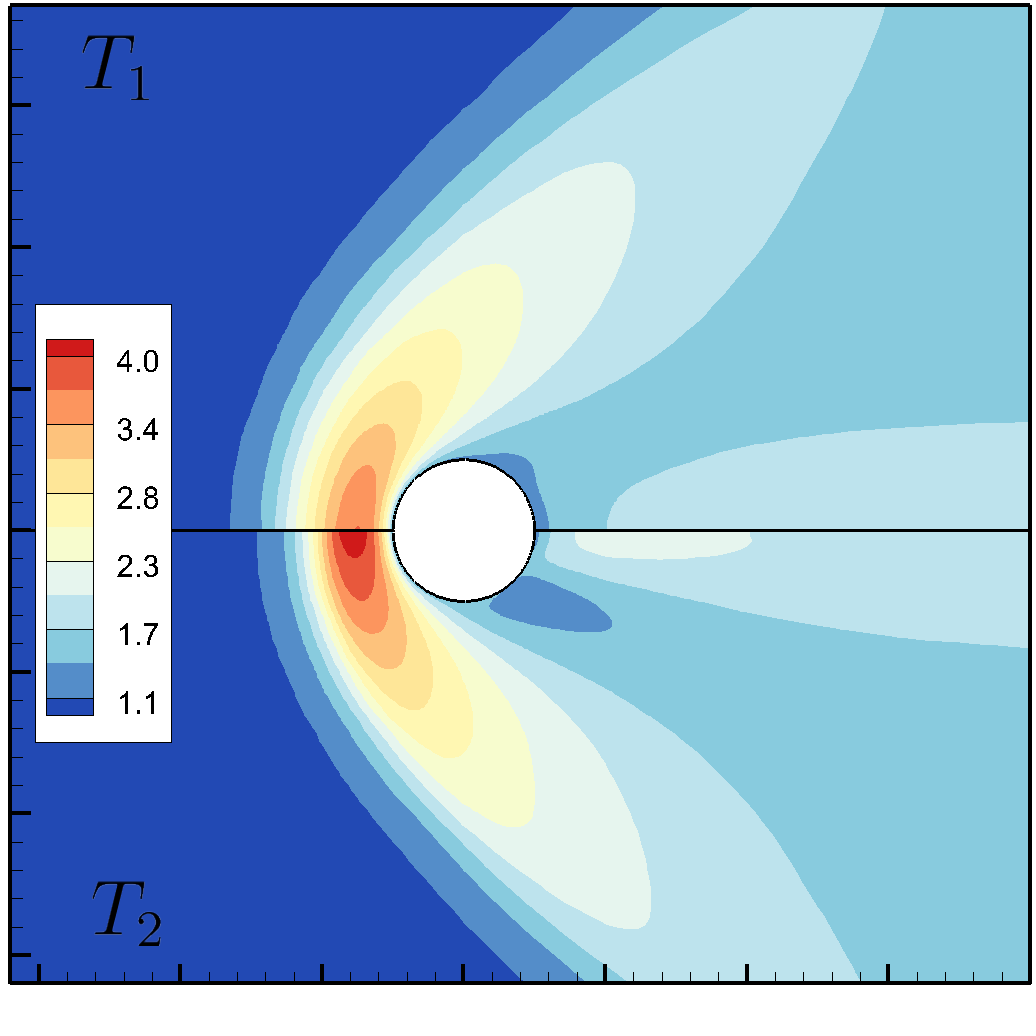}
}\\
\subfloat[$\text{Kn}=0.005$]{
\includegraphics[scale=0.14,clip=true]{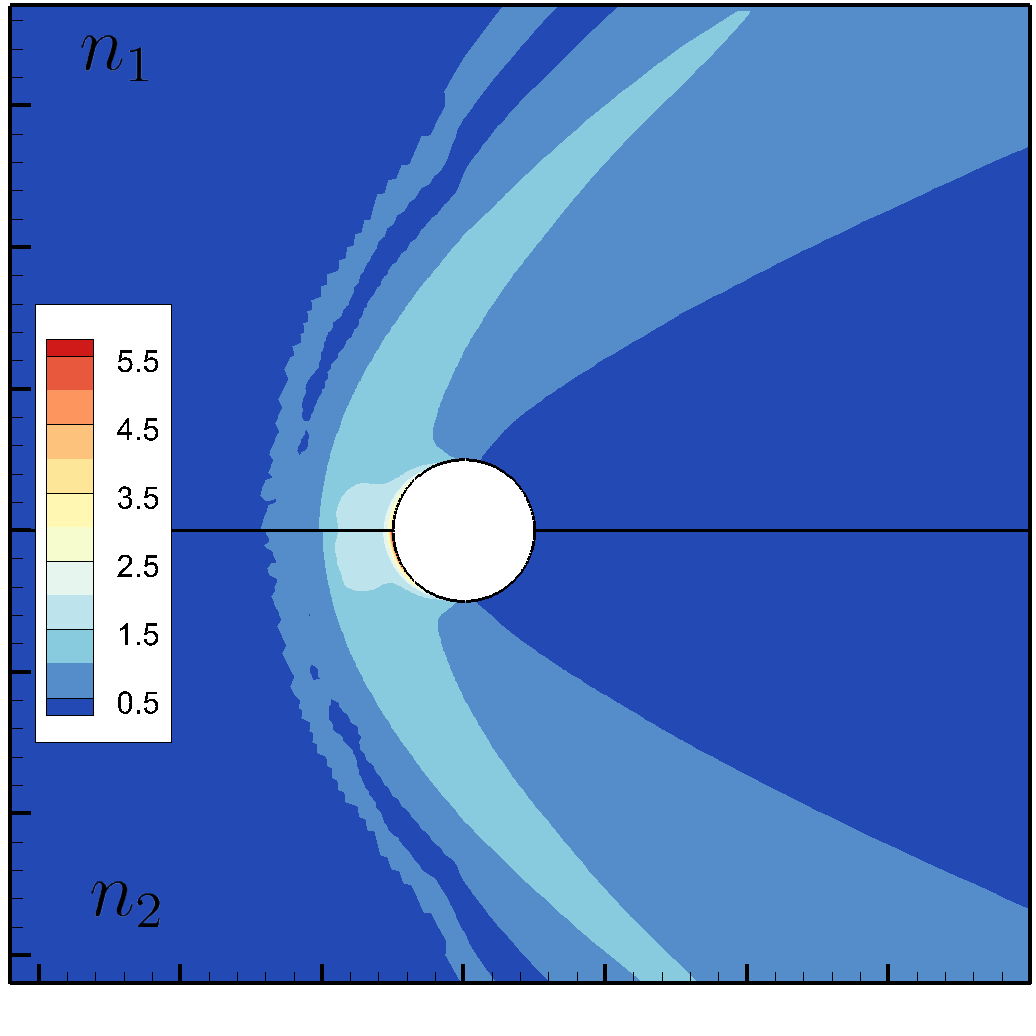}\;
\includegraphics[scale=0.14,clip=true]{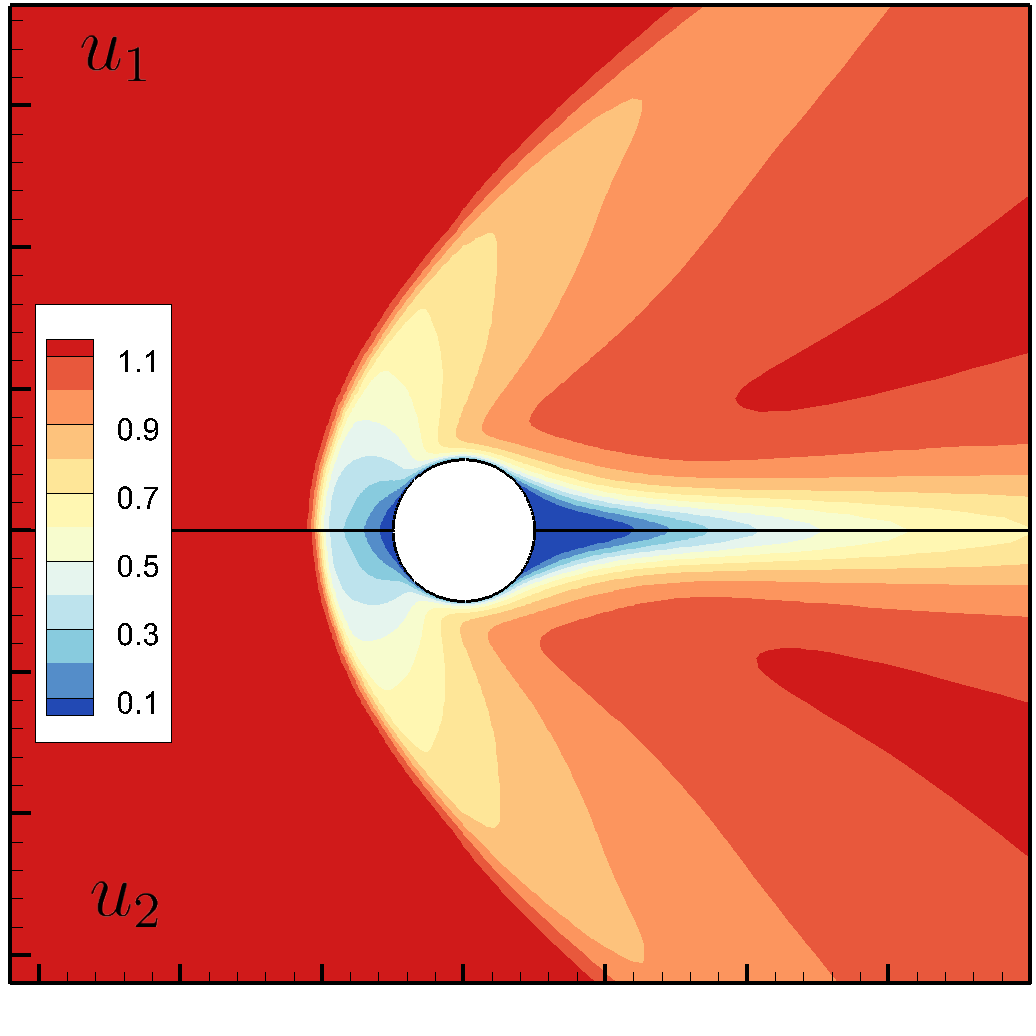}\;
\includegraphics[scale=0.14,clip=true]{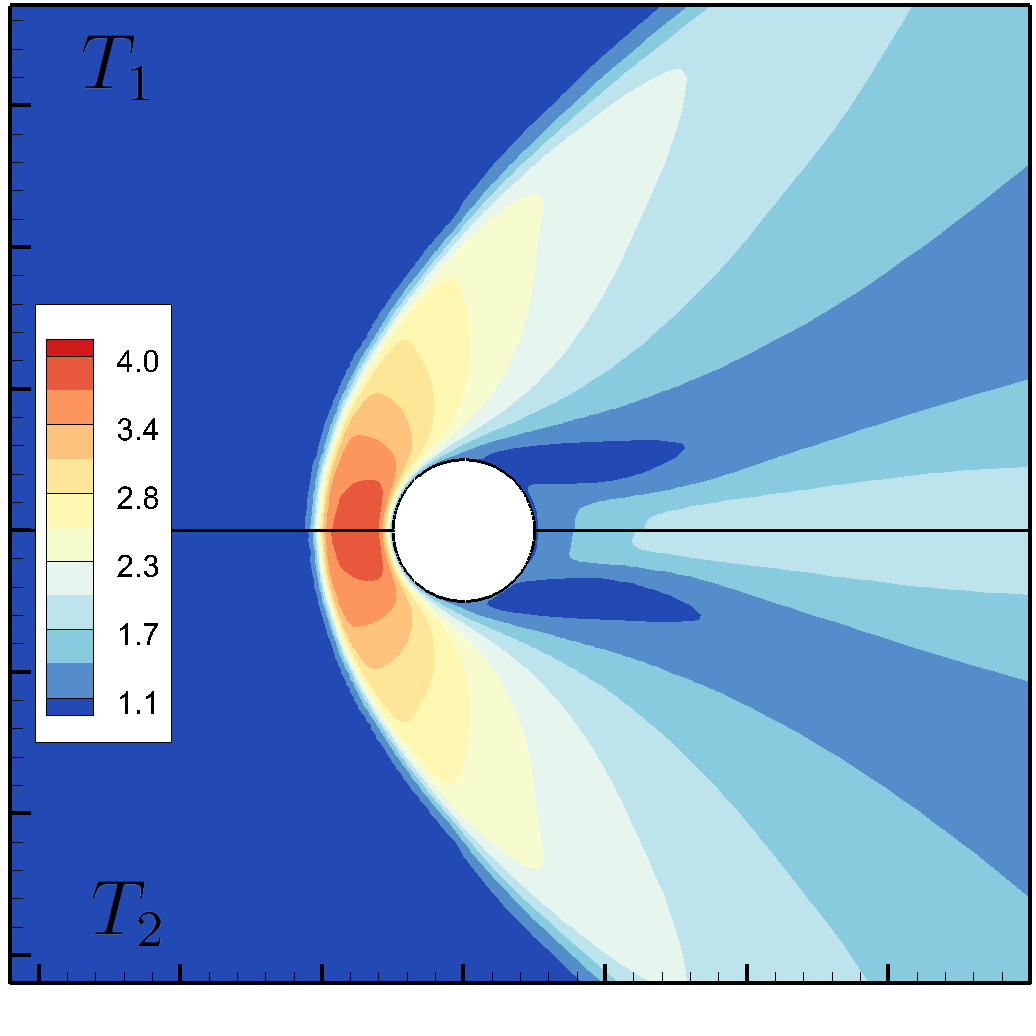}
}
\caption{Contours of number density (left column), horizontal velocity (middle column) and temperature $T$ (right column) for the binary mixture flow over cylinder at different Knudsen numbers. The velocity is uniformly normalized using the most probable velocity of the light species. }
\label{fig04:cydMa3}
\end{figure}

It is noted that the acceleration ratio of the computational time is smaller than that  of the iteration step, which is due to the additional time in solving the macroscopic synthetic equations. Nevertheless, the computational gain of GSIS is tremendous, e.g., when $\text{Kn}=0.005$,  GSIS requires about 145 less wall time than CIS.

%This is because the DVM leverages ten cores for velocity space parallelism, whereas the macroscopic solver utilizes only a single core.  %For problems with large number of spatial discretizations, both DVM and macroscopic, what is relation between the speedup ratio in steps and time? 

% We also compare results along the stagnation line obtained from GSIS and DSMC for Mixture 4 where the mass ratio is 100, when $\text{Ma}_{\infty}=3, \text{Kn}_1=0.5$. Fig.~\ref{fig04:cyd_ms100} illustrates the density, velocity, temperature, and heat flux distributions along the stagnation line. In the heavy gas computed by GSIS, a certain degree of temperature overshoot is observed at upstream locations $x = [-2.5, -1.2]$, similar to the simulation of a single-species gas. This phenomenon arises due to the fact that the collision frequency is independent of molecular velocities
% in the kinetic model, making it difficult to recover all the relaxation process~\cite{Yuan2022JFM}. However, upon neglecting the temperature overshoot, the overall distribution of macroscopic quantities exhibits good agreement with the DSMC results.

Using the efficient and accurate GSIS, we provide in Fig.~\ref{fig04:cydMa3} a comprehensive visualization of the supersonic flow over the cylinder at $\text{Kn}=0.005, 0.05$ and $0.5$. As the Knudsen number decreases, the shock thickness of each species narrows, and their positions gradually move towards the wall.
In the transition flow regime of $\text{Kn}=0.5$, the heavy species exhibits supersonic flow due to its lower characteristic sound speed, while the light species remains in subsonic flow. As the Knudsen number further decreases into the slip flow regime with $\text{Kn}=0.05$, the light species also enters supersonic flow. The shock thickness of the heavier component gas is slightly narrower than that of the lighter component gas, and the peak temperature within the shock layer of the heavier component gas is higher than that of the lighter component gas.
For the results in the near-continuum flow regime ($\text{Kn}=0.005$), the distributions of macroscopic quantities for both species are almost identical. Only slight differences can be observed in the velocity distribution upstream of the shock layer, which is attributed to velocity diffusion effects arising from the difference in concentration. Despite sufficient exchange of momentum and energy among the two species gases, these differences persist.

\begin{figure}[!t]
    \centering
    \subfloat[axial view]{\includegraphics[width=0.45\textwidth]{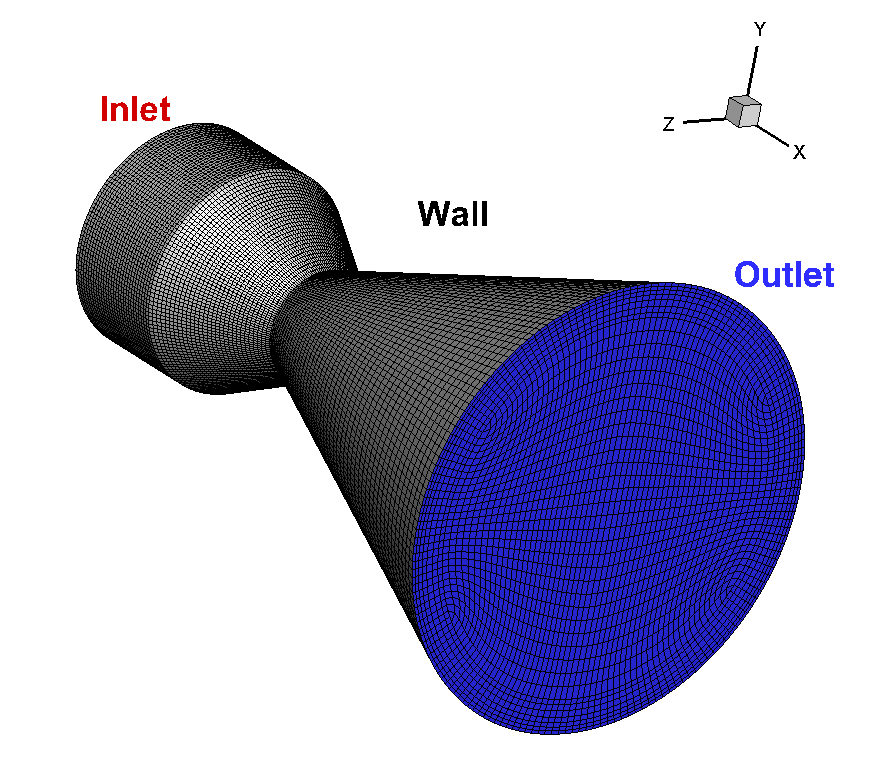}}
    \subfloat[cross-sectional view]{\includegraphics[width=0.5\textwidth]{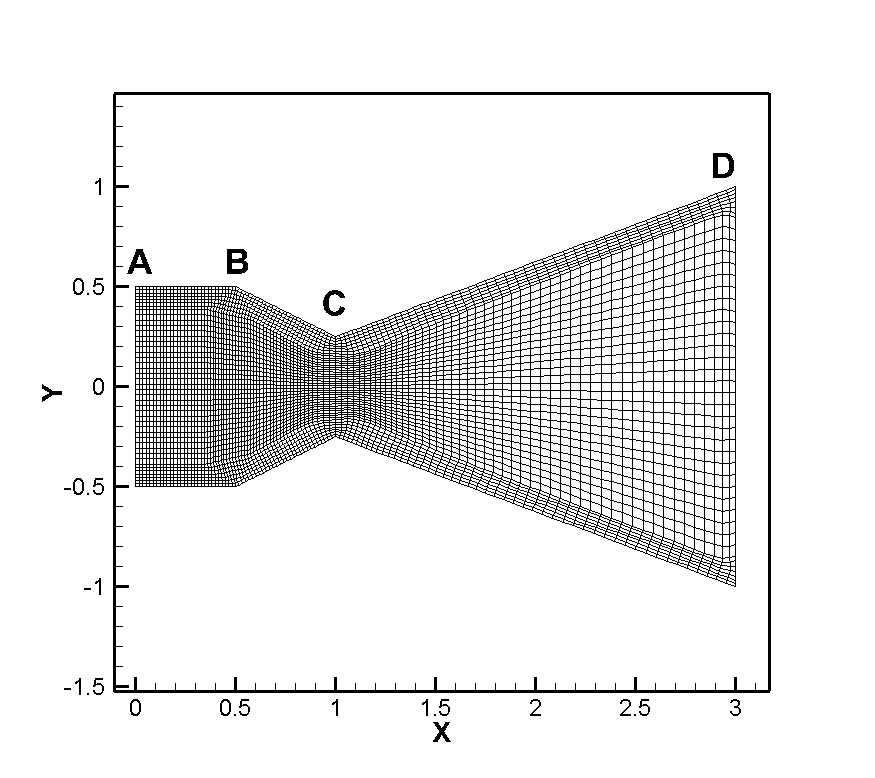}}\\
    \subfloat[velocity mesh for light gas]{\includegraphics[width=0.45\textwidth]{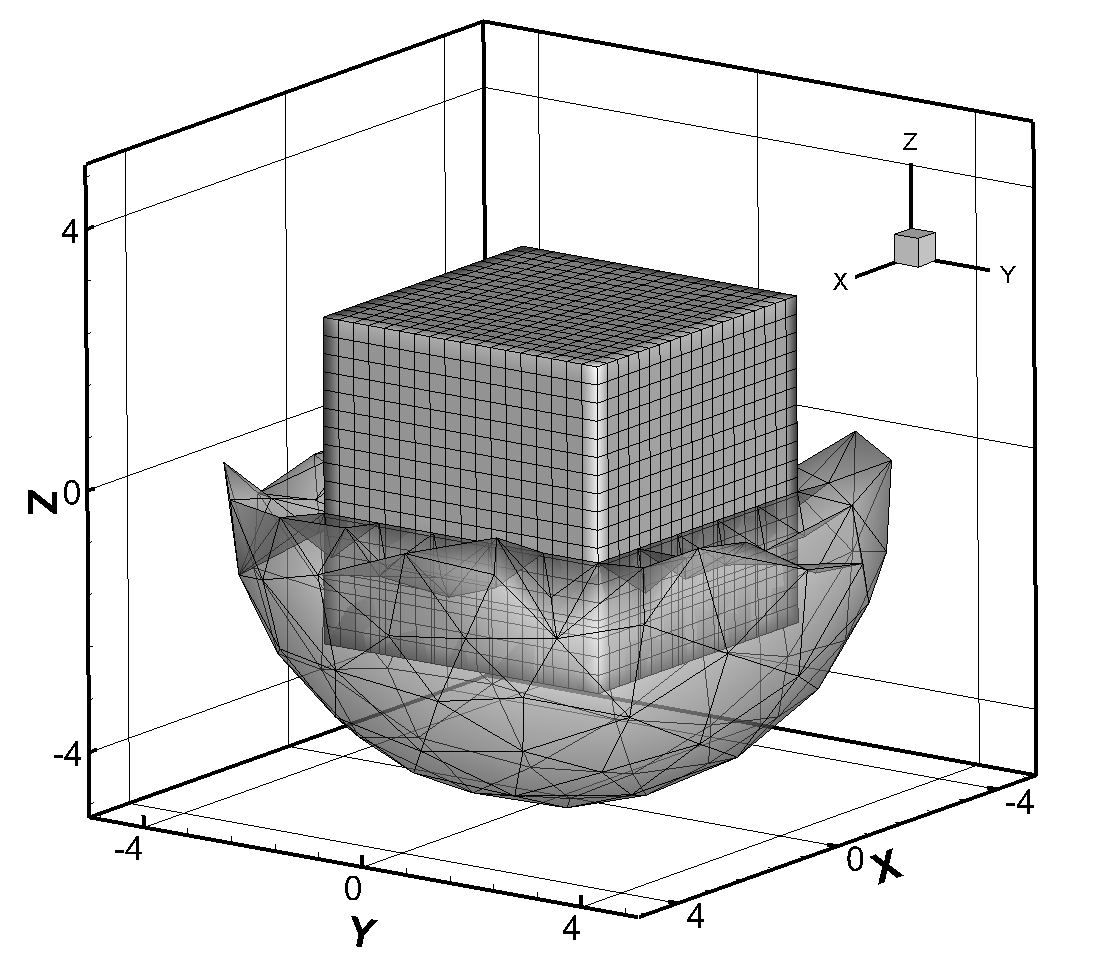}}
    \subfloat[hybrid velocity mesh for heavy gas]{\includegraphics[width=0.45\textwidth]{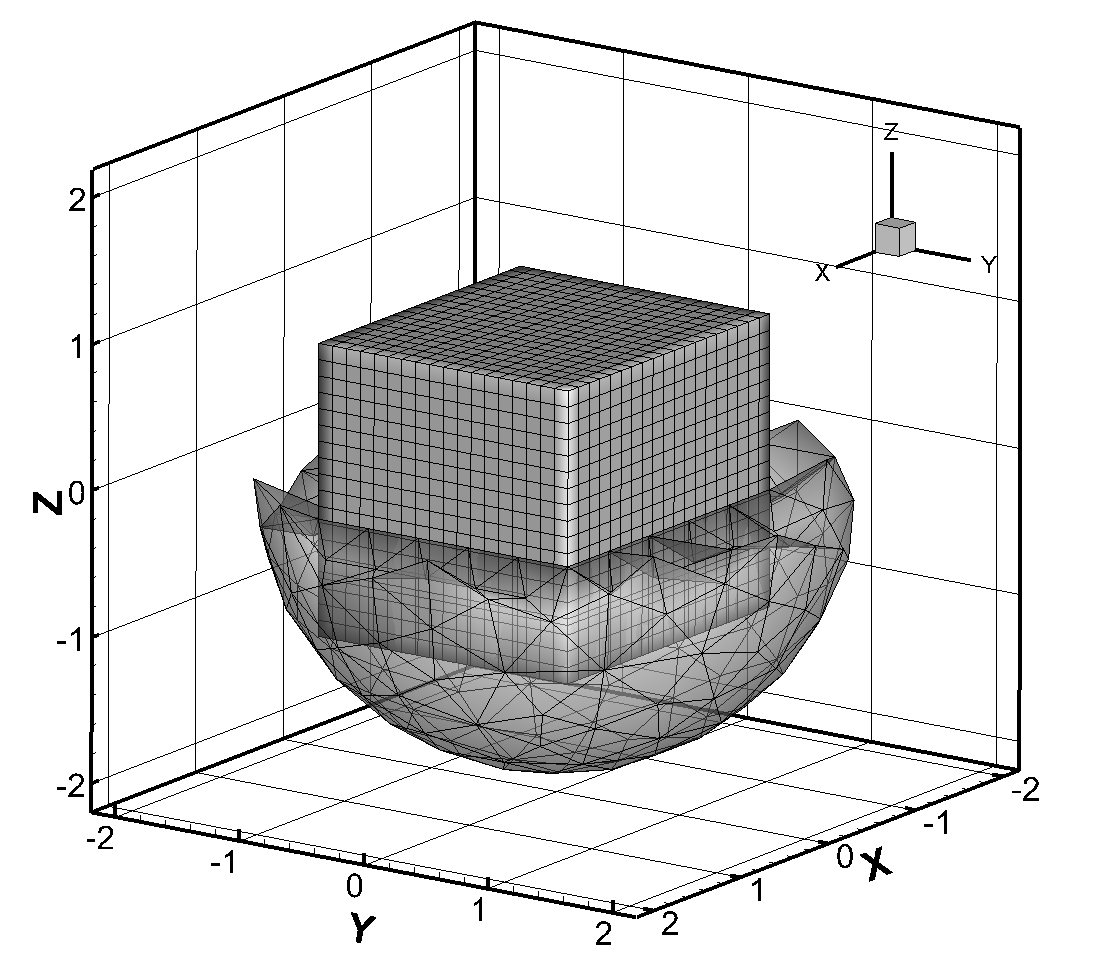}}
    \caption{Meshes in the physical and velocity spaces for the nozzle flow. Velocity meshes are refined around the stagnation velocities, which is normalized by $\sqrt{RT_0/m_1}$.}
    \label{fig:nozzle_mesh}
\end{figure}

\subsection{Three-dimensional nozzle flow}

The internal flow in a nozzle is considered to assess the GSIS in three-dimensional simulation, where the structure of the nozzle is shown in Fig.~\ref{fig:nozzle_mesh}(a). The flow domain encompasses a cylindrical chamber with a length of $3L_0$, where the inlet circle's center is positioned at the origin. Gas ingress occurs from the left channel with a diameter of $L_0$, traverses a contraction-expansion section (throat diameter of $0.5L_0$), and egresses from the right outlet. Geometric dimensions of the cylindrical chamber are delineated by points $A(0, 0.5, 0)$, $B(0.5, 0.5, 0)$, $C(1, 0.25, 0)$, and $D(3, 1, 0)$.
At the inlet ($x=0$), the flow is assumed to be significantly subsonic with $\text{Ma}_{in} = 0.05$, featuring Maxwell velocity distributions at temperature $T_0$. Gas molecules undergo reflection on the cold walls of the nozzle ($T_{w}=T_0/2$) with complete thermal accommodation before venting into vacuum at the outlet ($x=3L_0$). The Knudsen number, defined in relation to the gas properties at the inlet and with the characteristic length as the inlet channel diameter, is $\text{Kn}_1 = 0.1$.

For physical space discretization, the primary flow direction is segmented into 108 sections, comprising 8 layers of hexahedral mesh uniformly distributed from the wall to the inner field, each with a height of $0.125L_0$. The circumferential direction is evenly divided into 176 segments, resulting in a total of 146,334 grids, see Fig.~\ref{fig:nozzle_mesh}(b). 
The Mixture 1 flow is considered. For the discretization of velocity space, a hybrid structured-unstructured grid approach was used~\cite{yuanPhd,zhang4724172implicit,zhang2023efficient}, as illustrated in Fig.~\ref{fig:nozzle_mesh}(c,d).
For the structured velocity grid, the cutoff cube ranges for the light and heavy species were $[\pm2.5, \pm2.5, \pm2.5]$ and $[\pm1.3, \pm1.0, \pm1.0]$, respectively, corresponding to $19\times18\times18=6156$ uniformly distributed velocity discretization points.
To expand the discretization range, spherical unstructured velocity space was added around the structured velocity space, with sphere ranges of $r_1 = 3.5 v_m$ and $r_2 = 2.0 v_m$ for the light and heavy species, respectively.
The total number of discretization points for the heavy species reached 8625, including 6156 structured grid points and 2469 unstructured grid points. Similarly, the total number of discretization points for the light species was 8509, including 6156 structured grid points and 2353 unstructured grid points.

\begin{figure}[!t]
    \centering
    \subfloat[]{
        \includegraphics[width=0.4\textwidth]{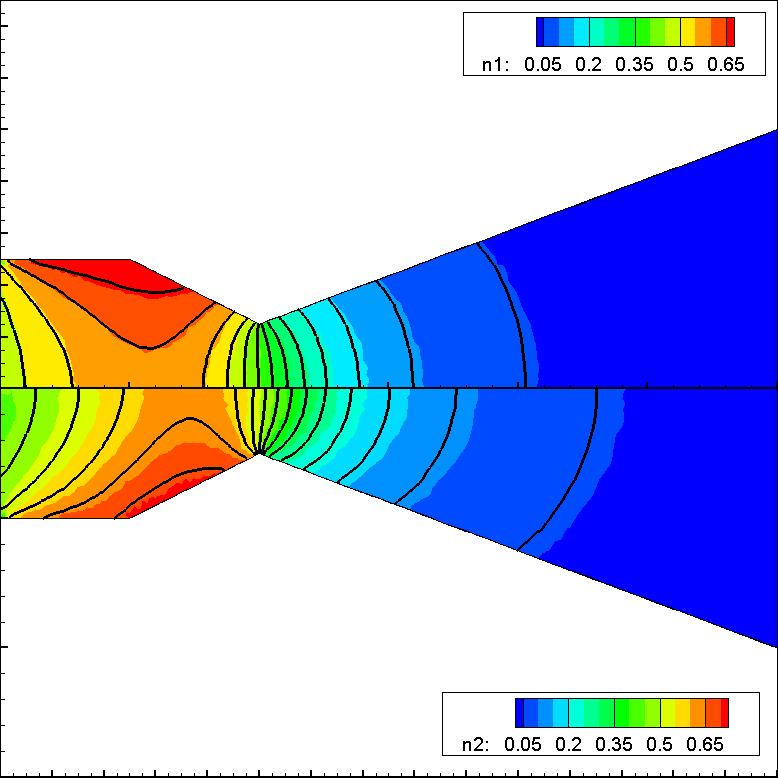}\label{fig04:nozzle_contour}
        \quad
        \includegraphics[width=0.4\textwidth]{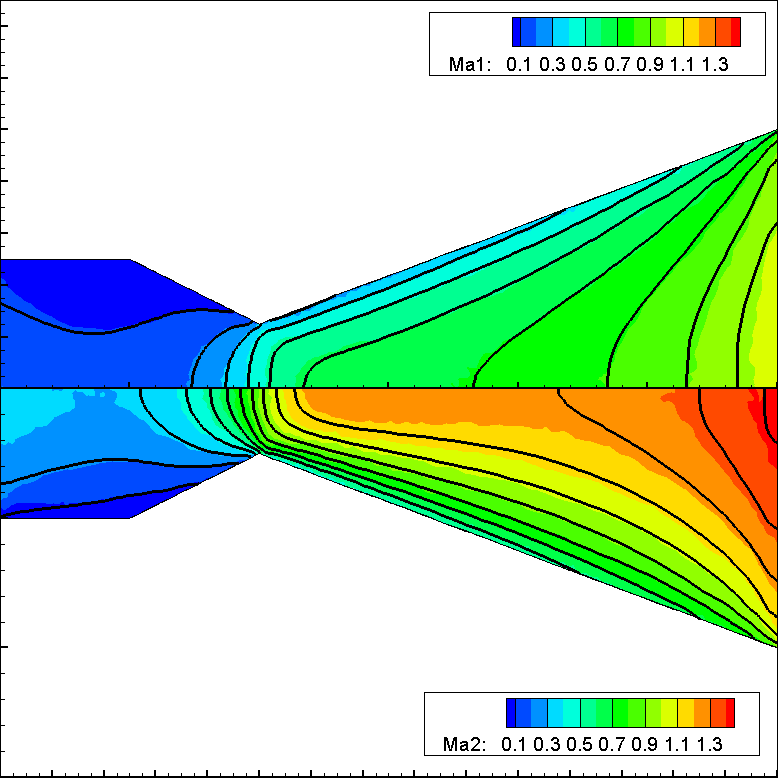}
        }\\
    \subfloat[]{
        \includegraphics[width=0.45\textwidth, trim=0 150 0 100, clip]{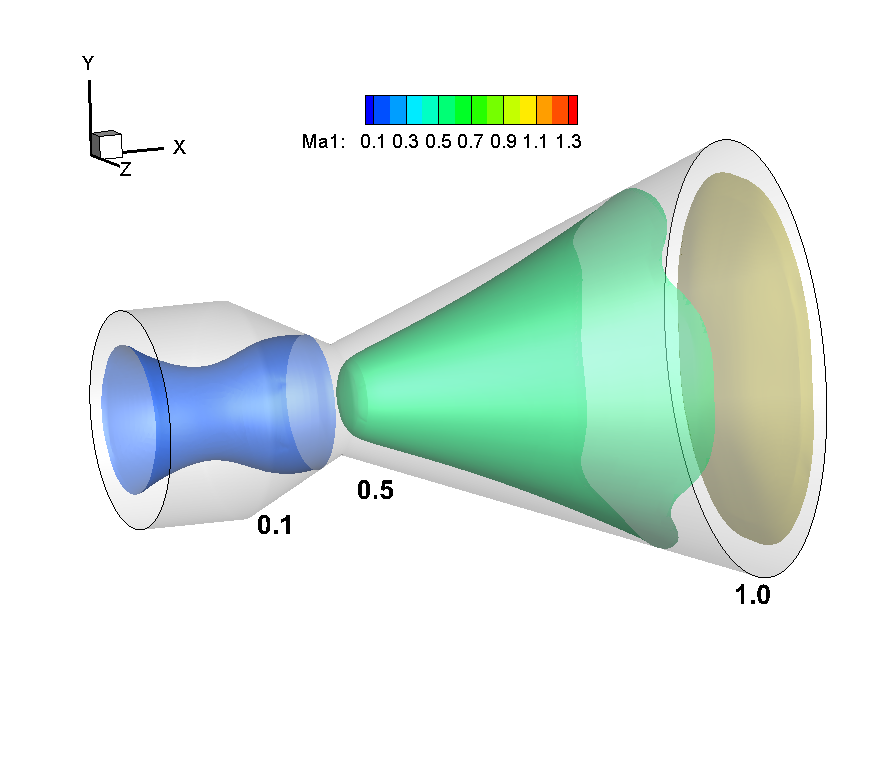}\label{fig04:nozzle_Ma}
        \includegraphics[width=0.45\textwidth, trim=0 150 0 100, clip]{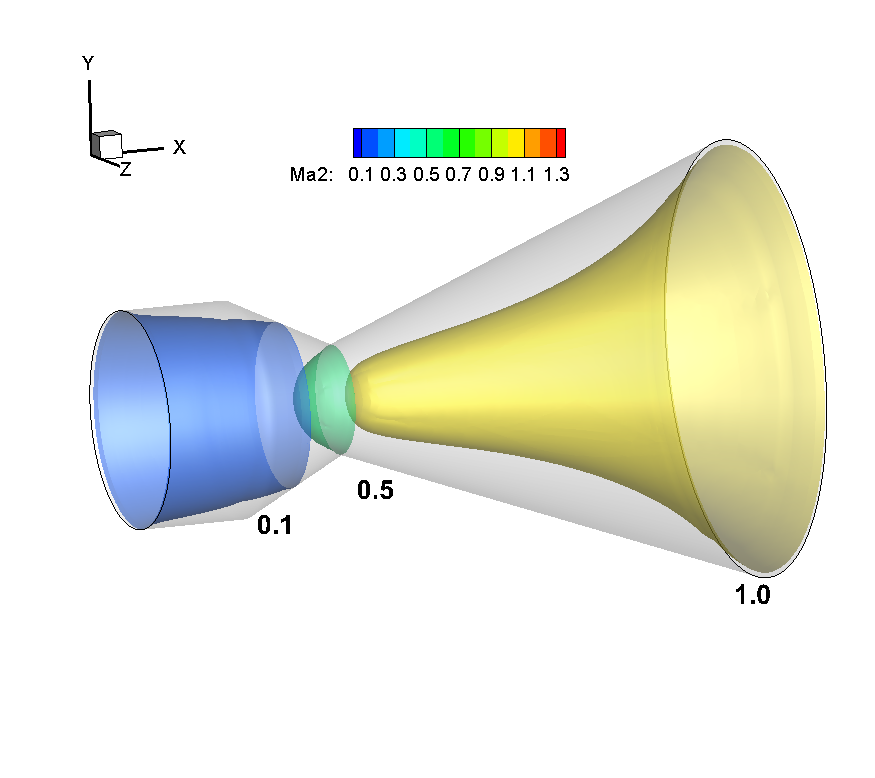}
        }
    \caption{The distribution of macroscopic variables for the Mixture 1, under the flow conditions are $\text{Kn}_1 = 0.1, \text{Ma}_{in}=0.05, \beta_n = 1$. (a) The distribution of number density (left) and velocity (right) in the $z=0$ slice. The upper and lower halves of the figure represent the light and heavy species, respectively. (b) Mach number iso-surfaces for the light (left) and heavy (right) species, respectively, which are defined
based on their own individual local sound speeds $\sqrt{5k_B T_s / 3m_s}$.
}
    % \label{fig04:nozzle_contour}
\end{figure}

The contour of macroscopic quantities obtained from GSIS and DSMC are depicted in Fig.~\ref{fig04:nozzle_contour}.
The density distribution reveals peak values of both species near the corner from the straight channel to the contraction section due to the prescribed cold wall conditions.  While the density distribution of the light species obtained from GSIS shows good agreement with the DSMC results, slight deviations are observed in the heavier species.
Examining the Mach distribution and subsequent axial velocity distribution, it is evident that although the macroscopic velocity of the light species exceeds that of the heavy species throughout the channel, it only achieves supersonic speeds near the outlet. In contrast, the heavy species enters supersonic flow near the lower side of the throat. This disparity arises from the different characteristic sound speeds. However, the temperature distribution of the two species is nearly identical.

To further illustrate the differences in the velocity fields of the two species, Fig.~\ref{fig04:nozzle_Ma} presents the contour plots of Mach number iso-surfaces for both species. Three Mach number iso-surfaces, $\text{Ma}=0.1, 0.5, 1.0$, are depicted. It can be observed that the flow structures of the two species are quite similar before the throat, characterized by nearly incompressible flow. The Mach number of the heavy species gas near the throat is around 0.5, and it rapidly increases after passing through the throat, reaching supersonic speeds near $x=1.5$ and attaining Mach numbers of approximately 1.4 near the outlet. In contrast, the light species remains in subsonic flow after passing through the throat, with only a small region near the outlet breaking into supersonic flow.

\begin{figure}[t]
    \centering
    \includegraphics[width=0.45\textwidth]{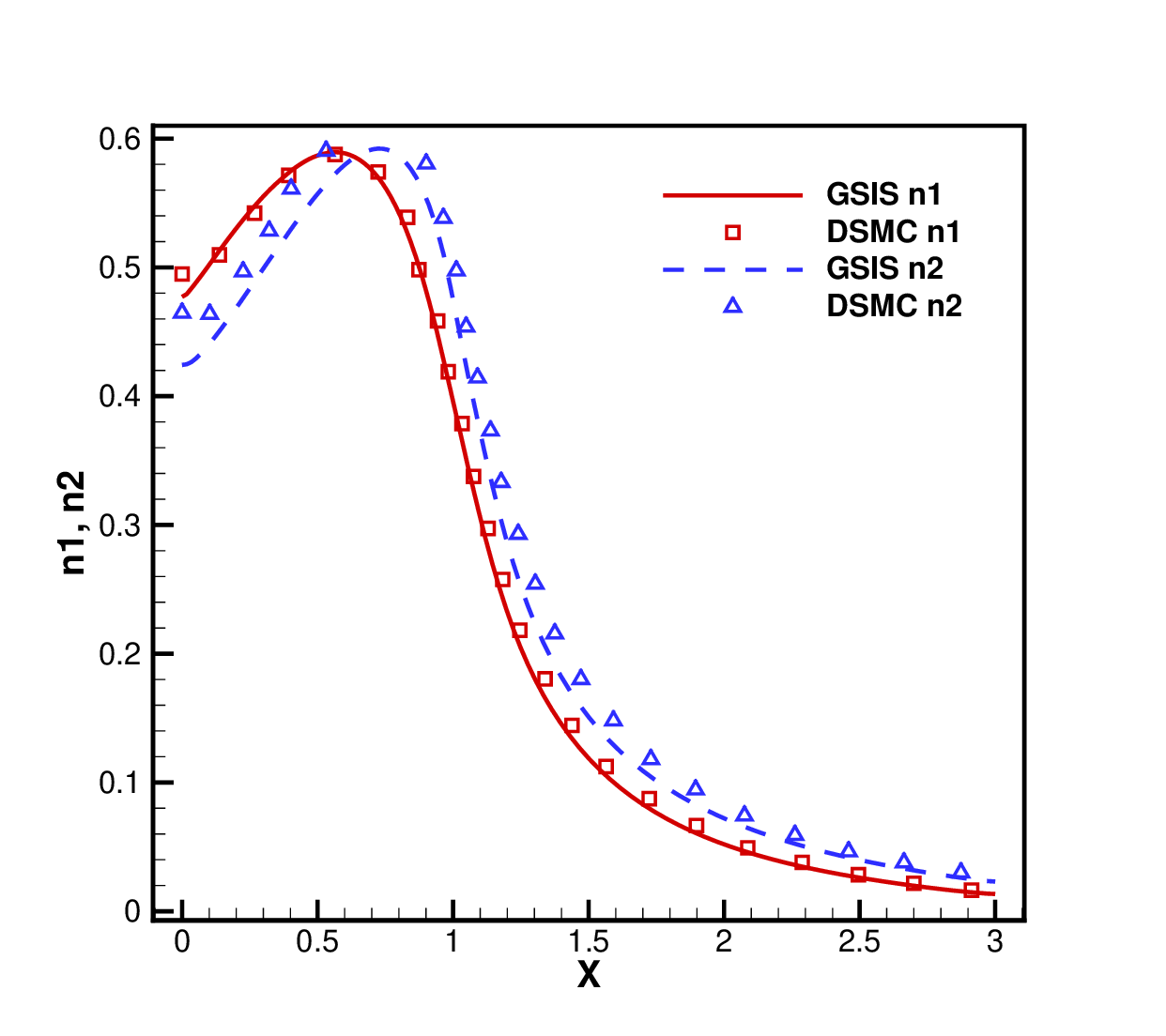}
   \includegraphics[width=0.45\textwidth]{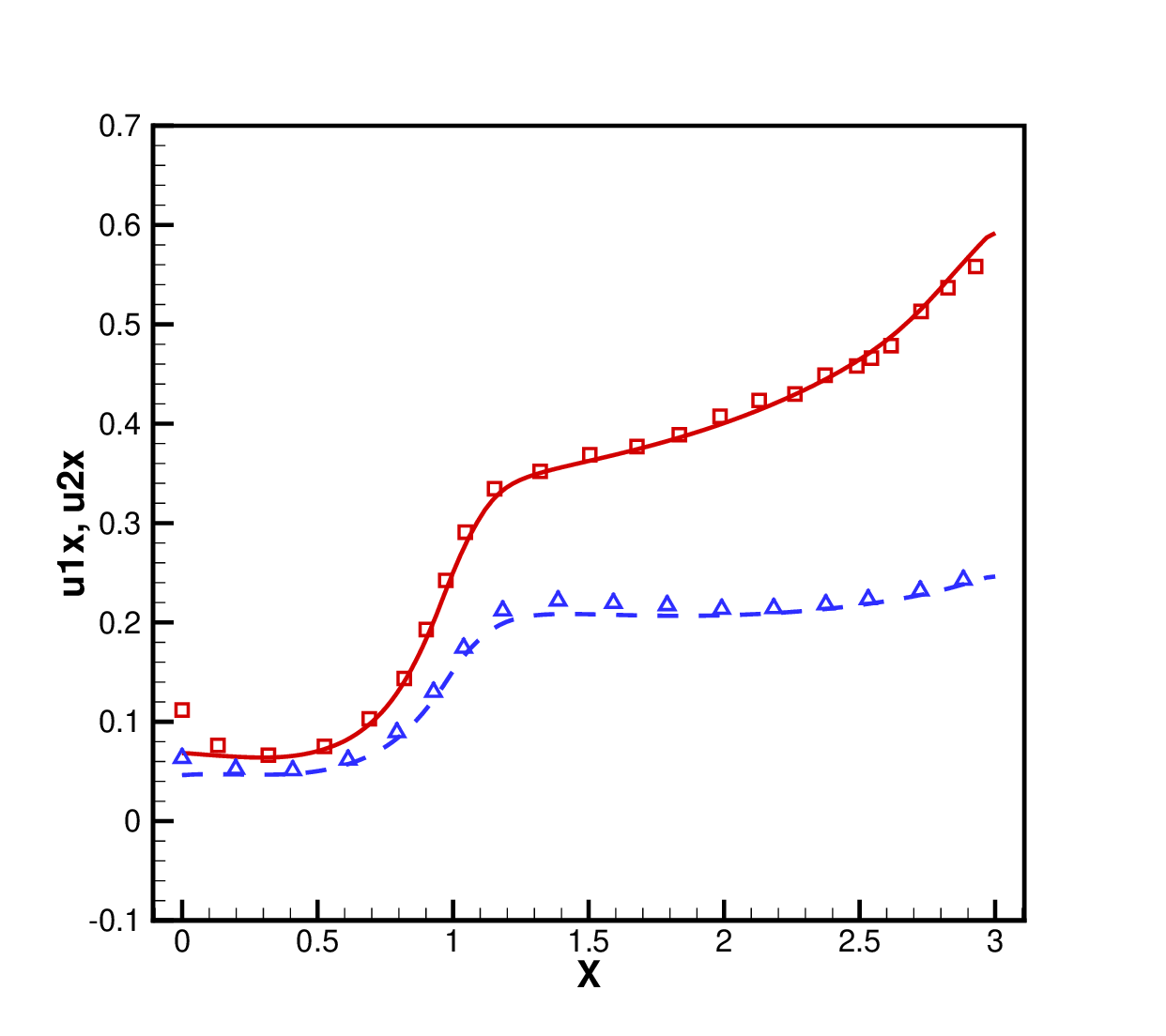}
    \caption{Distribution of the number density and horizontal velocity along the central line. Lines and symbols  represent the GSIS and DSMC results, respectively. Squares and triangles denotes the lighter and heavier species, respectively. The velocity is uniformly normalized using the most probable velocity of the light species.}
    \label{fig04:nozzle_line}
\end{figure}

Figure~\ref{fig04:nozzle_line} compares the distributions of macroscopic quantities along the central line. The trends in the density distribution for both species are similar, showing an initial increase followed by a decrease. There is a notable extremum near the throat, with the density decreasing by nearly an order of magnitude near the outlet. The peak densities of both species are nearly identical, with the peak position of the heavier species slightly trailing behind. The velocity distribution of the light species exhibits a monotonically increasing trend, with two acceleration processes near the contraction section and the vacuum outlet. In contrast, although the trend of the velocity distribution for the heavier species is similar, its maximum velocity at the outlet is only about one-third that of the light species, owing to its larger mass. %The temperature distributions of the two species are exactly opposite. 

Finally, Table~\ref{tab05:nozzle_compare_time} compares the iteration steps and computational costs. GSIS achieves a steady-state solution within 60 steps, while CIS requires 3939 and 303 steps for $\text{Kn}=0.01$ and 0.1, respectively. Consequently, GSIS is approximately 38.6 and 2.5 times faster. Additionally, CIS has not converged after 15,000 steps at $\text{Kn}=0.001$. Therefore, in near-continuum flow conditions, the acceleration ratio of GSIS exceeds 148 times. Table~\ref{tab05:nozzle_compare_time} also presents the computational costs of DSMC, which increase by two orders of magnitude when the Knudsen number decreases by one order of magnitude. It is noted that a two-dimensional axisymmetric DSMC calculation is conducted. When $\text{Kn}=0.01$, the computational cost of the three-dimensional GSIS can be lower than that of a the two-dimensional DSMC calculation. These results demonstrate the accuracy and efficiency of extending GSIS to binary mixture gases.

% Finally, under the condition of $\text{Kn}=0.1$, the convergence is reached after 59 iterations, while the computational cost with 256 cores is only 29.9 cores-hours. These results are sufficient to demonstrate that the extension of GSIS to binary mixture gas is accurate and efficient.

% Finally, convergence is achieved after 61 iterations under the condition of $\text{Kn}_1=0.1$, with a computational cost of only 30 core-hours using 256 cores. These results are sufficient to demonstrate that the extension of GSIS to binary mixture gas is accurate and efficient.  Note that the two-dimensional axisymmetric DSMC calculation is conducted. \leir{DSMC time? CIS time and iteration steps? ratio of speedup in iteration step and time??}

\begin{table}[t]
\begin{threeparttable}  
 \centering
 \caption{Computational overhead of CIS and GSIS for nozzle flow at different Knudsen numbers for the Mixture 1. Simulations are run on 256 cores. Time is given in core hours.  Note that the the DSMC runs on a two-dimensional axisymmetric geometry, while the GSIS simulation is fully three-dimensional in spatial domain and three-dimensional in molecular velocity domain.}
\label{tab05:nozzle_compare_time}
  \begin{tabular}{c c c c  c c c c}\toprule
 \multirow{2}{*}{\text{Kn}} & \multicolumn{1}{c}{DSMC\tnote{1}} & \multicolumn{2}{c}{CIS} & \multicolumn{2}{c}{GSIS} & \multicolumn{2}{c}{Speedup ratio}\\ \cmidrule(r){3-4} \cmidrule(r){5-6}
~  & time (core hours) & steps & time & steps & time  & in steps & in time \\ \hline
0.1  & 5 & 303 & 75.8 & 61 & 29.9 & 4.9 & 2.5\\ 
0.01 & 1030 & 3939 & 1007.6 & 50 & 26.1 & 78.8 & 38.6\\ 
0.001 & - & $> 15,000$ & $> 3837$ & 50 & 25.8 & $> 300$ &  $> 148.7$\\
\bottomrule
\end{tabular}
\begin{tablenotes}
 \footnotesize
 \item[1] In the two-dimensional axisymmetric DSMC simulations, 6165 grid cells and a total of 73,000 particles were used when $\text{Kn}=0.1$. For $\text{Kn}=0.01$, 373,761 grid cells were employed, with a total of 2,962,500 particles.
\end{tablenotes}
 \end{threeparttable}
\end{table}

\subsection{Pressure-driven flow in the channel}

\begin{figure}[!h]
    \centering
    \subfloat[pressure at $\beta_n = 10^{-3}$]{\includegraphics[scale=0.2,clip = true]{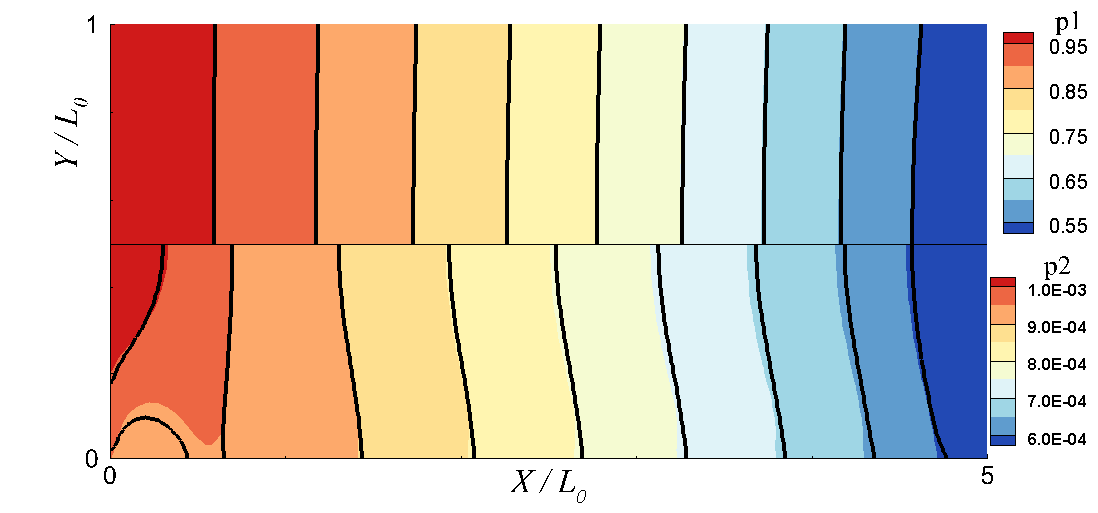}}\;
    \subfloat[velocity at $\beta_n = 10^{-3}$]{\includegraphics[scale=0.2,clip = true]{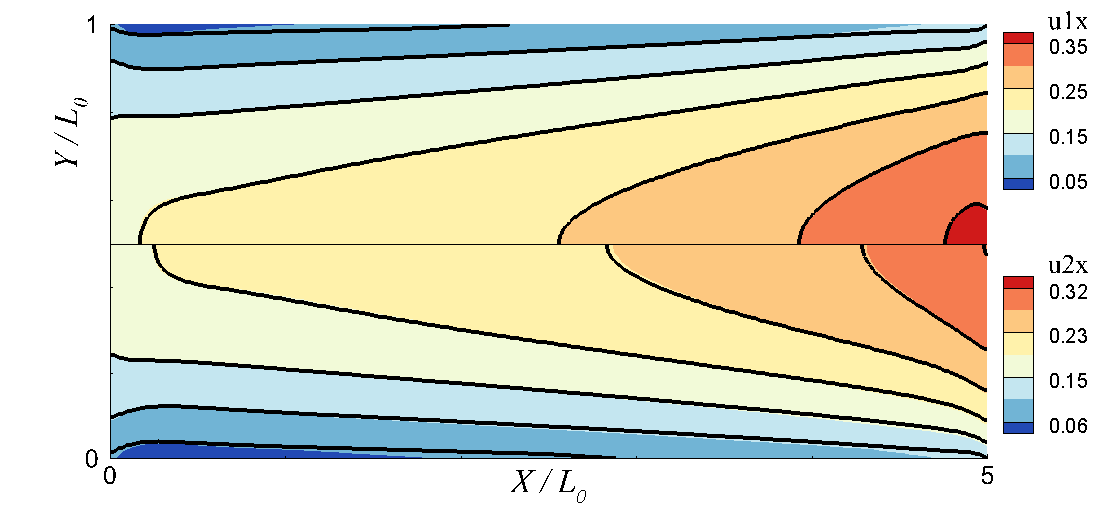}}\\
    \subfloat[pressure at $\beta_n = 1$]{\includegraphics[scale=0.2,clip = true]{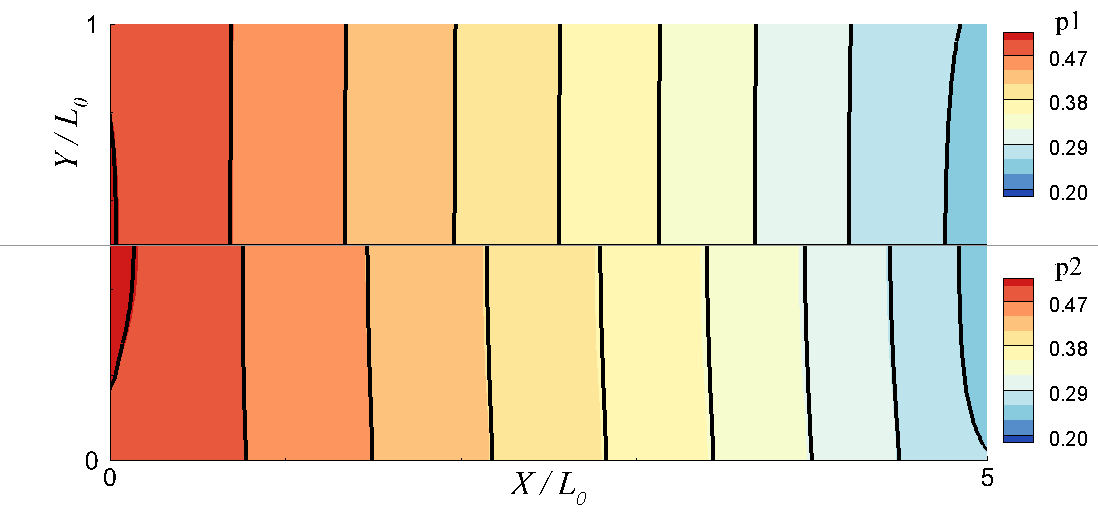}}\;
    \subfloat[velocity at $\beta_n = 1$]{\includegraphics[scale=0.2,clip = true]{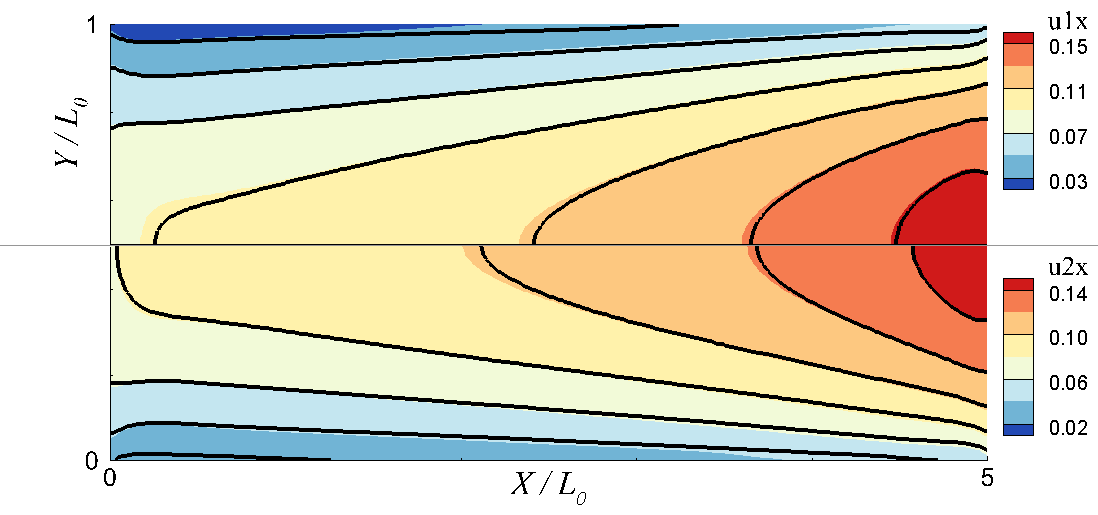}}\\
    \subfloat[pressure at $\beta_n = 10^{3}$]{\includegraphics[scale=0.2,clip = true]{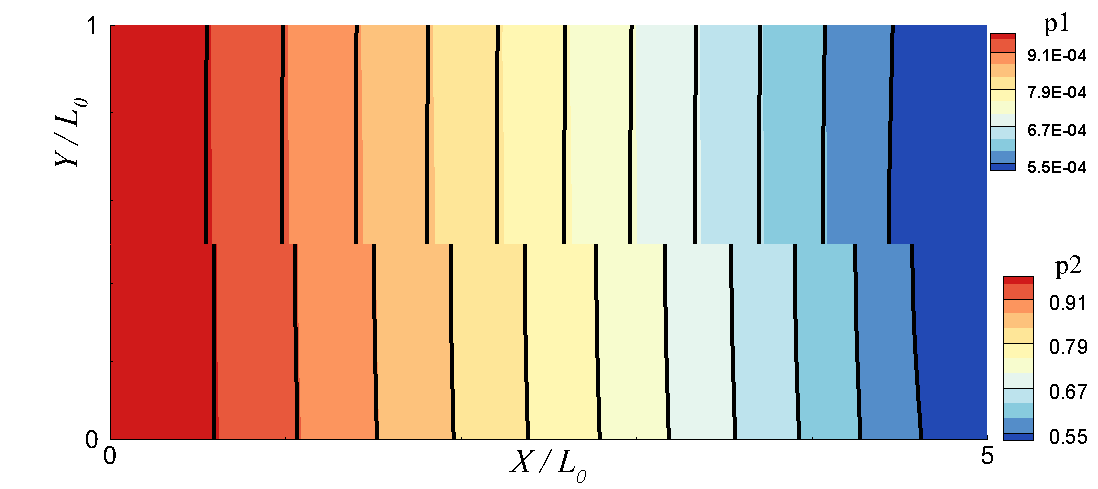}}\;
    \subfloat[velocity at $\beta_n = 10^{3}$]{\includegraphics[scale=0.2,clip = true]{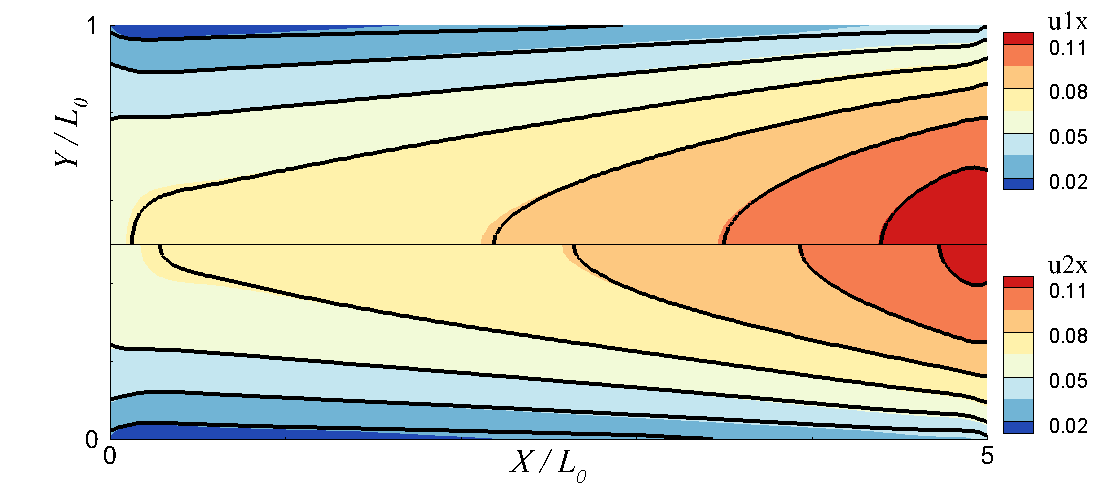}}
    % \subfloat[velocity]{\includegraphics[scale=0.3,clip = true]{fig/nc1000_ms_1.1_Kn_0.1_ux.eps}}
    % \subfloat[shear stress $\sigma_{xx}$]{\includegraphics[scale=0.3,clip = true]{fig/nc1000_ms_1.1_Kn_0.1_sxx.eps}}
    \caption{Contours of pressure (left column) and velocity (right column) in the pressure-driven Mixture 1 flow, when $\text{Kn}_1=0.05$ and $p_{in}/p_{out}=2$. The upper and lower halves of the figure represent the light and heavy species, respectively.  Background contour: GSIS. Black lines: CIS.  } % The species concentration is (a-b) $\beta_n = 10^{-3}$, (c-d) $\beta_n = 1$, (e-f) $\beta_n = 10^3$.
    \label{fig:pre_nc1000}
\end{figure}

Having validated the accuracy and efficiency of the GSIS in hypersonic flows, here we test the GSIS in pressure-driven low-speed gas mixture flows. The flow domain is a rectangular with a size of $L_0 \times 5L_0$. The left and right boundaries are designated as pressure inlet $p_{in}=p_0$ and pressure outlet $p_{out}=0.5p_0$, respectively. The Maxwell diffuse boundary condition in applied at the top and bottom walls with the temperature $T_0$. In the pressure inlet boundary condition, to promptly respond to the pressure information, the macroscopic quantities on the right side of the interface are specified as~\cite{wang2004simulations}:
\begin{equation}
    \begin{aligned}
   \bu_{s}^R  &=\bu_{s}^L + \frac{p_{in}-p^{L}}{\rho_{s}^L c_{s}^L}\bm{n}_{ij},\quad
    \rho_{s}^R  = \frac{p_{in}}{T_s^R}m_s, \quad 
      T_{s}^R = T_0,
    \end{aligned}
\end{equation}
where the superscript $L$ and $R$ are respectively the left and right side of the cell interface $ij$, and $ c_s = \textstyle \sqrt{\frac{\gamma_s T_0}{m_s}}$ is the local sound speed of the species $s$. A similar treatment is applied to the pressure outlet boundary:
\begin{equation}
    \begin{aligned}
    \bu_{s}^R &=\bu_{s}^L + \frac{p^{L}-p_{out}}{\rho_{s}^L c_{s}^L}\bm{n}_{ij},\quad
    \rho_{s}^R  = \rho_s^L + \frac{p_{out}-p^L}{c_s^2},\quad
    T_{s}^R = \frac{m_s p_{out}}{\rho_s^R}. 
    \end{aligned}
\end{equation}

We consider three types of mixtures as shown in Table~\ref{tab:pre_cmp_step}, with a wide range of the concentration ratio. In all cases, the physics mesh is set as $101\times211$.
For the light gas the velocity space in each direction is truncated to $[-5, 5]$. For the heavy species, the velocity regions in each direction are truncated to $[-2,2]$, $[-1.5,1.5]$ when the mass ratios are $10, 100$, respectively.
The discrete velocity points for the light and heavy species are $48 \times 48$ and $64 \times 64$, respectively. 

Figure~\ref{fig:pre_nc1000} compares the velocity and pressure contours of the Mixture 1. For concentration ratios $\beta_n = 10^{-3},1,10^{3}$, the GSIS results are essentially the same as the CIS. The outlet velocity decreases as the concentration of the heavy species increases.
It is noteworthy that in cases of large concentration ratios (e.g., differences exceeding two orders of magnitude), DSMC struggles to sample low-concentration species, leading to significant statistical noise and rendering it unsuitable for low-speed flow problems with such large concentration disparities. In contrast, GSIS excels in accurately computing the flow field of low-concentration species, and may have potential applications in problems with low-concentration species, such as in the EUV lithography~\cite{teng2023pollutant}.

% \begin{figure}[!h]
%     \centering
%     \subfloat[density]{\includegraphics[scale=0.3,clip = true]{fig/nc10_ms_1_Kn_0.1_rho.eps}}
%     \subfloat[pressure]{\includegraphics[scale=0.3,clip = true]{fig/nc10_ms_1_Kn_0.1_pre.eps}}\\
%     \subfloat[velocity]{\includegraphics[scale=0.3,clip = true]{fig/nc10_ms_1_Kn_0.1_ux.eps}}
%     \subfloat[shear stress $\sigma_{xx}$]{\includegraphics[scale=0.3,clip = true]{fig/nc10_ms_1_Kn_0.1_sxx.eps}}
%     \caption{$Kn=0.1, P_{in}/P_{out}=10, \beta_n = 10, \beta_m = 1.1$.}
%     \label{fig:pre_nc10}
% \end{figure}

\begin{table}[t]
% \vspace{0.1em}
% \hspace{0.1em}
\begin{threeparttable}  
\centering
\caption{Convergence step and speedup ratios in CIS and GSIS for the pressure driven flow at different concentration ratio $\beta_n={n_2}/{n_1}$, when the Knudsen number is $\text{Kn}_1 = 0.05$. All simulations are executed on a 64-core CPU.
}
\label{tab:pre_cmp_step}
\resizebox{\linewidth}{!}{
\begin{tabular}{c c c c c c c c c c}
\hline
\multirow{3}{*}{$\beta_n$}  & \multicolumn{3}{c}{Mixture 1 } & \multicolumn{3}{c}{Mixture 4} & \multicolumn{3}{c}{Mixture 2 } \\
~  &\multicolumn{3}{c}{ ($\beta_d=1, \beta_m=10$)}  &\multicolumn{3}{c}{ ($\beta_d=2, \beta_m=100$)} &\multicolumn{3}{c}{ ($\beta_d=3, \beta_m=100$)}\\ \cmidrule(r){2-4} \cmidrule(r){5-7} \cmidrule(r){8-10}
~  & CIS\tnote{1} & GSIS\tnote{2} & ratios\tnote{3} & CIS & GSIS & ratios &CIS &GSIS & ratios\\ \hline 
$10^3$ & 3040  &20 &152 (122)& 10051 &32 &314 (184) & 14487  &33 & 439 (252) \\ 
$10^2$ & 3013  &20 &151 (121)& 9934  &31 &320 (191) & 14280  &33 & 432 (249) \\ 
$10^1$ & 2767  &19 &146 (123)& 8840  &33 &268 (155) & 12410  &32 & 388 (226)\\ 
$1$& 1572      &21 &75 (57) & 3906  &37  &106 (58) & 4174    &37 & 113 (62)\\ 
$10^{-1}$& 547 &24 &23 (16) & 839    &37 & 23 (13) & 1082    &37 & 29 (16)\\ 
$10^{-2}$& 534 &25 &21 (14) & 604    &32 & 19 (11) & 643     &32 & 20 (12) \\ 
$10^{-3}$& 530 &26 &20 (13) & 562    &31 & 18 (11) & 562     &31 & 21 (11)\\
\hline
\end{tabular}
}
\begin{tablenotes}
 \footnotesize
 \item[1] The computational time for one step of CIS iteration is about 0.472 seconds. Therefore, for the Mixture 2 with $\beta_n=10^3$, the total simulation time of CIS is about $0.472\times14487=6838$ (seconds), while that of GSIS is about $6838/252=27$ (seconds). The simulation time in other cases can be calculated in a similar manner.
 \item[2] The iteration steps here  include the 10-step CIS. Each iteration of GSIS involves solving one mesoscopic equation and performing 1000 macroscopic inner iterations.
 \item[3] The ratios comprise the acceleration ratios of step and computational time. For example, 439 (252) means that GSIS needs 630 times less iterations steps, and 252 times less wall time. The computation cost does not include the time required for memory allocation and initialization of the velocity distribution functions. 
\end{tablenotes}
 \end{threeparttable}
\end{table}

Table~\ref{tab:pre_cmp_step} summarizes the computational costs, where the concentration ratio spans six orders of magnitude. It is seen that, across different types of gas mixtures, the iteration steps of CIS increase with the concentration of the heavy species. This phenomenon is attributed to the reduction of the effective Knudsen number of the gas mixture as the concentration of the heavy species increases, where the CIS becomes harder to converge. In the case of Mixture 2, the variation in iteration steps between different concentrations can reach up to 30-fold. Remarkably, GSIS can get the converged solutions within 30 iteration steps in the wide range of concentration ratio, demonstrating a significant computational advantage over the CIS for low-speed mixture flows with disparate concentrations. For example, for the Mixture 2 where the mass ratio is 100 and the Knudsen number ratio is 9, when the concentration ratio is $\beta_n=10^3$, GSIS is 252 times faster than CIS. 

\begin{figure}[t]
    \centering
   {\includegraphics[scale=0.25,clip = true]{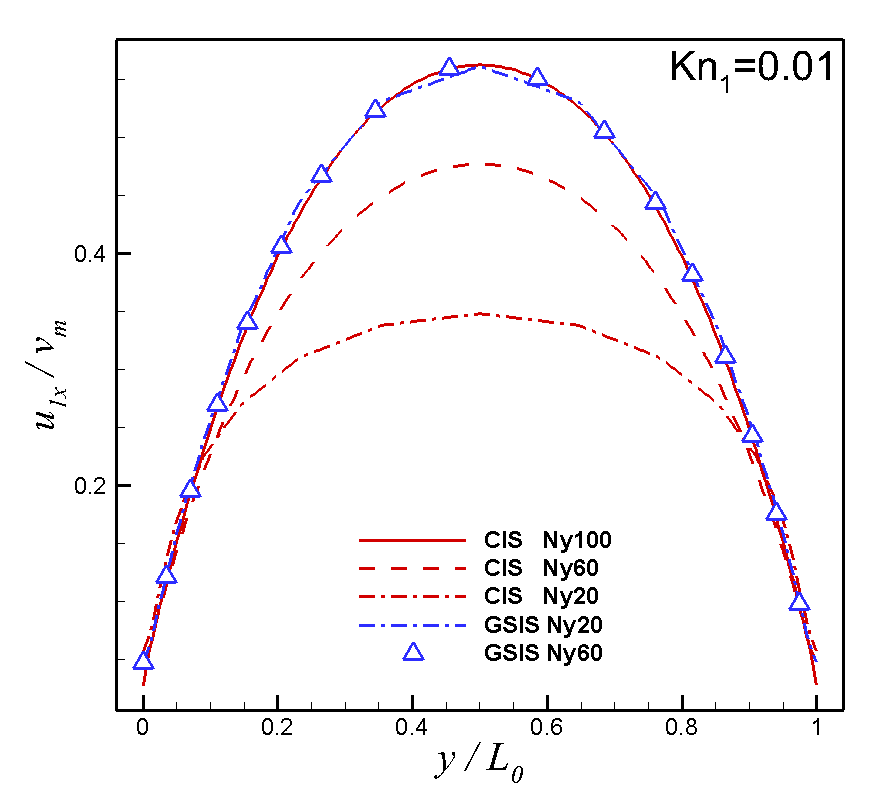}}\quad
  {\includegraphics[scale=0.25,clip = true]{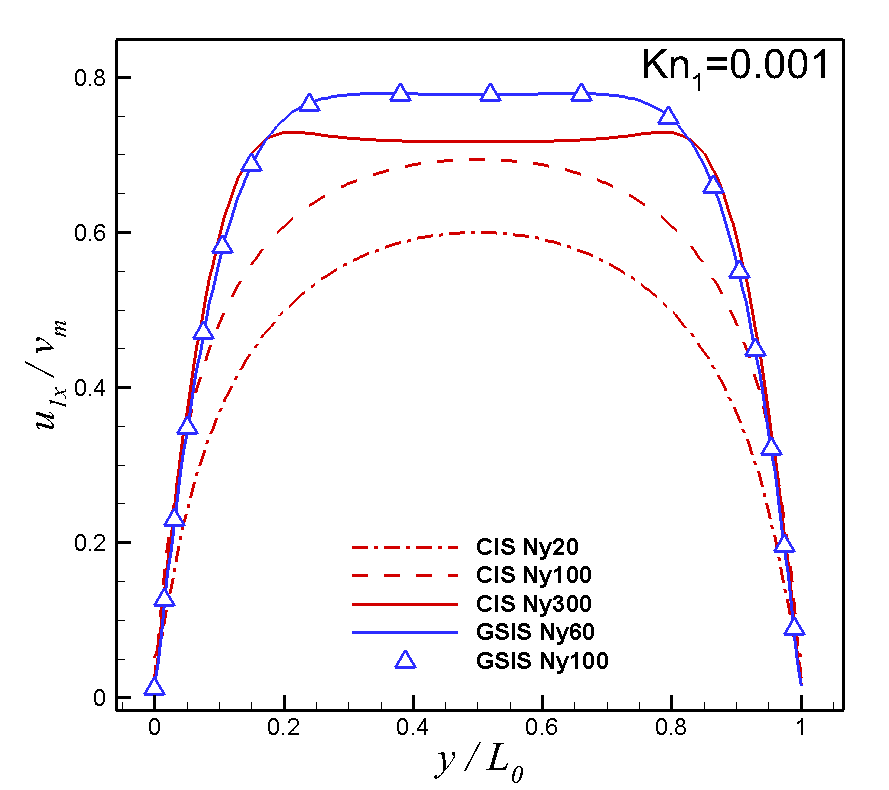}}
    \caption{Velocity profiles of the lighter gas at the channel cross-section $x=2.5 L_0$, obtained from the GSIS and CIS simulations with different spatial discretizations. Specifically, Ny100, Ny60, and Ny20 refer to simulations conducted with 100, 60, and 20 cells along the channel cross-section, respectively. In all mesh configurations, three layers of grids are allocated near the wall with a spacing of $\Delta y=0.01 L_0$, while the interior grids are generated with a growth rate of 1.2 from the wall to the channel center.}
    \label{fig:AP}
\end{figure}

To test the asymptotic-preserving property of the GSIS, that is, to see whether or not the GSIS requires very small number of spatial cells in the (near) continuum flow regime, we simulate the channel flow when $\text{Kn}_1 = 10^{-2}$ and $\text{Kn}_1= 10^{-3}$. This channel Poiseuille flow is a very good test case to validate the asymptotic-preserving property, since the velocity profile is sensitive to the numerical dissipation in the $y$ direction. Therefore, the  Cartesian grids are used to discretize the spatial domain, and we vary the number of cells (i.e., $N_y$) in the $y$ direction. From Fig.~\ref{fig:AP} we see that, when $\text{Kn}_1=0.01$, the CIS with $N_y=20$ has huge numerical dissipation, so that the computed velocity profile is significantly lower than the true solution; only when $N_y=100$ can the CIS get the correct solution. In contrast, the GSIS with $N_y=20$ finds the correct solution. When the Knudsen number is reduced to $\text{Kn}_1=0.001$, the numerical dissipation becomes more severe, and the CIS with $N_y=300$ cannot even get the right solution. However, the GSIS with $N_y=60$ is adequate to find the converged solution.

\section{Conclusion}\label{sec:5}

The GSIS has been successfully developed to simulate the binary gas flow in all flow regimes, while the Li model is used in the implicit finite-volume scheme. Benefiting from the coupled iteration of the macroscopic and mesoscopic equations, consistent accuracy have been observed in four challenging numerical cases. Moreover, compared to the conventional iteration scheme, the efficiency of GSIS can be increased significantly for steady-state solution, even for problems with high concentration ratio and high mass ratio. 
In fact, such a fast convergence has been rigorously proven by the Fourier stability analysis, where the error decay rate can be controlled to be smaller than 0.5 over a wide range of the Knudsen number, mass ratio, concentration ratio, and the viscosity ratio, such that the error can be reduced by three orders of magnitude after 10 iterations. Indeed, in the numerical simulation of nonlinear flows, the GSIS can find the converged solutions after dozens of iterations. Finally, it is worth to mention that the GSIS has the asymptotic-preserving property, where the spatial cell size can be much larger than the mean free path in the (near) continuum flow regime, demonstrating a much less numerical dissipation that the CIS. Combining the fast-converging and asymptotic-preserving properties, GSIS can be much efficient and accurate than CIS.

%Actually, the proposed GSIS for binary mixtures in this work can naturally be extended to the multi-species rarefied flows, providing an important foundation for understanding chemically reacting non-equilibrium flows. GSIS may have potential application in simulating re-entry problems, as chemical reactions in high-speed environments cannot be ignored.
%Since the collision terms employed in this paper involving auxiliary macroscopic quantities, the velocity distribution functions for binary mixtures deviate further from equilibrium, compared to single-species models. Consequently, more discrete velocity points are required to accurately describe the velocity distribution functions. In future work, we will employ velocity space adaptive techniques to reduce memory overhead.

We emphasis that the simulation of rarefied gas mixtures with disparate mass ratio is a huge challenge. For instances,  in the simulation of plasma generated in a rarefied hypersonic shock layer~\cite{Farbar2010}, the mass of the electron is artificially increased by three orders of magnitude to yield a ion-electron mass ratio of about 25 to increase the simulation efficiency. A similar mass ratio is chosen in the simulation of two-dimensional   Orszag-Tang vortex and magnetic reconnection via the unified gas kinetic scheme~\cite{Liu2017CiCP}. The GSIS developed here may have the potential applications in efficiently and accurately simulating the disparate mass mixture flows without the mass rescaling.
Also, we believe the current methodology can be extended straightforwardly to simulate time-dependent multi-species rarefied gas flows~\cite{zeng2023CICP}.

\section*{Acknowledgments}

This work is supported by the National Natural Science Foundation of China (12172162, 12202177), as well as the ``Climbing Program''  for Scientific and Technological Innovation in Guangdong (pdjh2024c10701). Special thanks are given to the Center for Computational Science and Engineering at the Southern University of Science and Technology.

\appendix

\section{Implicit treatment of source terms in macroscopic equations}\label{apdex:source_term}

Taking the two-dimensional two-species macroscopic equation \eqref{eq:macro_equation} as an example, the conservative variables of each species can be obtained as:
\begin{equation}\label{macro_var_define}
\begin{aligned}
\bm{W}&=
\begin{bmatrix}
\rho_s \\
\rho_s u_{s,x} \\
\rho_s u_{s,y} \\
E_s
\end{bmatrix},\quad 
\bm{F}_c=
\begin{bmatrix}
\rho_s u_n\\
\rho_s u_{s,x}u_{s,n}+n_xp_{s}\\
\rho_s u_{s,y}u_{s,n}+n_yp_{s}\\
u_{s,n} (E_{s}+p_{s})
\end{bmatrix}, \quad 
\bm{F}_v=
\begin{bmatrix}
0\\
n_x\sigma_{s,xx}+n_y\sigma_{s,xy}\\
n_x\sigma_{s,yx}+n_y\sigma_{s,yy}\\
n_x\Theta_{s,x}+n_y\Theta_{s,y}
\end{bmatrix},\\
\bm{Q}&=
\left[
0,
{\frac{\rho_s(u_{sr,x}-u_{sr,x})}{\tau_{sr}}},
{\frac{\rho_s(u_{sr,y}-u_{sr,y})}{\tau_{sr}}},
{\frac{E_{sr}-E_{s}}{\tau_{sr}}}
\right]^{\intercal},
\end{aligned}
\end{equation}
with 
$\Theta_{s,x} =u_{s,x}\sigma_{s,xx}+u_{s,y}\sigma_{s,xy}+q_{s,x}$ and
$\Theta_{s,y} =u_{s,x}\sigma_{s,yx}+u_{s,y}\sigma_{s,yy}+q_{s,y}$. 
Here, $u_n = u_{s,x} n_x + u_{s,y} n_y$ is defined as the scalar product of the macro velocity vector and the unit normal vector of the face. The total energy in the macroscopic system is obtained by adding its internal energy to its kinetic energy. Thus we can define those speciess as follows,
\begin{equation}
    E_{s}=\frac{3}{2}n_s T_{s}+\frac{1}{2}\rho_s u_s^2,\quad
    E_{sr}=\frac{3}{2}n_s T_{sr}+\frac{1}{2}\rho_s u_{sr}^2.
\end{equation}
we provide an approximation procedure for the source term Jacobi matrix in the Cartesian coordinate system. The Jacobi matrix $T$ represents the derivative of the source term $Q$. The matrix can be further approximated by only keeping its main diagonal elements, which leads to diagonalization of the matrix as follows:
\begin{equation}
\mathbf{T}_s = \frac{\partial \mathbf{Q}_s}{\partial \mathbf{W}_s} = 
\left[
\begin{matrix}
0 & 0 & 0 & 0  \\
\frac{\partial Q_s(2)}{\partial \rho_s} & \frac{\partial Q_s(2)}{\partial \rho_s u_s} & \frac{\partial Q_s(2)}{\partial \rho_s v_s} & \frac{\partial Q_s(2)}{\partial E_s}  \\
\frac{\partial Q_s(3)}{\partial \rho_s} & \frac{\partial Q_s(3)}{\partial \rho_s u_s} & \frac{\partial Q_s(3)}{\partial \rho_s v_s} & \frac{\partial Q_s(3)}{\partial E_s}  \\
\frac{\partial Q_s(4)}{\partial \rho_s} & \frac{\partial Q_s(4)}{\partial \rho_s u_s} & \frac{\partial Q_s(4)}{\partial \rho_s v_s} & \frac{\partial Q_s(4)}{\partial E_s}
\end{matrix}
\right],
% \approx
% \left[
% \begin{matrix}
% 0 & 0 & 0 & 0  \\
% 0 & \frac{\partial Q_s(2)}{\partial \rho_s u_{s,x}} & 0 & 0  \\
% 0 & 0 & \frac{\partial Q_s(3)}{\partial \rho_s u_{s,y}} & 0  \\
% 0 & 0 & 0 & \frac{\partial Q_s(4)}{\partial E_s}
% \end{matrix}
% \right],
\end{equation}
where
\begin{equation}
\begin{aligned}[b]
\frac{\partial Q_s(2)}{\partial \rho_s u_{s,x}}&=\frac{\partial }{\partial \left(\rho_s u_s\right)} \left(\frac{\rho_s(u_{sr} - u_s)}{\tau_{sr}}\right)\\
&=\frac{1}{\tau_{sr}}\frac{\partial}{\partial \left(\rho_s u_s\right)}\left(-\frac{\rho_s\rho_r\tau_{sr}}{\rho_s\tau_{rs}+\rho_r\tau_{sr}}a(u_s - u_r)-b{\nabla \ln T}\right)\\
% &\approx-\frac{\rho_r}{\rho_s\tau_{rs}+\rho_r\tau_{sr}}a,
&=-\frac{\rho_r}{\rho_s\tau_{rs}+\rho_r\tau_{sr}}a,
\end{aligned}
\end{equation}
\begin{equation}
\begin{aligned}[b]
\frac{\partial Q_s(3)}{\partial \rho_s u_{s,y}}&=\frac{\partial }{\partial \left(\rho_s v_s\right)}\left(\frac{\rho_s(v_{sr} - v_s)}{\tau_{sr}}\right)\\
&=\frac{1}{\tau_{sr}}\frac{\partial}{\partial \left(\rho_s v_s\right)}\left(-\frac{\rho_s\rho_r\tau_{sr}}{\rho_s\tau_{rs}+\rho_r\tau_{sr}}a(v_s - v_r)-b{\nabla \ln T}\right)\\
% &\approx-\frac{\rho_r}{\rho_s\tau_{rs}+\rho_r\tau_{sr}}a,
&=-\frac{\rho_r}{\rho_s\tau_{rs}+\rho_r\tau_{sr}}a,
\end{aligned}
\end{equation}
\begin{equation}
\begin{aligned}[b]
\frac{\partial Q_s(4)}{\partial E_s}&=\frac{\partial }{\partial E_s}\left(\frac{E_{sr} - E_s}{\tau_{sr}}\right)\\
&=\frac{1}{\tau_{sr}}\frac{\partial}{\partial  E_s}\left[\frac{3}{2}n_s(T_{sr}-T_s)+\frac{\rho_s}{2}(u_{sr}^2-u_s^2)\right] \\
&=\frac{1}{\tau_{sr}}\frac{\partial}{\partial  E_s}\left[\frac{3}{2}n_s\frac{n_r\tau_{sr}}{n_s\tau_{rs}+n_r\tau_{sr}}c(T_s-T_r)+\frac{\rho_s}{2}(-2u_s\frac{\rho_r\tau_{sr}}{\rho_s\tau_{rs}+\rho_r\tau_{sr}}a(u_s-u_r))\right] \\
&=\frac{1}{\tau_{sr}}\frac{\partial}{\partial  E_s}\left[\frac{3}{2}n_s\frac{n_r\tau_{sr}}{n_s\tau_{rs}+n_r\tau_{sr}}c(T_s-T_r)-{\rho_s}u_s^2\frac{\rho_r\tau_{sr}}{\rho_s\tau_{rs}+\rho_r\tau_{sr}}a\right]\\
% &\approx-\frac{n_r}{n_s\tau_{rs} + n_r\tau_{sr}}c - \frac{2\rho_r}{\rho_s\tau_{rs}+\rho_r\tau_{sr}}a.
&=-\frac{n_r}{n_s\tau_{rs} + n_r\tau_{sr}}c - \frac{2\rho_r}{\rho_s\tau_{rs}+\rho_r\tau_{sr}}a.
\end{aligned}
\end{equation}
Note that in the derivation of the above three equations, the following relations are used:
\begin{equation}
\begin{aligned}
T &= \frac{2E-\rho u^2}{3n},\quad
\frac{\partial T}{\partial E}=-\frac{2}{3n},\\
u^2 &=\frac{2E}{\rho} - \frac{3T}{m}, \quad
\frac{\partial u^2}{\partial E} = \frac{2}{\rho}, \quad \frac{\partial u}{\partial E} = \frac{2/\rho}{\sqrt{\frac{2E}{\rho}-\frac{3T}{m}}},\\
\tau &= \frac{\mu}{p}=\frac{T^{\omega-1}}{n},\quad\frac{\partial \tau}{\partial T}=(\omega-1)\frac{T^{\omega-2}}{n}.
\end{aligned}
\end{equation}

%The approximate Jacobian matrix indicates that all off-diagonal elements are zero.

\section{Dimensional reduction}\label{sec:reduction}

In two-dimensional problems, the velocity space can be reduced from three dimensions to two dimensions to save computational costs. To achieve this, reduced velocity distribution functions, denoted as $\hat{f}_s$ and $\hat{f}_{s,z}$, are introduced:
\begin{equation}
        \left(\hat{f}_s, \hat{f}_{s,z}\right) = \int \left(f_s, \xi_z^2 f_s\right) \myd\xi_z,
\end{equation}
and the dimensionless macroscopic quantities can be calculated based on these reduced velocity distribution functions:
\begin{equation}
    \begin{aligned}
    \left(n_s, \rho_s,\rho_s\bm{u}_s\right)&=\int\left(\frac{1}{m_s},1,\bm{\xi}_s\right) \hat f_s \mathrm{d}\bm{\xi}_s,\quad
\bm{\sigma}_s=\int\left(\bm{c}_s\bm{c}_s\hat f_s-\frac{1}{3}\left(c_s^2\hat f_s+\hat f_{s,z}\right)\mathrm{I}\right) \mathrm{d}\bm{\xi}_s,\\
\frac{3}{2}n_s T_s&=\int\frac{1}{2}(c_s^2\hat f_s+\hat f_{s,z})\myd \bxi_s,
\quad \bq_s=\int\bc_s \frac{1}{2}(c_s^2\hat f_s+\hat f_{s,z})\myd \bxi_s.
\end{aligned}
\end{equation}
where all vectors and tensors $\bm{\xi},~\bm{u}_s,~\bc_s = \bxi - \bu_s,~\bm{q}_s,~\bm{\sigma}_s$ are in two-dimensional space. 
The dimensionless model equations of the reduced velocity distribution functions can be written as:
\begin{equation}
    \begin{aligned}[b]
    \frac{\partial \hat f_s}{\partial t}+\bm{\xi}\cdot \frac{\partial \hat f_s}{\partial \bm{x}} &= \sum_{r} \frac{g_{sr}-\hat f_s}{\tau_{sr}}, \\
    \frac{\partial \hat f_{s,z}}{\partial t}+\bm{\xi}\cdot \frac{\partial \hat f_{s,z}}{\partial \bm{x}} &= \sum_{r} \frac{g_{sr,z}-\hat f_{s,z}}{\tau_{sr}}, \\
    \end{aligned}
\end{equation}
where the reduced reference velocity distribution functions are given by:
\begin{equation}\label{eq:reference_velocity distribution function_reduced}
    \begin{aligned}
g_{sr}=&m_s\hat{n}_{sr}\left(\frac{m_s}{2\pi T_{s}}\right)\exp\left(-\frac{m_s\left(\bm{\xi}-\hat{\bm{u}}_{sr}\right)^2}{2T_{s}}\right) \\
&\times\left[1+\frac{\hat T_{sr} - T_s}{T_s}\left(\frac{m_s\left(\bm{\xi}-\hat{\bm{u}}_{sr}\right)^2}{2 T_{s}}-1\right)+\frac{2m_s\bm{\hat q}_{sr} \cdot \left(\bm{\xi}-\hat{\bm{u}}_{sr}\right)}{5 \hat n_{sr}T_{s}^2}\left(\frac{m_s \left(\bm{\xi}-\hat{\bm{u}}_{sr}\right)^2}{2 T_{s}}-2\right)\right],\\
g_{sr,z}=&\hat{n}_{sr}T_{s}\left(\frac{m_s}{2\pi T_{s}}\right)\exp\left(-\frac{m_s\left(\bm{\xi}-\hat{\bm{u}}_{sr}\right)^2}{2T_{s}}\right) \\
&\times\left[1+\frac{\hat T_{sr} - T_s}{T_s}\left(\frac{m_s\left(\bm{\xi}-\hat{\bm{u}}_{sr}\right)^2}{2 T_{s}}\right)+\frac{2m_s\bm{\hat q}_{sr} \cdot \left(\bm{\xi}-\hat{\bm{u}}_{sr}\right)}{5 \hat n_{sr}T_{s}^2}\left(\frac{m_s \left(\bm{\xi}-\hat{\bm{u}}_{sr}\right)^2}{2 T_{s}}-1\right)\right].
% g_{sr}&=\rho_{s}\left(\frac{m_s}{2\pi T_{sr}}\right)\exp\left(-\frac{m_sc_{sr}^2}{2 T_{sr}}\right)\left[1+\frac{2m_s\bm{q}_{sr} \cdot \bm{c}_{sr}}{5 \rho_{s}T_{sr}^2}\left(\frac{m_sc_{sr}^2}{2 T_{sr}}-2\right)\right],\\
% g_{sr,z} &=n_{s}T_s\left(\frac{m_s}{2\pi T_{sr}}\right)\exp\left(-\frac{m_sc_{sr}^2}{2 T_{sr}}\right)\left[1+\frac{2m_s\bm{q}_{sr} \cdot \bm{c}_{sr}}{5 \rho_{s}T_{sr}^2}\left(\frac{m_sc_{sr}^2}{2 T_{sr}}-1\right)\right].
    \end{aligned}
\end{equation}

\bibliographystyle{elsarticle-num}
\bibliography{ref}
\end{document}